\documentclass[twocolumn, linenumbers]{aastex631}

\usepackage{natbib}           
\usepackage{graphicx}	      
\usepackage{amsmath}	      
\usepackage{amssymb}	      
\usepackage{multirow}

\usepackage{booktabs}
\usepackage{threeparttable}
\usepackage{placeins}
\usepackage{tabularx}
\usepackage{siunitx}
\usepackage{float}
\usepackage{ragged2e}

\newcommand{\mum}{\ifmmode{\rm \mu m}\else{$\mu$m }\fi}             

\newcommand{\chisq}{\ifmmode{\chi^{2} }\else{$\chi^2$}\fi}
\newcommand{\rchisq}{\ifmmode{\chi^{2} }\else{$\chi^2_\nu$}\fi}

\received{July 30, 2025}
\accepted{September 24, 2025}

\shorttitle{JWST NGC 346 MIRI MRS}   
\graphicspath{{./}{figures/}}              

\begin{document}
\nolinenumbers

\title{A Mid-Infrared Spectroscopic Study of Young Stellar Objects in the SMC Region NGC 346: JWST Detects Dust, Accretion, Ices and Outflows}

\author[0000-0002-2667-1676]{Nolan Habel}
\affil{Jet Propulsion Laboratory, California Institute of Technology, 4800 Oak Grove Dr., Pasadena, CA 91109, USA}
\correspondingauthor{Nolan Habel}
\email{nolan.m.habel@jpl.nasa.gov}

\author[0000-0001-6576-6339]{Omnarayani Nayak}
\affil{NASA Goddard Space Flight Center, 8800 Greenbelt Road, Greenbelt, MD, USA}

\author[0000-0001-6872-2358]{Patrick \ J.\ Kavanagh}
\affil{Department of Experimental Physics, Maynooth University, Maynooth, Co. Kildare, Ireland}

\author[0000-0003-4870-5547]{Olivia C.\ Jones}
\affil{UK Astronomy Technology Centre, Royal Observatory, Blackford Hill, Edinburgh, EH9 3HJ, UK}

\author[0000-0002-0522-3743]{Margaret Meixner}
\affil{Jet Propulsion Laboratory, California Institute of Technology, 4800 Oak Grove Dr., Pasadena, CA 91109, USA}

\author[0000-0001-7906-3829]{Guido De Marchi}
\affil{European Space Research and Technology Centre, Keplerlaan 1, 2200 AG Noordwijk, The Netherlands}

\author[0000-0003-4023-8657]{Laura Lenki\'{c}}
\affil{IPAC, California Institute of Technology, 1200 East California Boulevard, Pasadena, CA 91125, USA}
\affil{Jet Propulsion Laboratory, California Institute of Technology, 4800 Oak Grove Dr., Pasadena, CA 91109, USA}

\author[0000-0002-2954-8622]{Alec S.\ Hirschauer}
\affil{Department of Physics \& Engineering Physics, Morgan State University, 1700 East Cold Spring Lane, Baltimore, MD 21251, USA}



\author[0000-0002-1892-2180]{Katia Biazzo}
\affil{1INAF, Astronomical Observatory of Rome, Via Frascati 33, Monteporzio Catone, I-00078, Italy}

\author[0000-0002-0577-1950]{J. Jaspers}
\affil{Dublin Institute for Advanced Studies, School of Cosmic Physics, Astronomy \& Astrophysics Sec. 31 Fitzwilliam Place, Dublin 2, Ireland}
\affil{Department of Experimental Physics, Maynooth University, Maynooth, Co. Kildare, Ireland}

\author[0000-0002-7512-1662]{Conor Nally}
\affil{Institute for Astronomy, University of Edinburgh, Royal Observatory, Blackford Hill, Edinburgh EH9 3HJ, UK}

\author[0000-0002-9573-3199]{Massimo Robberto}
\affil{Space Telescope Science Institute, 3700 San Martin Drive, Baltimore, MD 21218, USA}
\affil{Department of Physics \& Astronomy, Johns Hopkins University, 3400 N.\ Charles St., Baltimore, MD 21218, USA}

\author[0000-0001-5742-2261]{C. Rogers}
\affil{Leiden Observatory, Leiden University,
PO Box 9513, 2300 RA Leiden, The Netherlands}

\author[0000-0003-2954-7643]{E.\ Sabbi}
\affil{Space Telescope Science Institute, 3700 San Martin Drive, Baltimore, MD 21218, USA}

\author[0000-0001-9855-8261]{B.\ A.\ Sargent}
\affil{Space Telescope Science Institute, 3700 San Martin Drive, Baltimore, MD 21218, USA}
\affil{Department of Physics \& Astronomy, Johns Hopkins University, 3400 N.\ Charles St., Baltimore, MD 21218, USA}

\author[0000-0002-6091-7924]{Peter\ Zeidler} 
\affil{AURA for the European Space Agency,
Space Telescope Science Institute, 3700 San Martin Drive, Baltimore, MD 21218, USA}



\begin{abstract}
\justifying We present mid-infrared spectroscopic observations of intermediate- to high-mass young stellar objects (YSOs) in the low-metallicity star-forming region NGC 346 located within the Small Magellanic Cloud (SMC). We conduct these integral-field-unit observations with the Mid-Infrared Instrument~/
Medium Resolution Spectroscopy instrument on board JWST. The brightest and most active star-forming region in the SMC, NGC 346 has a metallicity of $\sim$1/5 $Z_{\odot}$, analogous to the era when star formation in the early Universe ($z$~$\simeq$~2) peaked. 
We discuss the emission and absorption features present in the spectral energy distributions (SEDs) of five YSOs with coverage from 4.9~-~27.9~$\mu$m and three other sources with partial spectral coverage. Via SED model-fitting, we estimate their parameters, finding masses ranging from 2.9-18.0 M$_{\odot}$.
These targets show dusty silicates, polycyclic aromatic hydrocarbons and ices of CO$_2$, CO, H$_2$O and CH$_3$OH in their protostellar envelopes. We measure emission from H$_2$ and atomic fine-structure lines, suggesting the presence of protostellar jets and outflows. We detect \ion{H}{1} lines indicating ongoing accretion and estimate accretion rates for each source which range from \num{2.50E-06}-\num{2.23E-04} M$_{\odot}~$yr$^{-1}$ based on \ion{H}{1}~(7-6) line emission. We present evidence for a $\sim$~30,000AU protostellar jet traced by fine-structure, \ion{H}{1} and H$_2$ emission about the YSO Y535, the first detection of a resolved protostellar outflow in the SMC, and the most distant yet detected.
\end{abstract}



\section{Introduction} 
\label{sec:intro}

Located in the Small Magellanic Cloud (SMC) at a distance of $\sim$62 kpc \citep{bib:deGrijs2015} is NGC 346, the brightest and most active star-forming region in the nearby dwarf galaxy. Also known as N66 and DEM 103 \citep{bib:Henize1956}, this region situated at the northern end of the SMC bar produces stars at a rate of $\sim$\num{4e-3}~M$_{\odot}\,$yr$^{-1}$ (\citealt{bib:Simon2007, bib:hony2015}), accounting for nearly one tenth of the total star formation rate of the SMC ($\sim$0.05~M$_{\odot}$\,yr$^{-1}$, \citealt{bib:wilke2004}). In the last 100 Myr, NGC 346 has birthed stars at a rate of \num{1.4E-8} M$_{\odot}~$yr$^{-1}$~pc$^{-2}$ \citep{bib:Cignoni2011}. The region is bright in H$\alpha$ emission, with a luminosity 60 times that of Orion \citep{bib:Kennicutt1984}. The \ion{H}{2} region within NGC 346 is relatively young ($\sim$3~Myr; \citealt{bib:Bouret2003}) and houses upwards of 30 O-type stars of masses 35-100 M$_{\odot}$ \citep{bib:Dufton2019}, and accounts for over half of those known in the SMC \citep{bib:Massey1989,bib:Evans2006}.

With a metallicity of  $\sim$1/5 Z$_{\odot}$ \citep{bib:Peimbert2000}, the interstellar medium  (ISM) of NGC 346 differs substantially from that of the Milky Way.
With such a low metallicity, NGC 346 resembles the star-forming environments that existed in galaxies in the early Universe, when star formation was at its peak ($z\sim2$; \citealt{bib:Dimaratos2015}). With its close proximity to the Milky Way within the Local Group, NGC 346 is an important proxy for galaxies of this early era, also called cosmic noon. Importantly, it offers the ability to probe low-metallicity star formation with subparsec resolution, a necessity for understanding formation at the scale of individual stars.

Previous ultraviolet (UV), optical and infrared (IR) imaging surveys have revealed the young population within NGC 346. In the UV and optical, Hubble Space Telescope (\textit{HST}) imaging has discovered thousands of low-mass pre-main-sequence (PMS) candidates, including low-mass (0.6-3 M$_{\odot}$) sources \citep{bib:Nota2006, bib:Sabbi2007, bib:Hennekemper2008, bib:DeMarchi2011}. IR studies of the region have sought to further characterize the youngest objects: young stellar objects (YSOs). Formed from the gravitational collapse of the surrounding cloud, these young objects are still enshrouded in gas and dust, the reservoir from which they draw material during their main initial accretion phase \citep{bib:mcke07}. This envelope reprocesses the UV radiation emitted from the central source by absorbing it and re-radiating it in the mid- and far-IR wavelengths \cite{bib:chur02}.   Approximately 100 candidate YSOs were identified in Spitzer and Herschel imaging surveys of NGC 346 \citep{bib:Bolatto2007, bib:Gordon2011, bib:meix13}. These candidates are estimated to have formed in the last $\sim$1 Myr, possess masses ranging from 1.5-17~M$_{\odot}$ and span a range of evolutionary states from stage I, II and III \citep{bib:Simon2007, bib:sewilo2013, bib:Seale2014, bib:Ruffle2015}.

Recently, the implementation of JWST has enabled new IR studies of NGC 346 which reveal the individual YSOs in unprecedented detail and number. Near-IR imaging with JWST's Near Infrared Camera (NIRCam) has revealed $\sim$ 500 new PMS and YSO candidates down to subsolar masses \citep{bib:jones2023, bib:habel24}. With the imaging mode of the new telescope's Mid-Infrared Instrument (MIRI; \citealt{bib:riek15, bib:wrig23}), \cite{bib:habel24} cataloged a population of $\sim$800 of the youngest, most-reddened YSO candidates, and from their photometry, inferred masses with a high confidence down to 1~M$_{\odot}$. Further efforts to characterize the stellar population in this region from this same dataset will continue to expand the number of candidate and confirmed PMS stars and YSOs, as well as extend the mass range of these detections further below subsolar (Jaspers et al. in prep.)

\begin{figure}
\centering
\includegraphics[width=\columnwidth]{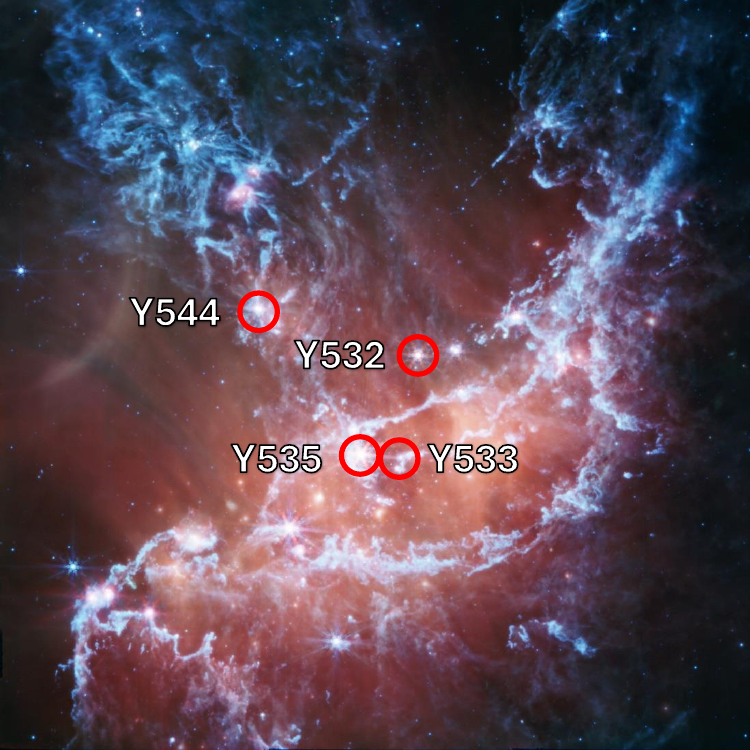}
 \caption{JWST multicolor MIRI image of NGC 346 combining the F770W (blue), F1000W (cyan), F1130W (green), F1500W (yellow), and F2100W (red) filters of NGC 346. The locations of the four MIRI/MRS pointings are marked with red circles. NASA, ESA, CSA, STScI, N. Habel (JPL, Caltech). Image processing: Alyssa Pagan (STScI) and Patrick Kavanagh (Maynooth University).}  
 \label{fig:observations}
 \end{figure}

While the basic picture of YSO formation is straightforward, the details of their formation are influenced by a range of factors, from their evolved and forming neighbors, to the metallicity of their environment to their mass regime. Purely photometric studies are limited in their ability to answer these questions. As a result, comprehensive spectroscopic studies in the IR are key tools for forming a complete and accurate picture of these objects and their evolution. IR spectra of YSOs, and the emission and absorption lines within, can reveal their mass, evolutionary state and accretion phenomena, as well as probe the conditions in the surrounding interstellar medium (ISM), such as the temperature and ionization conditions \citep{bib:boom03.2, bib:oliveira2009,bib:seal09,bib:rigl15}.

Only a few dozen YSOs in the SMC have been confirmed via spectroscopic observation previous to JWST \citep{bib:oliveira2011,bib:oliveira2013,bib:oliveira2019, bib:Ruffle2015, bib:ward2017, bib:Rubio2018, bib:reit19, bib:jones2022}. NGC 346 contains roughly 15 such sources as confirmed by Spitzer's Infrared Spectrograph (IRS), and the Very Large Telescope's (VLT) Infrared Spectrometer And Array Camera (ISAAC) and K-Band Multi-Object Spectrograph (KMOS) instruments \citep{bib:Ruffle2015, bib:Rubio2018, bib:jones2022}. 
The most detailed past spectroscopic surveys in the Magelanic Clouds are limited to the near-IR \citep{bib:ward2016, bib:ward2017, bib:Rubio2018, bib:reit19, bib:vanG2020, bib:sewilo2022, bib:jones2022}.
YSO spectra from these observations are rich in features such as molecular hydrogen (H$_2$), recombination lines (\ion{H}{1}) and fine-structure atomic lines. 
Integral-field spectroscopy in the near-IR in the SMC and LMC have examined the relationship between morphological and spectral emission features, including extended H$_2$ emission suggestive of outflows in early-stage YSOs \citep{bib:ward2016, bib:ward2017}.

In this work, we examine the mid-IR spectra of several early-stage YSOs in NGC 346 using the Medium Resolution Spectroscopy (MRS) mode of the MIRI instrument. As with the near-IR, the mid-IR is host to numerous potential atomic and molecular emission lines, but also encompasses unique polycyclic aromatic hydrocarbon (PAH) emission, silicate features and molecular ice absorption.  H$_2$ emission in the mid-IR is excited by UV radiation or collisional excitation from shock heating in the molecular gas \citep{bib:tiel93, bib:holl97}. UV radiation also impacts PAH molecules, both collisionally exciting their stretching and bending modes, but also breaking down larger molecules into smaller ones \citep{bib:tiel93, bib:peet17}. Accretion rates can be inferred from the strength of mid-IR \ion{H}{1} lines, which is linked to accretion luminosity in YSOs \citep{bib:Calvet_2004, bib:Herczeg_2008}. Recently, \cite{bib:nayak2024} used MIRI MRS to infer accretion rates between \num{1.44e-4}-\num{1.89e-2} M$_{\odot}$yr$^{-1}$ for four YSOs in the Large Magelanic Cloud (LMC) region N79 based on their \ion{H}{1} (7-6) line luminosities. The same observations also revealed silicate and ice absorption, characteristic of mid-IR YSO spectra.

In this work, we present the first mid-IR spectroscopic study of intermediate- to high-mass YSOs in NGC 346 with JWST. The objective of this work is to use these data to characterize their evolutionary state via their mid-IR continua and spectral features, and to detect and measure molecular ices, dusts and protostellar outflows.
Our target selections and observation setup are described in Section~\ref{sec:observations}. Section~\ref{sec:data} details the data reduction. In section~\ref{sec:results} we describe the IFU cubes, the extracted YSO spectra and the spectral features we detect. In section~\ref{sec:analysis}, we derive the physical parameters of our targets based on their SEDs, describe evidence for an extended protostellar jet, discuss the detection of molecular ices, investigate the dust content in our targets and derive their accretion properties. Finally, in section~\ref{sec:conclusion} we summarize our conclusions. 


\section{Source Selection and Observations}
\label{sec:observations}
\subsection{Selection Criteria}
\label{sec:source_selection}

Several factors influenced our choice of targets and our observing scheme. Overall, the choices were made to fulfill the objectives of GTO program \#1227 (PI: Meixner), which was designed to use a host of JWST's instruments and abilities to characterize star formation in the SMC region NGC 364 over multiple epochs. This program builds upon previous IR studies of the region, including the Spitzer SAGE and Herschel HERITAGE surveys (e.g., \citealt{bib:Blum2006}; \citealt{bib:meix06}; \citealt{bib:whit08};  \citealt{bib:grue09}; \citealt{bib:sewilo2013}; \citealt{bib:Carlson2012}; \citealt{bib:Seale2014}) which revealed the existence of active star formation. 
This program was intentionally separated over two JWST observing cycles. The first epoch conducted imaging of the star-forming region using a combined total of 11 filters from 1~$\mu$m to 21~$\mu$m using the Near Infrared Camera (NIRCam) and MIRI instruments (\citealt{bib:jones2023}, \citealt{bib:habel24}) and spectroscopic observations with the Near Infrared Spectrograph (NIRSpec)'s Multi-Object Spectroscopy (MOS) mode of pre-main sequence (PMS) sources \citep{bib:demarchi24_pms} identified in \cite{bib:Sabbi2007} and \cite{bib:DeMarchi2011} with \textit{HST}. The second epoch was dedicated to follow-up spectroscopic observations of embedded, lower-mass YSO sources identified or confirmed by resolved imaging and photometric analysis of the first epoch's data.
The second epoch of NIRSpec's MOS data of $\sim$15 embedded YSOs is published in De Marchi et al.~submitted. In this work, we focus the four YSO fields obtained with the MRS in the second epoch.

The observing time earmarked for the MRS observations was $\sim$6 hrs. 
Candidate YSOs considered for these MRS observations were drawn from several sources. \cite{bib:habel24} cataloged 833 YSO candidates in NGC 346 using a combination of NIRCam and MIRI photometry. 
From this, a subset of sources showing detections across both NIRCam and MIRI and an SED rising toward the mid-IR was identified. This candidate set was subjected to the same YSO fitting routine \citep{bib:robitaille06, bib:robi17} as described in \cite{bib:habel24} to reassess their nature as YSOs and to derive estimates of their bolometric temperature (T$_{bol}$), stellar radius, luminosity, and mass. Finally, each source was visually investigated and excluded if nearby ($\sim$0.25") sources were found which could confuse or contaminate the MRS observations. These final candidates were then ordered by their mid-IR luminosity. 
The brightest of these sources was selected as a target. The remaining candidates required substantially more exposure time to achieve a signal-to-noise ratio $>10$, than what was allocated to this second epoch of observations.

We considered YSO candidates previously identified with Spitzer, including several bright sources that appear as saturated in JWST NIRCam and MIRI imaging. Using a combination of \textit{Spitzer} IRAC (3.6-8.0~$\mu$m) and MIPS (24 and 70~$\mu$m) catalogs from the SAGE-SMC survey \citep{bib:meix06, bib:whit08, bib:Blum2006}, \cite{bib:sewilo2013} identified $\sim$1000 intermediate to high-mass YSOs, including 26 in NGC 346.
We considered only those sources that showed a relatively uncrowded field surrounding them within the footprint of an MRS pointing. We chose three additional sources to observe, the maximum number possible based on the remaining allocated observation time. These threes all have complimentary ground-based  near-infrared spectroscopy \citep{bib:Rubio2018, bib:jones2022}. We consulted these existing observations to confirm their nature as YSOs. 

\subsection{Description of Targets}
\label{sec:targets}
Here we summarize our four MRS observations and the sources observed therein. 
We introduce the observation numbers used in GTO program \#1227 for these observations to differentiate them and adopt the names introduced for these sources by \cite{bib:sewilo2013}, who positively identified them as YSOs based on Spitzer photometry. In  \autoref{tab:photom_arhival} and \autoref{tab:Y533_photom}, we list their coordinates and near- and mid-IR photometry. In \autoref{fig:observations} we show the location of our observations in NGC 346.

\begin{table}
\setlength{\tabcolsep}{1.2pt}
    \centering
    \begin{tabular}{||l|ccc||}
    \hline
            Source ID  & \textbf{Y535}  & \textbf{Y544A, B} & \textbf{Y532}  \\
    \hline
    \hline
    \multicolumn{4}{|c|}{Coordinates (J2000)}\\
    \hline
    RA   & 14.77241  & 14.80079/14.8013 & 14.76073 \\
    Dec  & -72.17649 & -72.16629/-72.16624 & -72.16867 \\
    \hline
    \hline
    \multicolumn{4}{|c|}{HST ACS Photometry (Magnitudes)}\\
    \hline
    F555W     & 16.61 $\pm$ 0.01  & 16.72 $\pm$ 0.01  & 18.37 $\pm$ 0.06  \\
    F814W     &  15.98 $\pm$ 0.01 & 15.97 $\pm$ 0.01 & 17.65 $\pm$ 0.02  \\
    \hline
    \hline
    \multicolumn{4}{|c|}{\textit{Spitzer} IRAC and MIPS Photometric Fluxes (mJy)}\\
    \hline
    3.6$\mu$m  & 25.6 $\pm$ 0.42 & 4.56 $\pm$ 0.45 & 3.32 $\pm$ 0.33  \\
    4.5$\mu$m  & 30.6 $\pm$ 0.36 & 5.05 $\pm$ 0.50 & 4.47 $\pm$ 0.44  \\
    5.8$\mu$m  & 37.6 $\pm$ 0.50 & 7.8 $\pm$ 0.78 & 5.92 $\pm$ 0.59 \\
    8.0$\mu$m  & 81.9 $\pm$ 0.93 & 19.0 $\pm$ 1.90 & 11.2 $\pm$ 1.12  \\
    24$\mu$m   & 2154 $\pm$ 8.48 & 284.6 $\pm$ 1.61 & 94.35 $\pm$ 9.43  \\
    70$\mu$m   & 3766 $\pm$ 51.51 & - & -  \\
    \hline
    \hline
    \multicolumn{4}{|c|}{\textit{Herschel} PACS and SPIRE Photometric Fluxes (mJy)}\\
    \hline
    100$\mu$m     & 1695 $\pm$ 225.9  & 1263.0 $\pm$ 155.5  & -  \\
    160$\mu$m     & 1724 $\pm$ 199.3  & 1204.0 $\pm$ 145.9  & -  \\
    250$\mu$m     & 822.1 $\pm$ 57.74 & 543.9 $\pm$ 40.09   & -  \\
    350$\mu$m     & 302.1 $\pm$ 48.99 & 347.3 $\pm$ 40.88   & -  \\
    \hline
    \end{tabular}
    \caption{Coordinates and archival photometry for three of our targets. Coordinates from \citealt{bib:jones2022}. \textit{HST} photometry from \citealt{bib:Sabbi2007}. \textit{Spitzer} IRAC photometric fluxes from \citealt{bib:sewilo2013}. \textit{Spitzer} MIPS fluxes from \citealt{bib:sewilo2013} and \citealt{bib:Seale2014}. \textit{Herschel} PACs and SPIRE fluxes from \citealt{bib:Seale2014}. 
    }
    \label{tab:photom_arhival}
\end{table}

\begin{table}
\setlength{\tabcolsep}{4pt}
\centering
    \begin{tabular}{||l|cccc||}
    \hline
    Source ID   & \textbf{Y533A} & \textbf{Y533B} & \textbf{Y533C} & \textbf{Y533D} \\
     \hline
     \hline
     \multicolumn{5}{|c|}{Coordinates (J2000)}\\
     \hline
     RA      &  14.76275   & 14.76339  & 14.76112  & 14.76038           \\
     Dec     &  -72.17649  & -72.17735 & -72.17622 & -72.17617        \\
     \hline
     \hline
     \multicolumn{5}{|c|}{NIRCam \& MIRI Photometry (AB Magnitudes)}\\
     \hline
     F115W   &  –     & 19.19 & 23.24 & 25.23     \\
     F187N   &  20.73 & 18.57 & 21.45 & 22.32     \\
     F200W   &  20.64 & 18.51 & 21.37 & 22.32      \\
     F277W   &  19.26 & -     & 20.80 & 21.21      \\
     F335M   &  18.86 & 18.35 & 20.50 & 20.81     \\
     F444W   &  17.64 & 18.16 & 20.07 & 19.72     \\
     F770W   &  16.82 & 17.21 & –     & 18.58      \\
     F1000W  &  16.77 & 16.86 & 19.07 & 18.61      \\
     F1130W  &  16.15 & 16.17 & 18.47 & 18.15      \\
     F1500W  &  15.54 & 16.13 & 18.55 & 17.60      \\
     F2100W  &  14.85 & -     & –     & 17.64      \\

    \hline
    \end{tabular}
    \caption{The coordinates and JWST NIRCam and MIRI photometry for five sources observed along with YSO target Y533, or observation O23. All data is from \cite{bib:habel24}. (Source Y533E, with coordinates: 14.76218, -72.17637, has only three photometric measurements and is omitted.)}
    \label{tab:Y533_photom}
\end{table}

\begin{figure*}
\centering
\includegraphics[width=\textwidth]{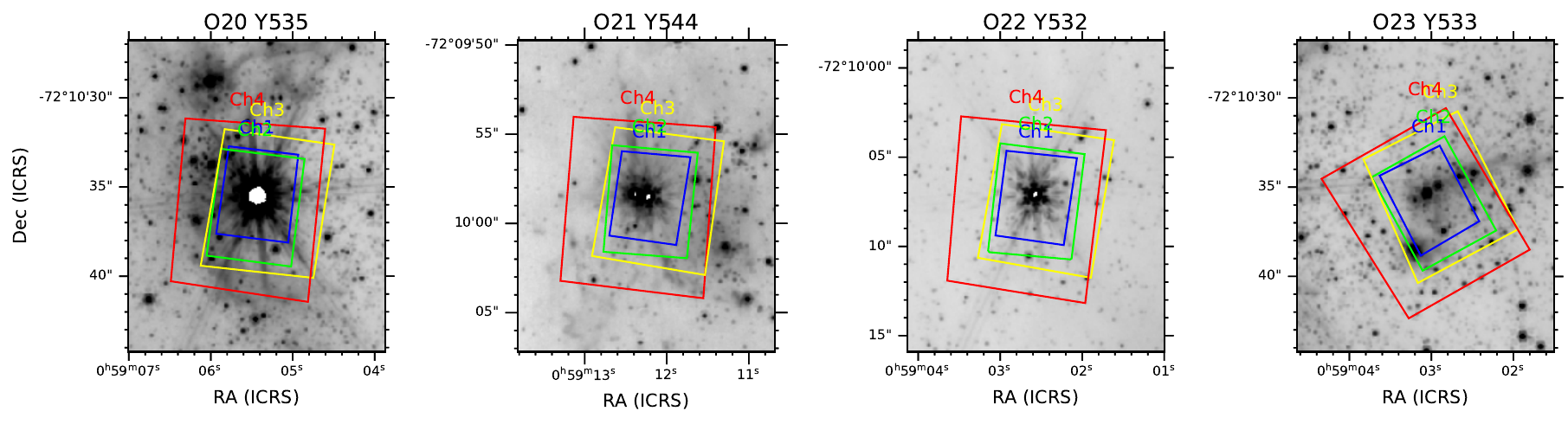}
 \caption{JWST NIRCam F444W imaging of the field surrounding our four MRS observations. The FOV of the four MIRI MRS channels (1, 2, 3, and 4) is shown in blue, green, yellow, and red, respectively.}  
 \label{fig:footprints}
 \end{figure*}

\subsubsection{O20: Y535}
\label{obs_y535}
Observation O20 targeted a single, bright YSO: ``Y535." This source was identified as a possible YSO based on its \textit{Spitzer} photometry by \cite{bib:sewilo2013}. The MRS pointing was centered at coordinates 00h59m5.4337, -72$^{\circ }$10'35.51''; (see \autoref{fig:footprints}.) Among the brightest sources in NGC 346 in the X-Ray \citep{bib:naze2002}, optical, near and mid-IR, this source was cataloged and observed in several past surveys. \cite{bib:Sabbi2007} observed Y535 within NGC 346 with \textit{HST} ACS in the V and I bands (F555W and F814W; \autoref{tab:photom_arhival}).
Y535 is also present in \textit{Spitzer} and \textit{Herschel} catalogs \citep{bib:Simon2007, bib:Seale2014}. \textit{Spitzer} IRAC (3.6-8~$\mu$m) MIPS (24-70~$\mu$m) and \textit{Herschel} (100-350~$\mu$m) photometry is shown in \autoref{tab:photom_arhival}. From mid-IR \textit{Spitzer} photometry, \citealt{bib:Simon2007} suggested that Y535 was a YSO of luminosity 33000~$\mathrm{L_{\odot}}$ and mass of 14.7~$\mathrm{M_{\odot}}$, or alternatively, a possible asymptotic giant branch (AGB) star.

Using the Infrared Spectrometer And Array Camera (ISAAC) on the Very Large Telescope (VLT), \cite{bib:Rubio2018} observed this source (identified in that work as Source C), and found a spectrum suggesting an embedded star with compact luminosity. They detected H$_2$ emission lines and compact \ion{He}{1}  emission at 2.058~$\mu$m, indicating a temperature in excess of 20,000~K. They suggested the compact nature of the \ion{He}{1}  emission, in contrast to the extended Br$\gamma$ and H$_2$ emission, indicated an early-type star experiencing ongoing accretion. Further, they argued that the absence of \ion{He}{2}  emission pointed to a later-O or early-B spectral type with an approximate temperature of $\sim$32,000~K based on the spectral-type effective temperature relation posed by \citet{bib:Trundle2007} at SMC metallicity. Fitting the IR SED of \cite{bib:Rubio2018} to the models of \cite{bib:robitaille06} yielded a rough mass estimate of 18~$\mathrm{M_{\odot}}$ and Stage I as a measure of evolutionary phase, (i.e., a YSO with a significant infalling envelope and a possible disk). From its position on the Hertzsprung-Russell diagram (HRD) along with PMS isochrones from \cite{bib:bressan2012}, they calculated the effective temperature, luminosity, mass and age of Y535 to be $\sim$31600~K, 125000~$\mathrm{L_{\odot}}$, 26~$\mathrm{M_{\odot}}$ and 17700~yr, respectively. \citeauthor{bib:Rubio2018} further suggested that the moderate extinction they observed toward Y535 (A$_V$~$\sim$~2.5) indicated it is not located behind the molecular clump of NGC 346 and instead must be part of the central cluster. Because isochrone-based age estimates place the age of this cluster at $\sim$1-3 Myr \citep{bib:Sabbi2007,bib:demarchi2013a}, they argue that the young age of Y535 points to another period of recent star formation. 

The near-IR spectrum of Y535 collected with the VLT's K-Band Multi-Object Spectrograph (KMOS) by \cite{bib:jones2022} shows that Y535 exhibits Br$\gamma$ emission from ionized gas, supporting its classification as a young YSO. They noted a P-Cygni profile in \ion{He}{1}  at 1.083~$\mu$m suggestive of an outflow, yet did not find extended Br$\gamma$ or H$_{2}$ emission.
They suggested that Y535 may be part of a young star complex, with a mid-IR spectrum indicating a massive YSO with silicates in emission \citep{bib:Whelan2013, bib:Ruffle2015, bib:Kraemer2017}

Y535 appears in the FOV of all 11 wavelength bands of JWST NIRCam and MIRI imaging described in \cite{bib:jones2023, bib:habel24} and is among the brightest sources in the region. However, it is heavily saturated in all bands and thus does not have measured photometry.

\begin{figure}
\centering
\vspace{.25cm}
\includegraphics[width=\columnwidth]{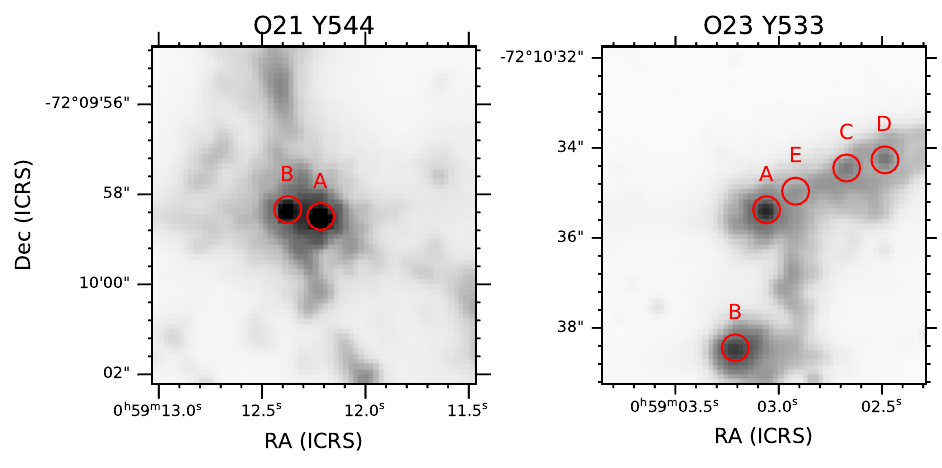}
 \caption{JWST MIRI F770W images of Y544 and Y533 showing their multiple components. Y544 is composed of two sources, Y544A \& B. Source Y533 is composed of Y533A, Y533B, Y533C, Y533D \& Y533E. Constituent sources are roughly ordered by decreasing mid-IR luminosity.}  
 \label{fig:constituents}
 \end{figure}

\subsubsection{O21: Y544A \& Y544B}
\label{obs_y544}
Observation O21 targeted two adjacent objects first collectively named Y544 by \cite{bib:sewilo2013}, who identified them as a singular high-reliability YSO. Both \textit{HST} and JWST NIRCam and MIRI imaging resolve Y544 into two components separated by $\sim$0.75\arcsec \citep{bib:Hennekemper2008, bib:habel24}. Through SED fitting, \cite{bib:sewilo2013} estimated the luminosity, temperature, and mass of the Y544 complex as 3240~$\mathrm{L_{\odot}}$, 7100~K, and 10.1~$\mathrm{M_{\odot}}$ and an evolutionary state of Stage I. Following \cite{bib:jones2023}, we adopt the names ``Y544A" for the brighter, western source and ``Y544B" for the comparatively-fainter eastern source, (see \autoref{fig:constituents}). Our observation was centered on the coordinates of Y544A: 00h59m12.2270,  -72$^{\circ }$09'58.51''. Called SMC
IRS 21 in \cite{bib:Ruffle2015}, Y544 was observed with \textit{Spitzer}'s Infrared Spectrograph (IRS) and classified as a YSO-4, a Herbig Ae/Be type object based on silicate emission \citep{bib:Whelan2013, bib:Ruffle2015}.

As with our other targets, Y544 was observed by \cite{bib:jones2022} with the VLT's KMOS, which spatially and spectrally resolved two components. They found fluorescent \ion{Fe}{2} and [\ion{Fe}{2}] lines toward Y544A consistent with a Herbig Oe/Be disk. They reported no dust excesses in the near-IR at $\lambda$$>$2.3~$\mu$m, and suggest that Y544B is cooler or has a greater extinction than its counterpart. Both components are present, but are saturated across NIRCam and MIRI in \cite{bib:habel24}.

\subsubsection{O22: Y532}
\label{obs_y532}
Observation O22 targeted the high-reliability YSO source identified as Y532 by \cite{bib:sewilo2013}. Like Y535, Y532 is revealed via near-IR and mid-IR JWST imaging as a bright, isolated source \citep{bib:habel24}. Our MRS observations of Y532 were centered at its coordinates: 00h59m2.5826, -72$^{\circ }$10'7.11''. SED fitting by \cite{bib:sewilo2013} yielded an estimated luminosity, temperature, and mass of 5188~$\mathrm{L_{\odot}}$, 12700~K, and 9.71~$\mathrm{M_{\odot}}$ with an evolutionary state of Stage I. Of the dozen sources observed by \cite{bib:jones2022}, only the spectra of Y532 and Y544B revealed a H$_2$/Br$\gamma$ ratio greater than unity, placing them as likely the youngest sources in their sample. As with Y535, Y544A, and Y544B, Y532 are also saturated in NIRCam and MIRI images \citep{bib:habel24}.

\subsubsection{O23: Y533A, Y533B, Y533C, Y533D \& Y533E}
\label{obs_y533}
The fourth MRS observation targeted the possible YSO denoted as Y533 by \cite{bib:sewilo2013}. This source appears to beto be fainter than the previous studies in near- and mid-IR JWST imaging \citep{bib:habel24}. It lies along a filament with several other relatively nearby ($\sim1-3''$) fainter sources captured at least partially in the MRS FOV. In this work, we refer to the brightest source as Y533A and assign similar names to the nearby sources captured in the same MRS FOV, as illustrated on \autoref{fig:constituents} and list the coordinates in \autoref{tab:Y533_photom}. We centered MRS observation O23 at 00h59m3.0637 -72$^{\circ }$10'35.61'', which is offset slightly from the coordinates of Y533A in order to capture Y533B more completely in MRS channels 2 and 3. We note that source Y533B was also cataloged by \citeauthor{bib:sewilo2013} as a high-reliability YSO and given its own name, ``Y534." In this work, we choose to refer to it as Y533B for consistency. Via SED fittng, \cite{bib:sewilo2013} estimated a luminosity, temperature and mass of 781.0~$\mathrm{L_{\odot}}$, 4630.0~K and 8.43~$\mathrm{M_{\odot}}$  and evolutionary state of Stage I for Y534/Y533B. The SED in that work for Y533A is poorly fitted, but an estimate of the luminosity based on available photometry yields 4940~$\mathrm{L_{\odot}}$ and a mass of  7.8~$\mathrm{M_{\odot}}$ \citep{bib:jones2022}.

KMOS observations of Y533A by \cite{bib:jones2022} showed absorption in Pa$\beta$ and Pa$\gamma$ and a tentative detection of \ion{He}{1}  in absorption at 1.083~$\mu$m, but showed poor S/N. They concluded that their near-IR spectrum of Y533A was consistent with the source being the O7 dwarf, NGC MPG 396, observed previously by \cite{bib:Massey1989} and \cite{bib:Dufton2019}. H$_2$ in absorption at 2.122~$\mu$m was also tentatively seen. Because such absorption lines are generally weak and need very hot or radiatively excited gas at large column densities to observe \citep{bib:lacy2017}, \citeauthor{bib:jones2022} suggested that Y533A is still deeply enshrouded in its natal dust and gas envelope, or alternatively, is located behind the molecular cloud. Additionally, they observed a P-Cygni-type profile of \ion{He}{1}  at 1.083~$\mu$m indicating the presence of outflowing gas.

Unlike the bright sources Y535, Y544A, Y544B, and Y532, the comparatively fainter source Y533A and its neighbors are not saturated in near- or mid-IR imaging seen by \cite{bib:habel24}. (See \autoref{tab:Y533_photom} for their photometry from that work.)

\subsection{Observing Setup}
\label{sec:observing}

We describe the observation setup for our four MIRI MRS fields. These observations each targeted one or multiple YSOs within the FOV of the integral field unit (IFU). Each observation employed all four channels (1,2,3 and 4) of MIRI MRS, which, respectively, span the wavelength ranges 4.90–7.65~$\mu$m, 7.51–11.70~$\mu$m, 11.55–17.98~$\mu$m and 17.70–27.9~$\mu$m. Each channel is divided into three bands (A, B and C), or ``SHORT," ``MEDIUM," and ``LONG" in reference to the portion of the wavelength range of the channel covered by the band. This results in a total of 12 wavelength bands (1A, 1B, 1C, 2A, 2B, 2C, 3A, 3B, 3C, 4A, 4B and 4C) which provide complete coverage over mid-IR wavelengths. The spectral resolution and FOV of the MRS instrument varies with channel. Channels 1 and 2 have the highest spectral resolution ($R$=2700-3700) whereas Channels 3 and 4 have lower spectral resolution (R=16-2800; \citealt{bib:jonesmiri2023}). Channels 1, 2, 3 and 4 have FOV of 10, 20, 32 and 51 arcsec$^2$ respectively. Thus, with three exposures per observation, we attain full spectral coverage for the area overlapped by each channel's FOV. 

We conducted all four observations using the \texttt{FASTR1} readout mode with a standard four-point dither pattern optimized for extended sources. Within any given observation, we selected identical exposure parameters for each of the wavelength subbands. For observations O20, O21, O22 and O23, we chose one integration per exposure and 7, 14, 18 and 125 groups per integration, respectively, amounting to integration times of 77.7, 155.4, 199.8 and 1387.5 seconds for each source in each subband. We did not collect separate off-source backgrounds as each observation includes enough empty field for background estimation. Additionally, the varying nebulosity across NGC 346 would result in improper background subtraction if a background were collected far from the target.

\section{Data Processing}
\label{sec:data}

Our observations were processed using the JWST pipeline version 1.14.0 \citep{10.5281/zenodo.10870758}, with CRDS context jwst\_1242.pmap. This pipeline implementation uses time-dependent photometric corrections, outlier detection with a kernel size and threshold that may be set, and is capable of residual fringe correction during spectral extraction. We follow a similar data reduction approach to \cite{bib:nayak2024}. Our Stage 1 and Stage 2 processing used the \textit{calwebb\_spec1} and \textit{calwebb\_spec2} steps \citep{bib:labi16}. For Stage 2 processing, we switched on the \textit{residual\_fringe} step, which corrects for fringing arising from the difference in the extended-source fringe pattern on the detector and the pipeline fringe flat \citep{bib:Gasman2023}. For Stage 3 processing (\textit{calwebb\_spec3} we use both the \textit{outlier\_detection} and \textit{ifu\_rfcorr} routines.  The outlier detection step compares the median of the stacked images to the original images to identify bad pixels and cosmic rays. Here we set 11 pixels as the kernel size used to normalize the pixel difference. The IFU residual fringe correction provides spectrum-level fringe correction. Though these correction steps mitigate the impact of fringing, some residuals can still be present in the extracted spectra. As noted by \cite{bib:nayak2024}, high-frequency fringing may be present in channels 3 and 4, which is thought to arise from the dichroics and is difficult to remove at the detector level. As a final measure in the data reduction process to reduce contrast from any remaining fringes, we applied the \textit{ifu\_rfcorr} on the extracted spectra (Kavanagh in prep.).

For observations O20 and O22, which contained only a single isolated point source (our target YSOs) within the FOV of each MRS pointing, we employed an automated method that both found the centroid of the emission in the cube and performed spectral extraction.

Observations O21 and O23 captured multiple sources in the FOV of the MRS IFU as shown in \autoref{fig:observations}. O21 captured two sources (Y544A and Y544B). O23 captured five sources visible in the mid-IR in one or more channel of the instrument.
Because of the multiple constituents in observations O21 and O23, we could not automatically locate each source as with O20 and O22. Instead, we examined the integrated emission across the cube for each channel. For each source, we provided an initial coordinate estimate, which was refined via a centroiding routine to find the peak of the integrated emission. These coordinates were then averaged across all available channels. 
These averaged coordinates were then adopted during the spectral extraction step.

\section{Results}
\label{sec:results}
Here we summarize the results of our four MRS observations. We discuss the four IFU data cubes and the emission morphology revealed at several wavelengths. We further describe the spectra extracted from sources within the FOV of each cube and the specific spectral features detected therein.

\subsection{IFU Cubes}
\label{sec:cubes}

\begin{figure*}
    \centering
    \includegraphics[width=.9725\linewidth]{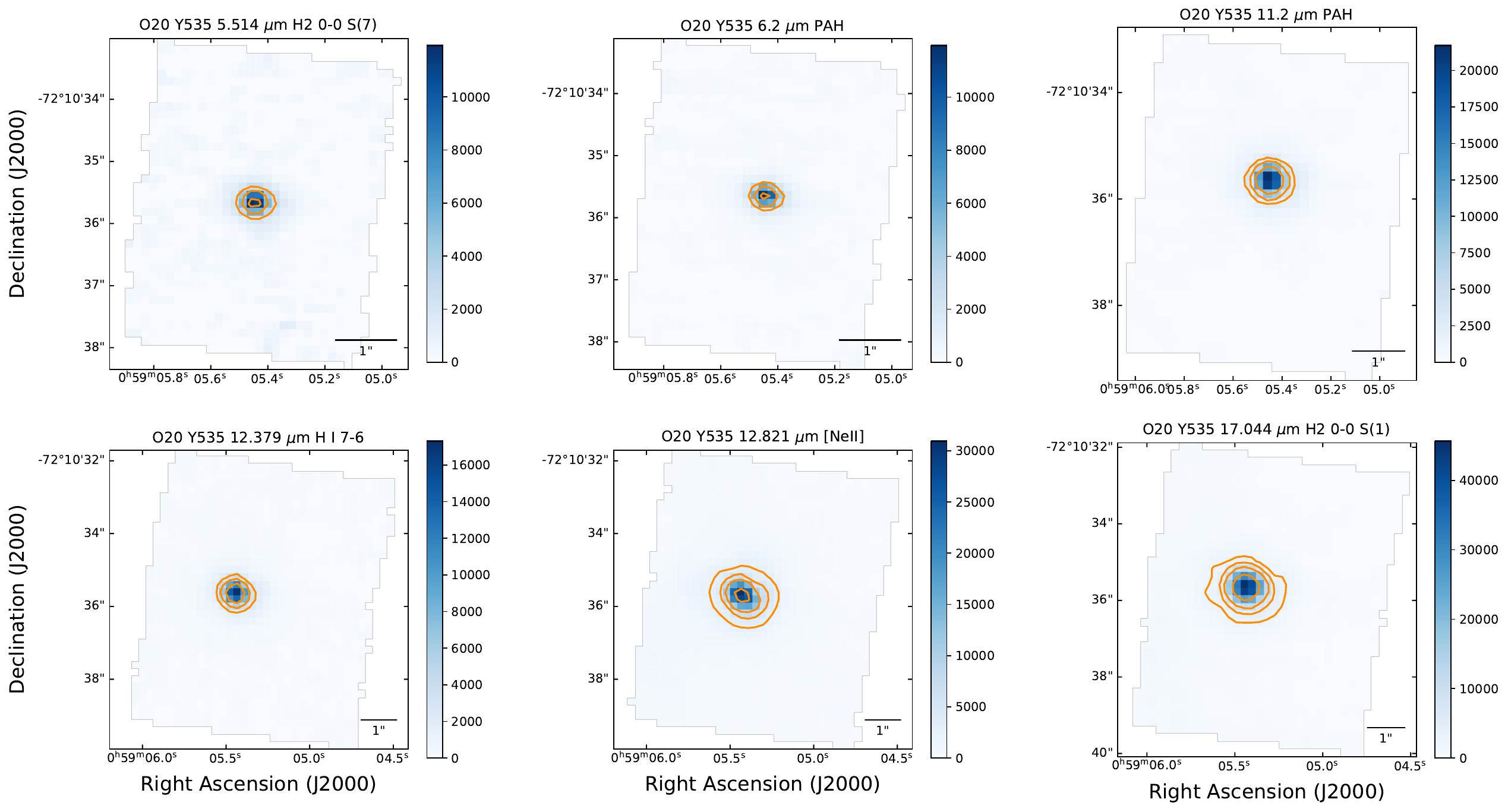}
    \caption{Six slices of the IFU cube targeting Y535: H$_2$ 0-0 S(7) emission at 5.514~$\mu$m (\textbf{top left}), 6.2~$\mu$m PAH continuum emission (\textbf{top middle}), 11.2~$\mu$m PAH continuum emission (\textbf{top right}), \ion{H}{1} 7-6 emission at at 12.379~$\mu$m (\textbf{bottom left}), {[\ion{Ne}{2}]} emission at 18.821~$\mu$m (\textbf{bottom middle}) and H$_2$ 0-0 S(1) emission at 17.044~$\mu$m (\textbf{bottom right}). Contours levels for all panels are at 2500, 5000, 10000, 25000, and 50000 MJy sr$^{-1}$. PAH slices are a width of 0.2~$\mu$m centered at the given wavelength.}
    \label{fig:slices_O20}
\end{figure*}

\begin{figure*}
    \centering
    \includegraphics[width=.9725\linewidth]{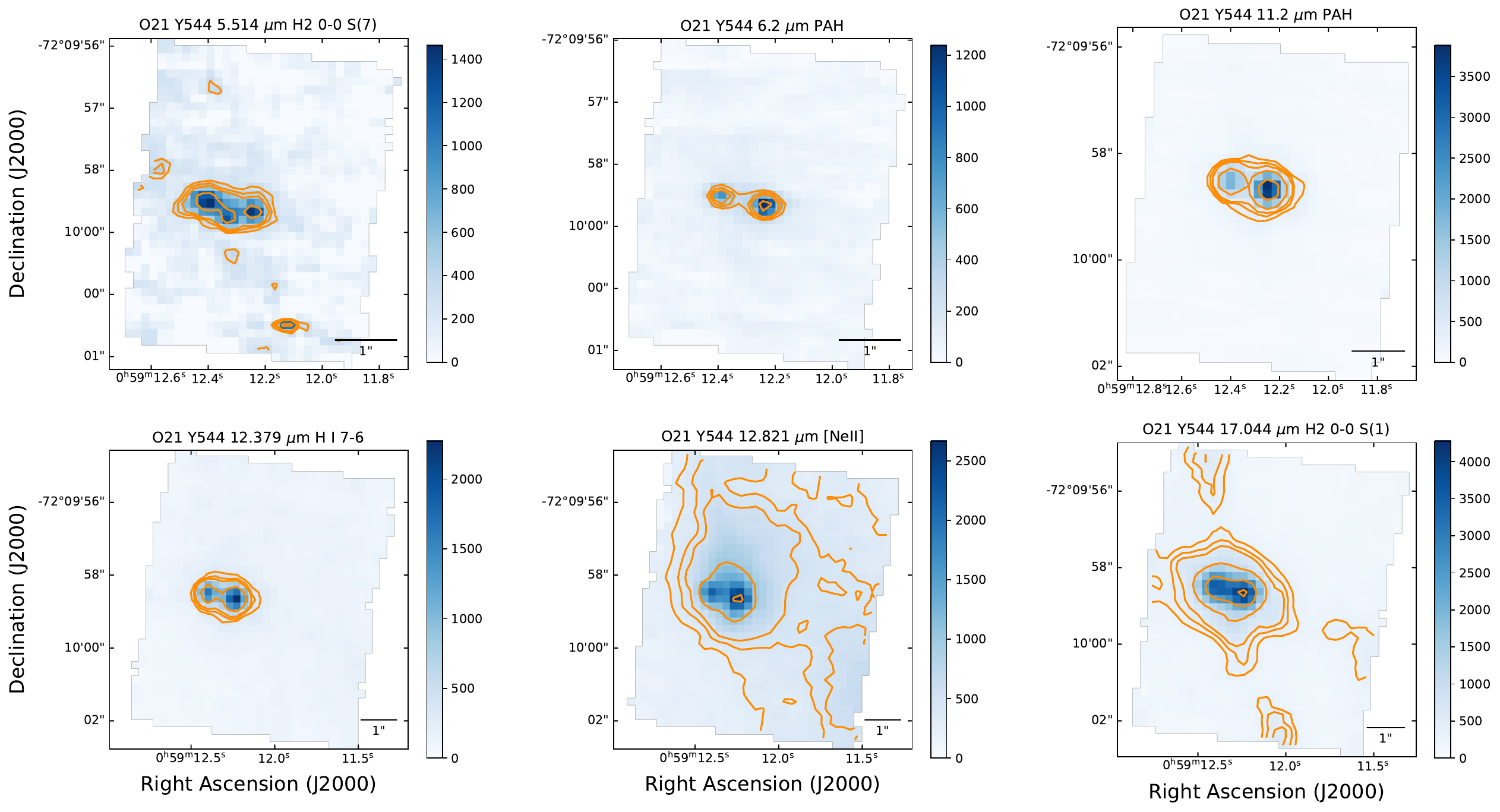}
    \caption{Six slices of the IFU cube targeting Y544: H$_2$ 0-0 S(7) emission at 5.514~$\mu$m (\textbf{top left}), 6.2~$\mu$m PAH continuum emission (\textbf{top middle}), 11.2~$\mu$m PAH continuum emission (\textbf{top right}), \ion{H}{1} 7-6 emission at at 12.379~$\mu$m (\textbf{bottom left}), {[\ion{Ne}{2}]} emission at 18.821~$\mu$m (\textbf{bottom middle}) and H$_2$ 0-0 S(1) emission at 17.044~$\mu$m (\textbf{bottom right}). Contours levels for all panels are at 300, 400, 500, 1000, 250 and 4000 MJy sr$^{-1}$. PAH slices are a width of 0.2$\mu$m centered at the given wavelength.}
    \label{fig:slices_O21}
\end{figure*}

\begin{figure*}
    \centering
    \includegraphics[width=.9725\linewidth]{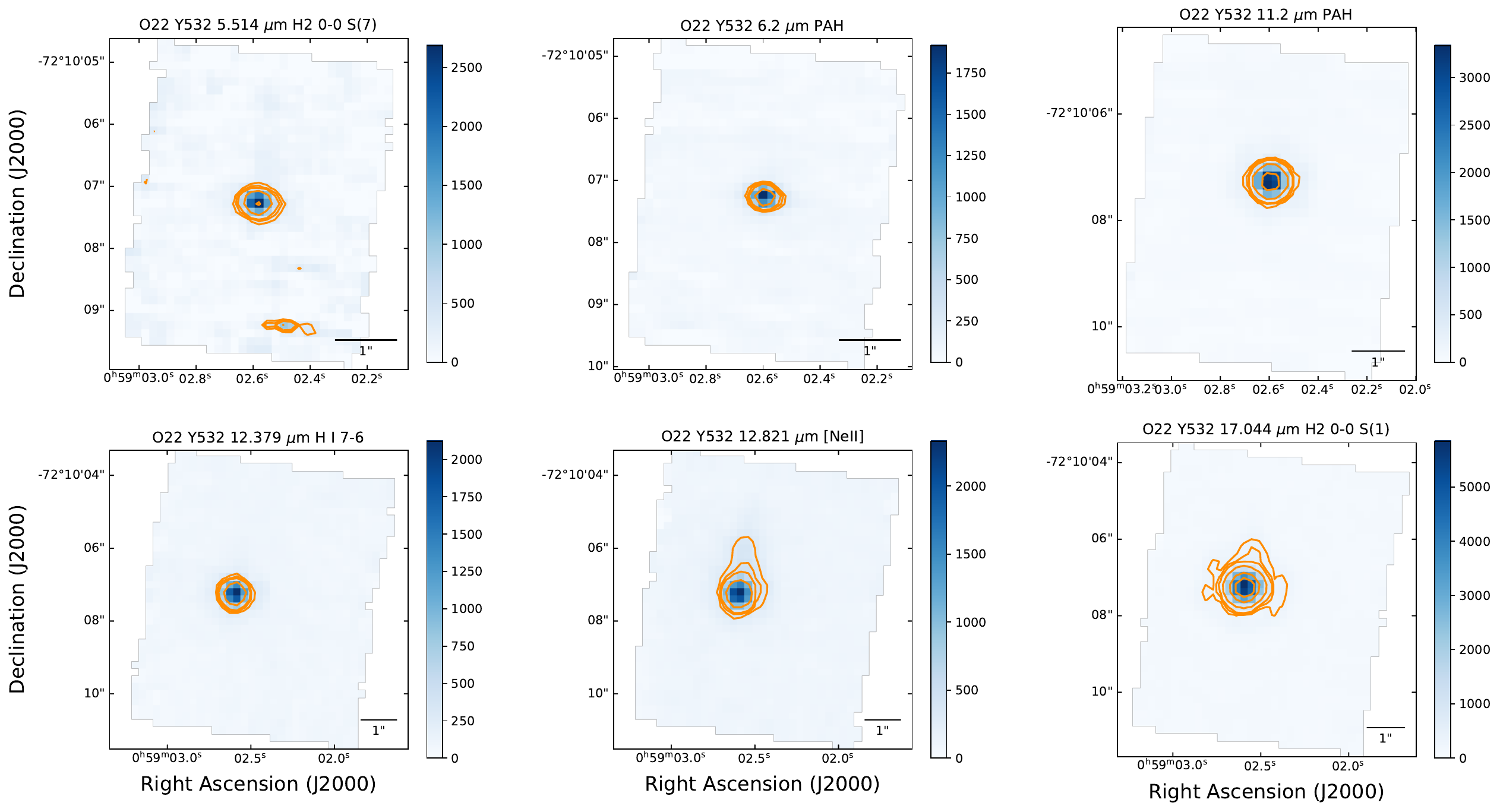}
    \caption{Six slices of the IFU cube targeting Y532: H$_2$ 0-0 S(7) emission at 5.514~$\mu$m (\textbf{top left}), 6.2~$\mu$m PAH continuum emission (\textbf{top middle}), 11.2~$\mu$m PAH continuum emission (\textbf{top right}), \ion{H}{1} 7-6 emission at at 12.379~$\mu$m (\textbf{bottom left}), {[\ion{Ne}{2}]} emission at 18.821~$\mu$m (\textbf{bottom middle}) and H$_2$ 0-0 S(1) emission at 17.044~$\mu$m (\textbf{bottom right}). Contours levels for all panels are at 300, 400, 500, 1000, 2500 and 4000 MJy sr$^{-1}$. PAH slices are a width of 0.2~$\mu$m centered at the given wavelength.}
    \label{fig:slices_O22}
\end{figure*}

\begin{figure*}
    \centering
    \includegraphics[width=.9725\linewidth]{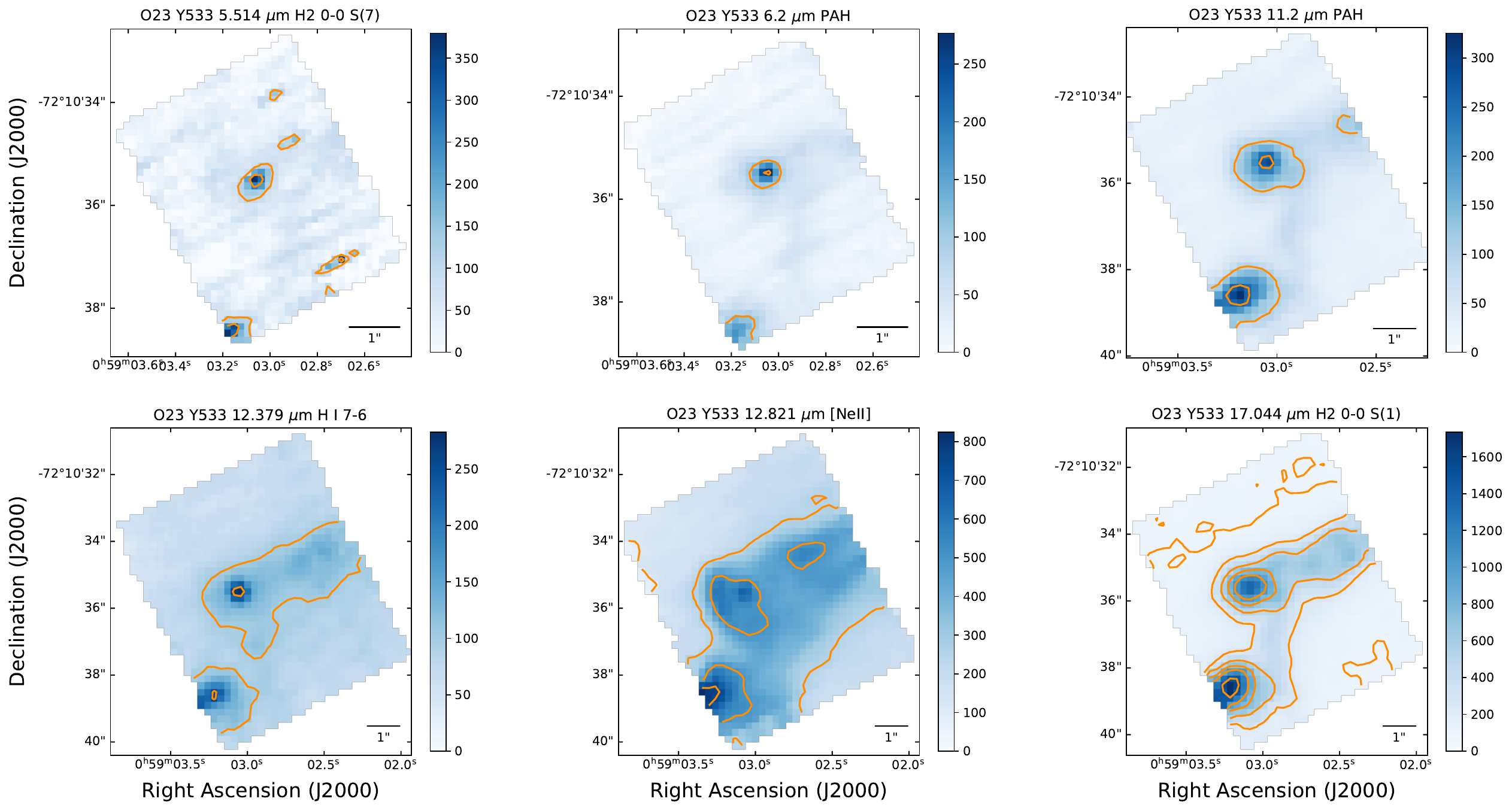}
    \caption{Six slices of the IFU cube targeting Y533: H$_2$ 0-0 S(7) emission at 5.514~$\mu$m (\textbf{top left}), 6.2~$\mu$m PAH continuum emission (\textbf{top middle}), 11.2~$\mu$m PAH continuum emission (\textbf{top right}), \ion{H}{1} 7-6 emission at at 12.379~$\mu$m (\textbf{bottom left}), {[\ion{Ne}{2}]} emission at 18.821~$\mu$m (\textbf{bottom middle}) and H$_2$ 0-0 S(1) emission at 17.044~$\mu$m (\textbf{bottom right}). Contours levels for all panels are at 100, 250, 500, 750, 1000 and 1500 MJy sr$^{-1}$. PAH slices are a width of 0.2~$\mu$m centered at the given wavelength.}
    \label{fig:slices_O23}
\end{figure*}

In \autoref{fig:slices_O20}, \autoref{fig:slices_O21}, \autoref{fig:slices_O22} and \autoref{fig:slices_O23} we show slices of the IFU spectral cubes for each observation at six given wavelengths. We inspect the wavelengths of two known H$_2$ emission lines (at 5.51 and 17.04 $\mu$m), the 12.82 $\mu$m [\ion{Ne}{2}] line and two PAH features at 6.2 and 11.2 $\mu$m.

Across the entire mid-IR spectral range, the IFU cube for observation O20 shows tightly-centered emission from a single source, Y535. We find no evidence of extended emission due to PAH features across the FOV (\autoref{fig:slices_O20}). Notably, we observe a bright, outflow-like structure closely adjacent to Y535 extending from its center toward the southwest at 6.989~$\mu$m, coincident with the [\ion{Ar}{2}] emission line. We further discuss this feature in \autoref{sec:jets}

The spectral cube of observation O21 reveals the two YSOs Y544A and Y544B centered in the FOV. The western source Y554A shows brighter emission than its companion across the IFU cube. At the shortest wavelengths, the sources remain distinct, but become increasingly blended at longer wavelengths. Beginning in channel 4 (at $\sim$~17.7$\mu$m), the two are blended almost entirely. We find an extended filamentary structure extending northward from the two sources. In \autoref{fig:slices_O21}, this structure can be seen brightly at 12.82~$\mu$m, coincident with the [\ion{Ne}{2}] emission line. However, we find that this feature persists broadly across the IFC cube, and thus is not consistent with narrow line emission. Additionally, this feature also seen in both NIRCam and MIRI as a large filament intersecting Y544A and Y544B, (see \autoref{fig:observations}, \autoref{fig:constituents}), indicating it is likely a dusty filament not directly associated with the YSOs \citep{bib:habel24}.

The spectral cube of observation O22 contains the single source Y532. Emission from this source is tightly centered about it across the IFU cube. Similarly to Y544A \& B, we find diffuse structure extending northward from the center of Y532 at several wavelengths, (ex. \autoref{fig:slices_O22}). As with the filament adjacent to Y544A \& B, a  similar structure to that seen toward Y532 is present in NIRCam and MIRI, though fainter \citep{bib:habel24}. Thus we assess this structure to be associated with the larger diffuse morphology in the surrounding region.

The IFU cube for observation O23 is the most complex of the four described in this work, containing up to five sources (see \autoref{fig:constituents} \& \autoref{fig:slices_O23}). The primary target, source Y533A, is centered. Y533E, closely adjacent to Y533A, is contained in the FOV of all channels, but is not distinguishable as a separate source. The remaining three sources are present incompletely across the IFU cube as a result of the differing FOV of the wavelength channels (see \autoref{fig:observations}). Y533B and Y533C, in the southeast and northwest corners respectively, are cut off by the smaller size of the channel 1 detector, but are present in the longer wavelength channels in the remainder of the cube. Y533D, the furthest source from the center, is present only in channels 3 and 4 of the IFU cube. NIRCam and MIRI imaging \citep{bib:jones2023, bib:habel24} previously detected a filament intersecting these five sources, extending from the northwest to the southeast, as shown in \autoref{fig:constituents}. In our MRS data cubes, we observe this structure across nearly the entire wavelength range, with it being most visible at wavelengths past $\sim$6.5~$\mu$m. As shown in \autoref{fig:slices_O23}, we observe a consistent shape of this filament at wavelengths characteristic of PAH, molecular hydrogen, atomic hydrogen and atomic fine-structure emission. Slicing the IFU cube at the [\ion{Ne}{2}] 12.82$\mu$m emission line reveals wider, more extended emission about the filament. We find strong excitation of this line in all of our sources (discussed below in \autoref{sec:ne_ar_cl}).

\begin{figure*}[h!]
    \centering
    \includegraphics[width=.79\linewidth]{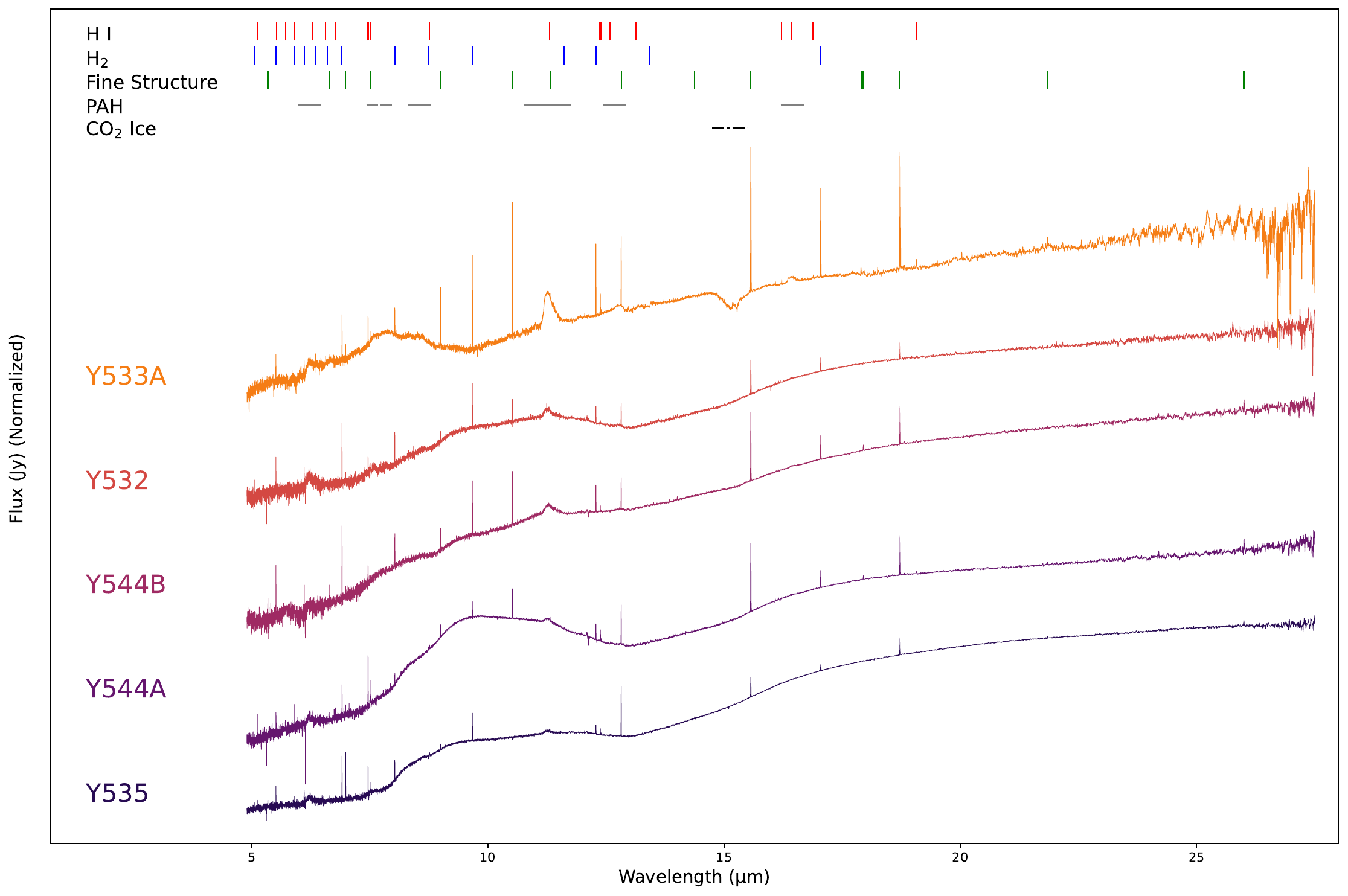}
    \caption{Normalized and offset spectra of five YSOs captured within our four MRS observations. Y544A and Y544B were observed in the same observation. Spectra of additional sources captured within the same observation as source Y533A are shown in \autoref{fig:targ_spec_compare_Y533}. We mark the location of various identified spectral features at the top.}
    \label{fig:targ_spec_compare}
\end{figure*}

\begin{figure*}[h!]
    \centering
    \includegraphics[width=.79\linewidth]{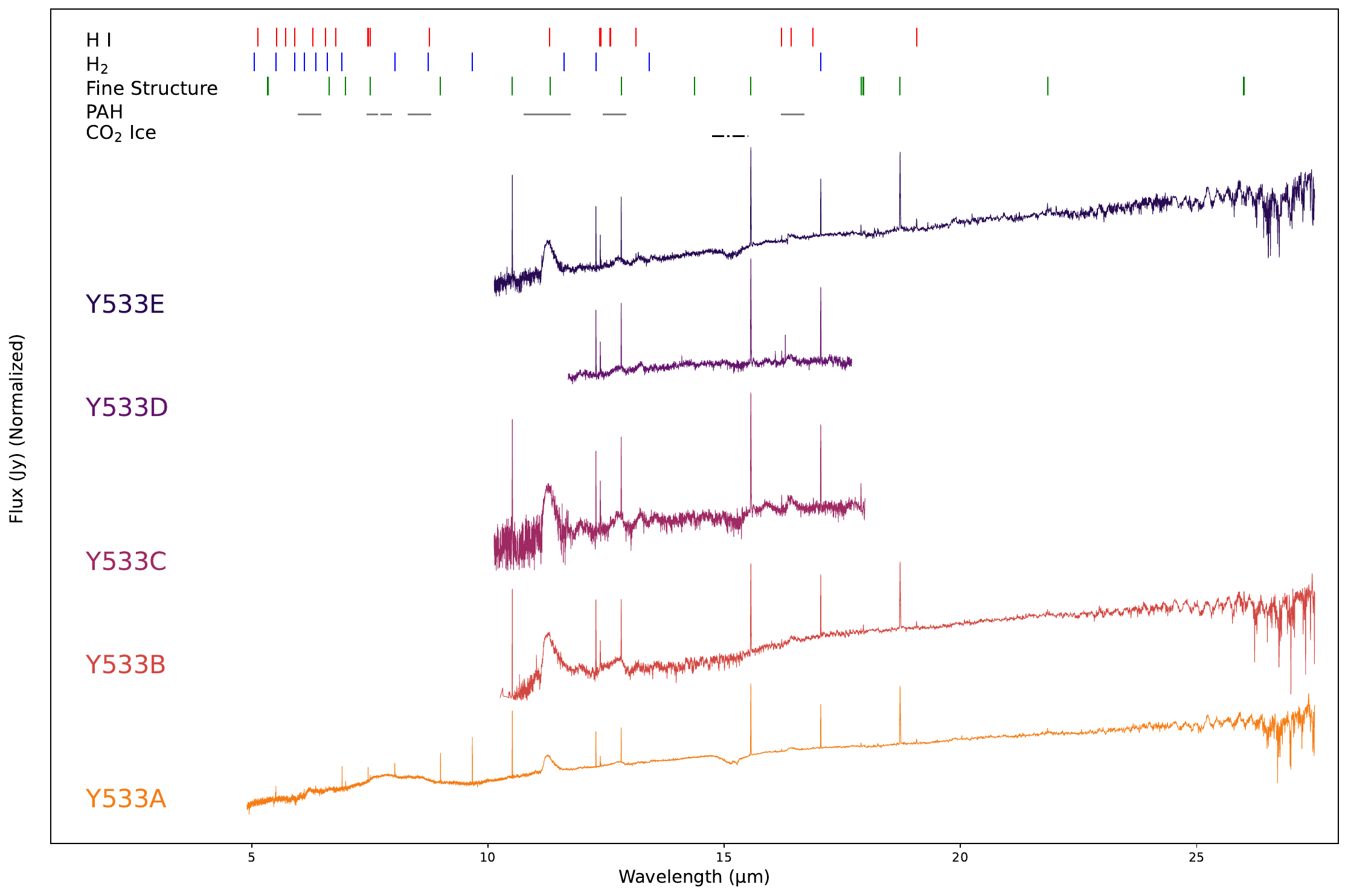}
    \caption{Normalized and offset spectra of the five sources (Y533A, B, C, D \& E) captured within the FOV of observation O23. Partial spectra are presented for sources B, C and D, which were only partially captured within the footprints of the four wavelength channels. Source Y533E is substantially blended with emission from nearby Y533A and exhibited poor signal-to-noise in the short-wavelength channels which are not shown. We mark the location of various identified spectral features at the top.}
    \label{fig:targ_spec_compare_Y533}
\end{figure*}

\subsection{Extracted YSO Spectra}
\label{sec:spectra}

Spectra for all targets were extracted using an aperture defined as 1.22~$\lambda/D$ where $\lambda$ and $D$ are the wavelength of the IFU cube and the beam size, respectively. For the singular sources Y535 and Y532, the background was determined and subtracted automatically in the \texttt{Extract1dStep} step with \texttt{subtract\_background} enabled. For observations O20 and O22 which contained a single source each in the IFU FOV, background subtraction was done automatically by the spectral extraction routine. Because observations O21 and O23 contained multiple sources within their FOV, the background was determined by extracting a spectrum off-source and away from the bright sources.  We implemented the pipeline spectral leak correction (\texttt{jwst.spectral\_leak}) for spectral segments extracted from channels 1B and 3A. Next, each of the 12 spectral cube segments per source was scaled consecutively using the median of the flux value between adjacent segments. 
Finally, we stitch the segments together with JWST pipeline's \texttt{combine\_1d} step. We observe some amount of residual fringing noise for the longest and shortest wavelengths ($<$ 5.5~$\mu$m and $>$ 25~$\mu$m), particularly for fainter sources Y533B, Y533C, Y533D and Y533E.

We find that each source in our sample shows a rising mid-IR SED characteristic of a YSO. Each source displays numerous narrow emission lines and broad emission corresponding to PAHs features at 6.2, 7.7, 8.6, 11.2, 12.7 and 16.4~$\mu$m (\autoref{fig:targ_spec_compare} \& \autoref{fig:targ_spec_compare_Y533}). The 10~$\mu$m silicate feature is present in several sources in emission (Y535, Y544A, and Y323) and likely in absorption in source Y533A, though emission from the 7.7~$\mu$m PAH feature makes this difficult to confirm.  
Source Y533A shows double-toughed absorption at $\sim$15~$\mu$m immediately identifiable as the CO$_2$ ice feature. Sources Y535A, Y533C, Y533D and Y533E also show evidence of absorption at 15~$\mu$m to varying extent, (see \autoref{tab:spec_features}). Several sources exhibit an absorption-like feature in their spectra at 5.3, 9.8, 12.1, and 16.2~$\mu$m, however investigation of the IFU cube reveals these features to be caused by instrumental artifacts.  


\begin{deluxetable*}{l|ccccccccccc}
\tabletypesize{\scriptsize}
\tablecaption{Summary of Spectral Features Observed in YSOs
\label{tab:spec_features}}
\tablehead{
\colhead{Name} & \colhead{6.2~$\mu$m} & \colhead{7.7~$\mu$m} & \colhead{8.6~$\mu$m}& \colhead{11.0~$\mu$m} & \colhead{11.2~$\mu$m} & \colhead{12.7~$\mu$m} & \colhead{16.4~$\mu$m} & \colhead{$\mathrm{CO_{2}}$ Absorp} & \colhead{No. of } & \colhead{No. of} & \colhead{No. of} \\
\colhead{} & \colhead{PAH} & \colhead{PAH} & \colhead{PAH}& \colhead{PAH} & \colhead{PAH} & \colhead{PAH} & \colhead{PAH} & \colhead{}  & \colhead{\ion{H}{1} Lines} & \colhead{Fine-Structure Lines} & \colhead{$\mathrm{H_{2}}$ Lines} 
}
\decimals
\startdata
\hline
Y535  & \checkmark  &                  &             &             &  \checkmark     &                &                     &  \checkmark    &  9  &  13 & 10    \\
\hline            
Y544A & \checkmark  &                  &             &             &  \checkmark     & \checkmark     &                     &                &  14  &  10 & 10    \\
Y544B & \checkmark  &                  &             &             &  \checkmark     & \checkmark     &  \checkmark         &                &  2  &  10 & 7    \\
\hline            
Y532  & \checkmark  &  \checkmark      &  \checkmark &             &  \checkmark     & \checkmark     &                     &                &  3  &  7 & 8    \\
\hline            
Y533A & \checkmark  &  \checkmark      &  \checkmark &             &  \checkmark     & \checkmark     &  \checkmark         &  \checkmark    &  8  &  8 & 9    \\
Y533B & -           &  -               &  -          &  -          &  \checkmark     & \checkmark     &  \checkmark         &                &  3  &  6 & 4    \\
Y533C & -           &  -               &  -          &  -          &  \checkmark     & \checkmark     &  \checkmark         &  \checkmark$^*$&  4  &  6 & 2    \\
Y533D & -           &  -               &  -          &  -          &      -          & \checkmark     &  \checkmark         &  \checkmark$^*$&  2  &  4 & 2    \\
Y533E$^{\dagger}$ &-&  -               &  -          &  -          & \checkmark      & \checkmark     &  \checkmark         &  \checkmark$^*$&  8  &  9 & 7    \\
\hline
\enddata
\tablecomments{``$*$"CO$_2$ absorption in these sources could not be reliably measured, (see \autoref{fig:targ_spec_compare_Y533}). $\dagger$: Y533E is closely adjacent to brighter Y533A and likely significantly contaminated. Spectra of sources Y533B, Y533C, Y533D \& Y533E are incomplete or exhibit excessive noise across the range spanning the first several PAH features, marked with ``-". For detected lines that may be blended between two emission lines, both constituents are tallied.}
\end{deluxetable*}

\subsection{Narrow Emission Line Identifications and Measurements}
\label{sec:narrow_lines}

To identify and characterize the narrow line emission in our spectra, we first fit a continuum to each spectrum using \textit{astropy}'s spline function and subtract this from the original spectrum. Then, using the \textit{find\_lines\_threshold} function of \textit{specutils}, we detect the narrow emission and absorption features and extract their measured wavelength, the uncertainty in the measured wavelength, their FWHM, their flux (in erg~s$^{-1}$~cm$^{-2}$) and the uncertainty in the flux. We match these detected lines to a list of mid-IR molecular hydrogen emission lines, hydrogen recombination lines, and fine-structure lines corrected for their wavelength using the systemic velocity of NGC 346 of 165.5 km s$^{-1}$ found by \cite{bib:zeidler2022}. The typical difference in expected and identified wavelength is 0.01~$\mu$m or less.

\begin{figure*}
\centering
\vspace{-.25in}
\includegraphics[width=.9\linewidth]{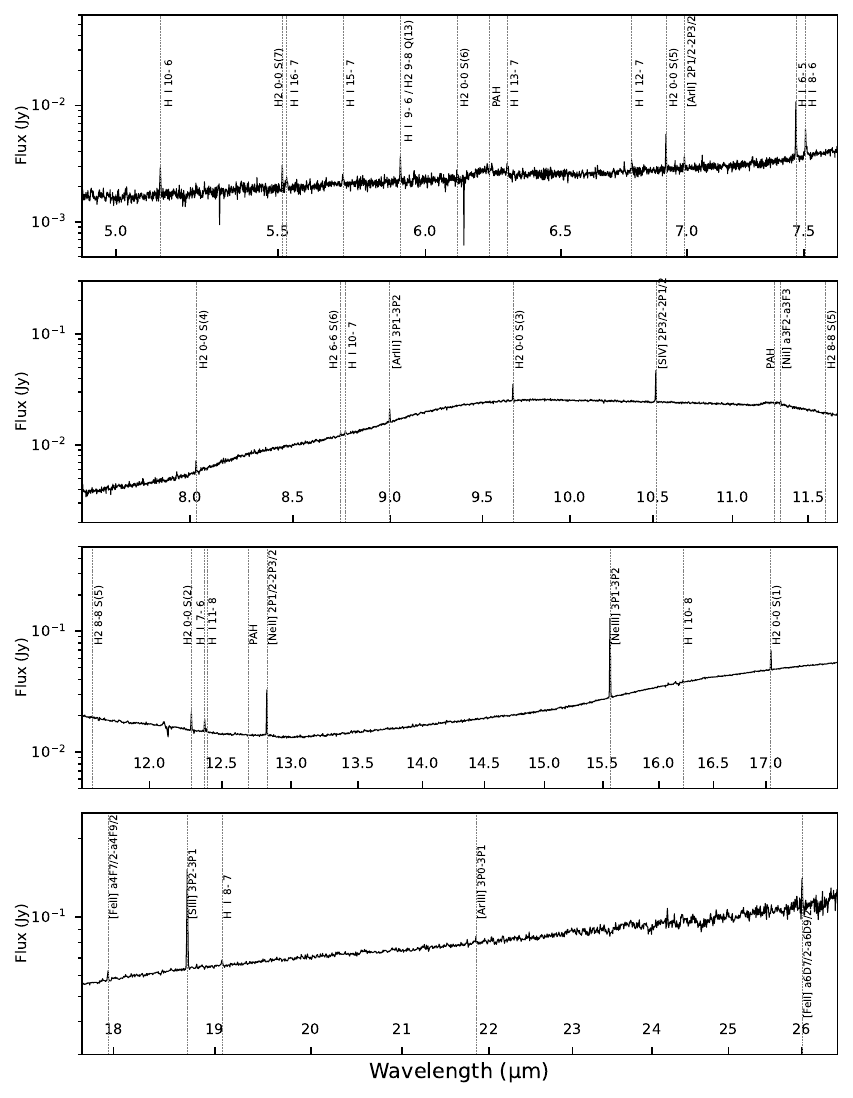}\hfill
\caption{Spectra for channels 1-4 of source \textbf{O21 Y544A}. Full line list is given in \textbf{\autoref{tab:e_lines_Y544A}}.}
\label{fig:spec_Y544A}
\end{figure*}

\begin{figure*}
\centering
\vspace{-.25in}
\includegraphics[width=.9\linewidth]{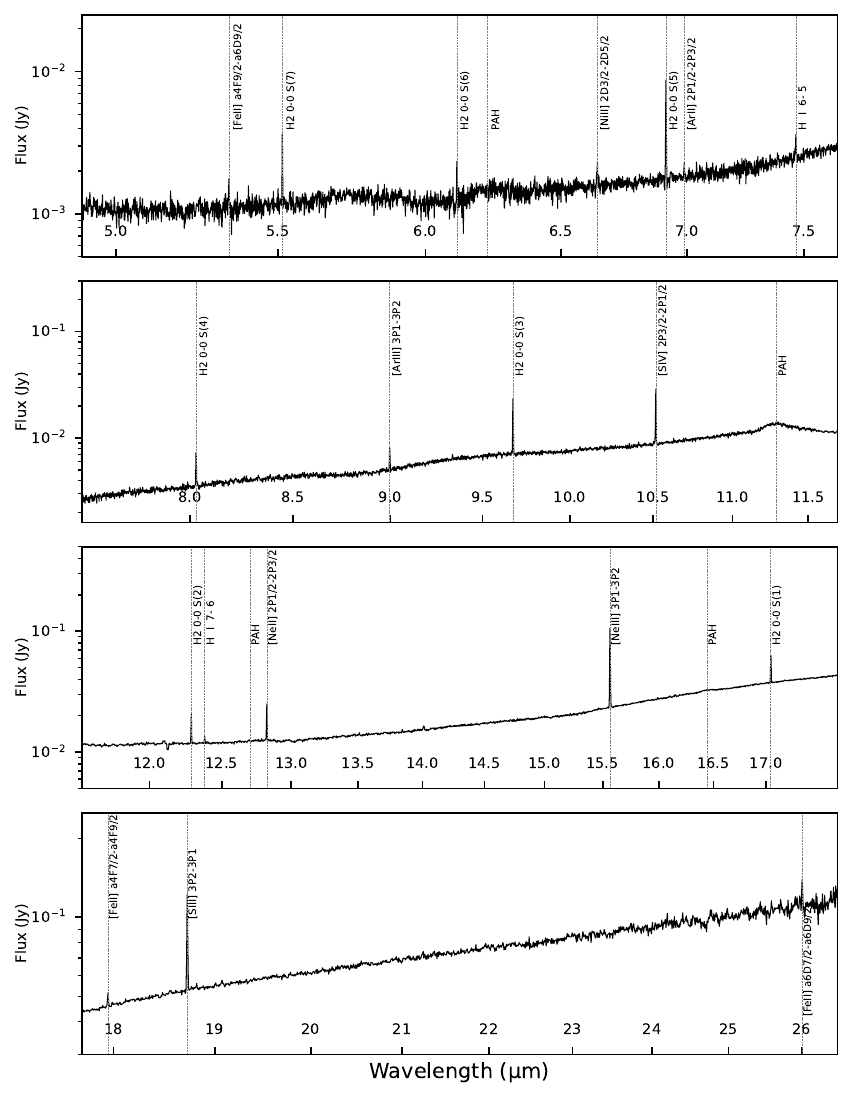}\hfill
\caption{Spectra for channels 1-4 of source \textbf{O21 Y544B}. The full line list is given in \textbf{\autoref{tab:e_lines_Y544B}}.}
\label{fig:spec_Y544B}
\end{figure*}

\begin{figure*}
\centering
\vspace{-.25in}
\includegraphics[width=.9\linewidth]{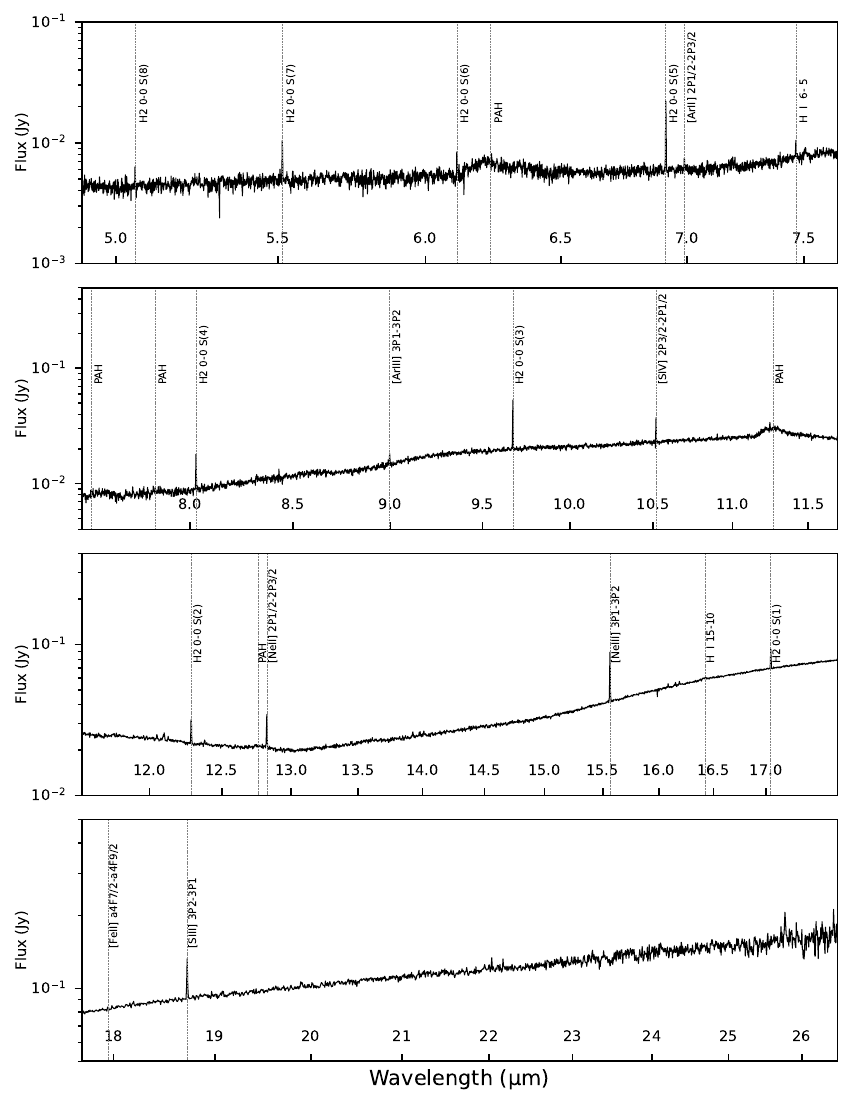}\hfill
\caption{Spectra for channels 1-4 of source \textbf{O22 Y532}. The full line list is given in \textbf{\autoref{tab:e_lines_Y532}}.}
\label{fig:spec_Y532}
\end{figure*}

A summary of the number and species of lines detected for each source is presented in \autoref{tab:spec_features}. Tables of the detected lines and their measured parameters are shown for sources Y535 (\autoref{tab:e_lines_Y535}), Y544A (\autoref{tab:e_lines_Y544A}), Y544B (\autoref{tab:e_lines_Y544B}), Y532 (\autoref{tab:e_lines_Y532}), Y533A (\autoref{tab:e_lines_Y533A}), Y533B (\autoref{tab:e_lines_Y533B}), Y533C (\autoref{tab:e_lines_Y533C}), Y533D (\autoref{tab:e_lines_Y533D}) and Y533E (\autoref{tab:e_lines_Y533E}).
We display the annotated spectra of Y535 (\autoref{fig:spec_Y535}), Y532 (\autoref{fig:spec_Y532}), Y544A (\autoref{fig:spec_Y544A}), Y544B (\autoref{fig:spec_Y544B}) and Y533A (\autoref{fig:spec_Y533A}), the five sources with complete mid-IR spectral coverage, and label the measured emission features.

\subsection{Atomic (\ion{H}{1}) and Molecular (H$_2$) Hydrogen}
\label{sec:hydrogen}

We detect narrow emission lines from hydrogen recombination in the spectra of each YSO. In the near- and mid-IR, \ion{H}{1} emission is commonly used to trace accretion luminosity, and derive mass accretion rates (e.g. \citealt{bib:Calvet_2004, bib:alca14, bib:rigl15, bib:ward2016, bib:ward2017, bib:reit19, bib:nayak2024}). Each of our sources show the emission at 12.37~$\mu$m from the \ion{H}{1} (7-6) line. In \autoref{sec:acc_rates}, we derive accretion luminosities and mass accretion rates based on the strength of this emission.

In addition to narrow-line \ion{H}{1}, we find molecular hydrogen emission in all of our YSO spectra. H$_2$ emission is often tied to processes in the local environment of the YSO, such as shock heating of the molecular gas by jets and outflows, or UV radiation heating nearby gas to temperatures of approximately several hundred Kelvin. 

Both the 5.51~$\mu$m (H$_2$ 0-0 S7) and 17.03~$\mu$m (H$_2$ 0-0 S1) lines are common to sources Y535, Y544A, Y544B, Y532 and Y533A. Spectra for Y533B-D are incomplete at 5.51~$\mu$m, but all show H$_2$ 0-0 S1 at 17.03~$\mu$m. The relative emission strength of these two lines is sensitive to gas temperature. Hotter (several hundred Kelvin) gas will more strongly excite the 5.51~$\mu$m line than the 17.03~$\mu$m. At cooler temperatures, as low as 100~K \citep{bib:Burton_1997}, 17.03~$\mu$m emission will dominate. We find that the 17.03~$\mu$m line is stronger with 17.03~$\mu$m/5.51$\mu$m line ratios of 1.5, 7.2, 4.2, 1.3, and 30.4, respectively, consistent with these sources being overall young and still embedded. This ratio is significantly higher in Y533A, consistent with it being the youngest of these sources. Future analysis of H$_2$ line emission, especially if combined with new near-IR measurements, will further probe the mixture of gas phases in these objects.

\subsection{Fine-Structure Lines}
\label{sec:fine}

In addition to both H$_2$ and \ion{H}{1} emission, we observe multiple fine-structure emission lines in each of our YSO spectra. As is the case in our sources, such fine-structure emission is commonly seen alongside PAH emission, with both indicating that the central YSO is emitting UV radiation. Both PDRs and shocks caused by jets and winds can be traced via fine-structure. UV photons ionize atoms with potentials below 13.6\,eV such as [\ion{Fe}{1}], [\ion{Fe}{2}] and [\ion{Si}{1}]. Shocks with velocities in excess of 70 km s$^{-1}$ can heat the gas to 10$^5$ K, exciting lines such as [\ion{Ni}{2}], [\ion{Ar}{2}],
[\ion{Ne}{2}], [\ion{Ar}{3}], and [\ion{Fe}{2}] with ionization energies $>$21eV \citep{bib:drai93, bib:holl97}.

\begin{deluxetable*}{lcccccc}
\tablecaption{Spectral Features for Source Y535}
\tablehead{
\colhead{Name} & \colhead{Lab Wave} & \colhead{Meas Wave}& \colhead{Meas Wave Err} & \colhead{FWHM} & \colhead{Flux} & \colhead{Flux Err} 
\\
\colhead{}  & \colhead{($\mu$m)} & \colhead{($\mu$m)} & \colhead{($\mu$m)} & \colhead{($\mu$m)} & \colhead{($\mathrm{erg\;s^{-1}\;cm^{-2}}$)} & \colhead{($\mathrm{erg\;s^{-1}\;cm^{-2}}$)}
}
\decimals
\startdata
H2 0-0 S(8) & 5.056 & 5.056 & 0.00128 & 0.001239 & 1.11E-15 & 8.45E-17 \\
H  I 10- 6 & 5.131 & 5.132 & 0.00195 & 0.000445 & 1.71E-15 & 8.35E-17 \\
{[FeII] a4F9/2-a6D9/2} & 5.343 & 5.343 & 0.00095 & 0.001504 & 6.02E-16 & 8.53E-17 \\
H2 0-0 S(7) & 5.514 & 5.514 & 0.00174 & 0.000587 & 3.57E-15 & 8.30E-17 \\
H  I  9- 6$^*$ & 5.911 & 5.912 & 0.00196 & 0.000678 & 1.96E-15 & 8.55E-17 \\
H2 9-8 Q(13)$^*$ & 5.913 & 5.912 & 0.00196 & 0.000678 & 1.96E-15 & 8.55E-17 \\
H2 0-0 S(6) & 6.112 & 6.112 & 0.00180 & 0.000433 & 2.03E-15 & 7.80E-17 \\
PAH & 6.2 & 6.226 & 0.00250 & 0.137892 & 1.15E-13 & 6.65E-15 \\
{[NiII] 2D3/2-2D5/2} & 6.640 & 6.639 & 0.00145 & 0.000578 & 6.76E-16 & 5.31E-17 \\
H  I 12- 7 & 6.776 & 6.775 & 0.00313 & 0.000657 & 8.18E-16 & 5.00E-17 \\
H2 0-0 S(5) & 6.913 & 6.914 & 0.00215 & 0.000286 & 1.03E-14 & 5.60E-17 \\
{[ArII] 2P1/2-2P3/2} & 6.989 & 6.989 & 0.00204 & 0.000293 & 1.05E-14 & 5.94E-17 \\
H  I  6- 5 & 7.464 & 7.464 & 0.00205 & 0.000134 & 5.28E-15 & 6.19E-17 \\
H  I  8- 6 & 7.507 & 7.506 & 0.00229 & 0.002601 & 1.41E-15 & 5.96E-17 \\
{[NiI] a3F3-a3F4}$^*$ & 7.511 & 7.512 & 0.00124 & 0.002526 & 8.15E-16 & 6.47E-17 \\
H  I 11- 7$^*$ & 7.512 & 7.512 & 0.00124 & 0.002526 & 8.15E-16 & 6.47E-17 \\
H2 0-0 S(4) & 8.029 & 8.029 & 0.00234 & 0.000258 & 3.54E-15 & 7.24E-17 \\
H  I 10- 7 & 8.764 & 8.763 & 0.00518 & 0.000283 & 1.33E-15 & 4.96E-17 \\
{[ArIII] 3P1-3P2} & 8.996 & 8.996 & 0.00309 & 0.000473 & 1.69E-15 & 5.65E-17 \\
H2 0-0 S(3) & 9.670 & 9.670 & 0.00232 & 0.000277 & 1.14E-14 & 6.67E-17 \\
PAH & 11.2 & 11.255 & 0.00207 & 0.138935 & 7.22E-14 & 3.42E-15 \\
{[NiI] a3F2-a3F3}$^*$ & 11.314 & 11.315 & 0.00305 & 0.000548 & 4.52E-16 & 1.04E-16 \\
H  I  9- 7$^*$ & 11.315 & 11.315 & 0.00305 & 0.000548 & 4.52E-16 & 1.04E-16 \\
H2 0-0 S(2) & 12.285 & 12.286 & 0.00438 & 0.001052 & 2.08E-15 & 6.40E-17 \\
H  I  7- 6 & 12.379 & 12.379 & 0.00361 & 0.000304 & 3.59E-16 & 6.08E-17 \\
{[NeII] 2P1/2-2P3/2} & 12.821 & 12.821 & 0.00468 & 0.000790 & 3.05E-14 & 6.39E-17 \\
H2 8-8 S(4) & 13.407 & 13.409 & 0.00294 & 0.000509 & 4.18E-16 & 6.80E-17 \\
{[ClII] 3P1-3P2} & 14.376 & 14.374 & 0.00396 & 0.000981 & 5.61E-16 & 5.93E-17 \\
CO$_2$  & 15.1 & 15.097 & 0.69324 & 0.001591 & 1.68E-15 & 1.99E-12 \\
{[NeIII] 3P1-3P2} & 15.564 & 15.564 & 0.00627 & 0.000653 & 1.53E-14 & 6.56E-17 \\
H2 0-0 S(1) & 17.044 & 17.044 & 0.00556 & 0.000362 & 5.49E-15 & 9.18E-17 \\
{[PIII] 2P3/2-2P1/2} & 17.895 & 17.899 & 0.00666 & 0.001331 & 1.88E-15 & 1.05E-16 \\
{[SIII] 3P2-3P1} & 18.723 & 18.723 & 0.00827 & 0.000126 & 4.15E-14 & 4.12E-16 \\
{[ArIII] 3P0-3P1} & 21.842 & 21.845 & 0.00548 & 0.000324 & 1.30E-15 & 3.74E-16 \\
{[FeII] a6D7/2-a6D9/2} & 26.003 & 25.999 & 0.01179 & 0.000542 & 9.81E-15 & 1.29E-15 \\
\enddata
\tablecomments{Column 1: Name of line. Column 2: Laboratory wavelength. Column 3: Measured wavelength. Column 4: Error in measured wavelength. Column 5: FWHM of line. Column 5: Measured flux. Column 6: Error in flux. Emission lines marked with ``$*$" are blended. All lines are in emission except CO$_2$ which is in absorption.}
\label{tab:e_lines_Y535}
\end{deluxetable*}

\begin{deluxetable*}{lcccccc}
\tablecaption{Spectral Features for Source Y544A}
\tablehead{
\colhead{Name} & \colhead{Lab Wave} & \colhead{Meas Wave}& \colhead{Meas Wave Err} & \colhead{FWHM} & \colhead{Flux} & \colhead{Flux Err} 
\\
\colhead{}  & \colhead{($\mu$m)} & \colhead{($\mu$m)} & \colhead{($\mu$m)} & \colhead{($\mu$m)} & \colhead{($\mathrm{erg\;s^{-1}\;cm^{-2}}$)} & \colhead{($\mathrm{erg\;s^{-1}\;cm^{-2}}$)}
}
\decimals
\startdata
H  I 10- 6 & 5.131 & 5.132 & 0.00247 & 0.000258 & 3.47E-16 & 4.11E-17 \\
H2 0-0 S(7) & 5.514 & 5.514 & 0.00168 & 0.000963 & 1.99E-16 & 2.76E-17 \\
H  I 16- 7 & 5.528 & 5.528 & 0.00240 & 0.000504 & 1.16E-16 & 2.74E-17 \\
H  I 15- 7 & 5.715 & 5.715 & 0.00241 & 0.000120 & 1.29E-16 & 2.68E-17 \\
H  I  9- 6$^*$ & 5.911 & 5.912 & 0.00255 & 0.000241 & 3.30E-16 & 3.84E-17 \\
H2 9-8 Q(13)$^*$ & 5.913 & 5.912 & 0.00255 & 0.000241 & 3.30E-16 & 3.84E-17 \\
H2 0-0 S(6) & 6.112 & 6.112 & 0.00193 & 0.005509 & 4.77E-17 & 2.77E-17 \\
PAH & 6.2 & 6.229 & 0.00284 & 0.114896 & 7.23E-15 & 5.67E-16 \\
H  I 13- 7 & 6.295 & 6.296 & 0.00256 & 0.001461 & 1.66E-16 & 2.98E-17 \\
H  I 12- 7 & 6.776 & 6.775 & 0.00269 & 0.000053 & 1.50E-16 & 2.42E-17 \\
H2 0-0 S(5) & 6.913 & 6.914 & 0.00228 & 0.000008 & 4.14E-16 & 2.42E-17 \\
{[ArII] 2P1/2-2P3/2} & 6.989 & 6.989 & 0.00228 & 0.000134 & 1.28E-16 & 2.33E-17 \\
H  I  6- 5 & 7.464 & 7.464 & 0.00223 & 0.000276 & 9.72E-16 & 3.61E-17 \\
H  I  8- 6 & 7.507 & 7.506 & 0.00254 & 0.000916 & 3.84E-16 & 2.68E-17 \\
H2 0-0 S(4) & 8.029 & 8.031 & 0.00280 & 0.000610 & 1.92E-16 & 5.37E-17 \\
H2 6-6 S(6) & 8.745 & 8.740 & 0.00255 & 0.000381 & 1.92E-16 & 1.04E-16 \\
H  I 10- 7 & 8.765 & 8.766 & 0.00453 & 0.000210 & 2.91E-16 & 8.50E-17 \\
{[ArIII] 3P1-3P2} & 8.996 & 8.996 & 0.00300 & 0.000448 & 6.89E-16 & 1.17E-16 \\
H2 0-0 S(3) & 9.670 & 9.670 & 0.00251 & 0.000040 & 8.82E-16 & 1.39E-16 \\
{[SIV] 2P3/2-2P1/2} & 10.516 & 10.516 & 0.00343 & 0.000422 & 2.39E-15 & 1.95E-16 \\
PAH & 11.2 & 11.276 & 0.00147 & 0.152547 & 1.37E-14 & 4.22E-16 \\
{[NiI] a3F2-a3F3}$^{*}$ & 11.314 & 11.316 & 0.00376 & 0.001851 & 1.02E-16 & 1.37E-16 \\
H  I  9- 7$^*$ & 11.315 & 11.316 & 0.00376 & 0.001851 & 1.02E-16 & 1.37E-16 \\
H2 8-8 S(5) & 11.614 & 11.616 & 0.00287 & 0.000194 & 7.57E-17 & 1.08E-16 \\
H2 0-0 S(2) & 12.285 & 12.286 & 0.00463 & 0.000405 & 7.37E-16 & 7.55E-17 \\
H  I  7- 6 & 12.379 & 12.379 & 0.00513 & 0.000204 & 5.00E-16 & 7.27E-17 \\
H  I 11- 8 & 12.394 & 12.394 & 0.00401 & 0.000441 & 1.31E-16 & 6.83E-17 \\
PAH & 12.7 & 12.685 & 0.01159 & 0.249994 & 3.88E-15 & 5.71E-16 \\
{[NeII] 2P1/2-2P3/2} & 12.821 & 12.821 & 0.00389 & 0.000382 & 1.64E-15 & 6.59E-17 \\
{[NeIII] 3P1-3P2} & 15.564 & 15.564 & 0.00611 & 0.000740 & 7.45E-15 & 9.61E-17 \\
H  I 10- 8 & 16.218 & 16.219 & 0.00454 & 0.000393 & 1.02E-16 & 7.73E-17 \\
H2 0-0 S(1) & 17.044 & 17.044 & 0.00594 & 0.000948 & 1.45E-15 & 8.73E-17 \\
{[FeII] a4F7/2-a4F9/2} & 17.946 & 17.946 & 0.00599 & 0.000619 & 3.47E-16 & 8.87E-17 \\
{[SIII] 3P2-3P1} & 18.723 & 18.723 & 0.00824 & 0.000456 & 7.38E-15 & 4.58E-16 \\
H  I  8- 7 & 19.072 & 19.077 & 0.00982 & 0.001698 & 2.22E-16 & 4.09E-16 \\
{[ArIII] 3P0-3P1} & 21.842 & 21.845 & 0.00656 & 0.000560 & 2.53E-16 & 2.42E-16 \\
{[FeII] a6D7/2-a6D9/2}& 26.002 & 26.004 & 0.00969 & 0.002028 & 6.70E-16 & 3.88E-16 \\
\enddata
\tablecomments{Column 1: Name of line. Column 2: Laboratory wavelength. Column 3: Measured wavelength. Column 4: Error in measured wavelength. Column 5: FWHM of line. Column 5: Measured flux. Column 6: Error in flux. Emission lines marked with ``$*$" are blended. All features are in emission.}
\label{tab:e_lines_Y544A}
\end{deluxetable*}

\begin{deluxetable*}{lcccccc}
\tablecaption{Spectral Features for Source Y544B}
\tablehead{
\colhead{Name} & \colhead{Lab Wave} & \colhead{Meas Wave}& \colhead{Meas Wave Err} & \colhead{FWHM} & \colhead{Flux} & \colhead{Flux Err} 
\\
\colhead{}  & \colhead{($\mu$m)} & \colhead{($\mu$m)} & \colhead{($\mu$m)} & \colhead{($\mu$m)} & \colhead{($\mathrm{erg\;s^{-1}\;cm^{-2}}$)} & \colhead{($\mathrm{erg\;s^{-1}\;cm^{-2}}$)}
}
\decimals
\startdata
{[FeII] a4F9/2-a6D9/2} & 5.343 & 5.344 & 0.00185 & 0.001174 & 1.32E-16 & 1.84E-17 \\
H2 0-0 S(7) & 5.514 & 5.514 & 0.00162 & 0.000140 & 3.97E-16 & 2.32E-17 \\
H2 0-0 S(6) & 6.112 & 6.112 & 0.00187 & 0.000405 & 1.30E-16 & 1.82E-17 \\
PAH & 6.2 & 6.223 & 0.00491 & 0.113533 & 3.47E-15 & 4.77E-16 \\
{[NiII] 2D3/2-2D5/2} & 6.640 & 6.640 & 0.00226 & 0.000135 & 1.79E-16 & 1.79E-17 \\
H2 0-0 S(5) & 6.913 & 6.914 & 0.00215 & 0.000173 & 9.68E-16 & 2.77E-17 \\
{[ArII] 2P1/2-2P3/2} & 6.989 & 6.989 & 0.00196 & 0.002027 & 3.96E-17 & 1.51E-17 \\
H  I  6- 5 & 7.464 & 7.464 & 0.00285 & 0.000566 & 1.94E-16 & 1.65E-17 \\
H2 0-0 S(4) & 8.029 & 8.031 & 0.00295 & 0.000452 & 5.58E-16 & 4.14E-17 \\
{[ArIII] 3P1-3P2} & 8.996 & 8.996 & 0.00285 & 0.000190 & 3.51E-16 & 3.46E-17 \\
H2 0-0 S(3) & 9.670 & 9.670 & 0.00267 & 0.000117 & 1.56E-15 & 5.03E-17 \\
{[SIV] 2P3/2-2P1/2} & 10.516 & 10.516 & 0.00360 & 0.000498 & 2.26E-15 & 6.70E-17 \\
PAH & 11.2 & 11.285 & 0.00150 & 0.163314 & 1.28E-14 & 3.75E-16 \\
H2 0-0 S(2) & 12.285 & 12.286 & 0.00398 & 0.000534 & 8.58E-16 & 6.56E-17 \\
H  I  7- 6 & 12.379 & 12.379 & 0.00468 & 0.000190 & 2.08E-16 & 5.86E-17 \\
PAH & 12.7 & 12.699 & 0.00641 & 0.183492 & 2.77E-15 & 3.07E-16 \\
{[NeII] 2P1/2-2P3/2} & 12.821 & 12.821 & 0.00411 & 0.000393 & 1.19E-15 & 6.04E-17 \\
{[NeIII] 3P1-3P2} & 15.564 & 15.564 & 0.00601 & 0.000454 & 5.99E-15 & 7.96E-17 \\
PAH & 16.4 & 16.446 & 0.00712 & 0.123778 & 2.36E-15 & 4.31E-16 \\
H2 0-0 S(1) & 17.044 & 17.044 & 0.00584 & 0.000909 & 1.68E-15 & 7.44E-17 \\
{[FeII] a4F7/2-a4F9/2} & 17.946 & 17.946 & 0.00572 & 0.000296 & 4.04E-16 & 7.56E-17 \\
{[SIII] 3P2-3P1} & 18.723 & 18.723 & 0.00836 & 0.000370 & 5.80E-15 & 3.87E-16 \\
{[FeII] a6D7/2-a6D9/2} & 26.003 & 26.005 & 0.00973 & 0.001840 & 8.27E-16 & 4.10E-16 \\
\enddata
\tablecomments{Column 1: Name of line. Column 2: Laboratory wavelength. Column 3: Measured wavelength. Column 4: Error in measured wavelength. Column 5: FWHM of line. Column 5: Measured flux. Column 6: Error in flux. All features are in emission.}
\label{tab:e_lines_Y544B}
\end{deluxetable*}


\begin{deluxetable*}{lcccccc}
\tablecaption{Spectral Features Lines for Source Y532}
\tablehead{
\colhead{Name} & \colhead{Lab Wave} & \colhead{Meas Wave}& \colhead{Meas Wave Err} & \colhead{FWHM} & \colhead{Flux} & \colhead{Flux Err} 
\\
\colhead{}  & \colhead{($\mu$m)} & \colhead{($\mu$m)} & \colhead{($\mu$m)} & \colhead{($\mu$m)} & \colhead{($\mathrm{erg\;s^{-1}\;cm^{-2}}$)} & \colhead{($\mathrm{erg\;s^{-1}\;cm^{-2}}$)}}
\decimals
\startdata
H2 0-0 S(8) & 5.056 & 5.056 & 0.00107 & 0.002419 & 3.61E-16 & 1.57E-17 \\
H2 0-0 S(7) & 5.514 & 5.514 & 0.00173 & 0.000170 & 9.28E-16 & 1.56E-17 \\
H2 0-0 S(6) & 6.112 & 6.112 & 0.00169 & 0.005807 & 3.20E-16 & 1.42E-17 \\
PAH & 6.2 & 6.233 & 0.00331 & 0.152637 & 2.81E-14 & 1.93E-15 \\
H2 0-0 S(5) & 6.913 & 6.913 & 0.00209 & 0.000239 & 2.24E-15 & 1.07E-17 \\
{[ArII] 2P1/2-2P3/2} & 6.989 & 6.989 & 0.00232 & 0.002420 & 1.51E-16 & 1.04E-17 \\
H  I  6- 5 & 7.464 & 7.464 & 0.00231 & 0.000328 & 3.55E-16 & 1.12E-17 \\
PAH & 7.7 & 7.549 & 0.00716 & 0.197960 & 3.88E-14 & 3.48E-15 \\
PAH & 7.7 & 7.848 & 0.01006 & 0.061608 & 5.12E-15 & 2.01E-15 \\
H2 0-0 S(4) & 8.029 & 8.029 & 0.00237 & 0.000246 & 1.01E-15 & 1.32E-17 \\
PAH & 8.6 & 8.556 & 0.00988 & 0.181472 & 5.49E-15 & 1.20E-15 \\
{[ArIII] 3P1-3P2} & 8.996 & 8.996 & 0.00307 & 0.000119 & 4.15E-16 & 9.81E-18 \\
H2 0-0 S(3) & 9.670 & 9.670 & 0.00246 & 0.000417 & 2.83E-15 & 1.22E-17 \\
{[SIV] 2P3/2-2P1/2} & 10.516 & 10.518 & 0.00169 & 0.000164 & 5.79E-16 & 1.69E-17 \\
PAH & 11.2 & 11.264 & 0.00151 & 0.155031 & 3.33E-14 & 1.03E-15 \\
H2 0-0 S(2) & 12.285 & 12.286 & 0.00504 & 0.001422 & 5.83E-16 & 1.09E-17 \\
H  I  7- 6 & 12.378 & 12.378 & 0.00503 & 0.000703 & 4.71E-17 & 1.03E-17 \\
PAH & 12.7 & 12.757 & 0.00583 & 0.309945 & 1.57E-14 & 9.43E-16 \\
{[NeII] 2P1/2-2P3/2} & 12.821 & 12.821 & 0.00441 & 0.000606 & 1.19E-15 & 9.80E-18 \\
{[NeIII] 3P1-3P2} & 15.564 & 15.564 & 0.00689 & 0.000389 & 3.13E-15 & 1.35E-17 \\
H  I 15-10 & 16.421 & 16.421 & 0.00755 & 0.001117 & 2.84E-16 & 1.17E-17 \\
H2 0-0 S(1) & 17.044 & 17.044 & 0.00519 & 0.000213 & 1.18E-15 & 1.25E-17 \\
{[FeII] a4F7/2-a4F9/2} & 17.946 & 17.946 & 0.00629 & 0.000987 & 1.43E-16 & 1.39E-17 \\
{[SIII] 3P2-3P1} & 18.723 & 18.723 & 0.00827 & 0.000013 & 3.45E-15 & 5.28E-17 \\
\enddata
\tablecomments{Column 1: Name of line. Column 2: Laboratory wavelength. Column 3: Measured wavelength. Column 4: Error in measured wavelength. Column 5: FWHM of line. Column 5: Measured flux. Column 6: Error in flux. All features are in emission.}
\label{tab:e_lines_Y532}
\end{deluxetable*}

\begin{deluxetable*}{lcccccc}
\tablecaption{Spectral Features Lines for Source Y533A}
\tablehead{
\colhead{Name} & \colhead{Lab Wave} & \colhead{Meas Wave}& \colhead{Meas Wave Err} & \colhead{FWHM} & \colhead{Flux} & \colhead{Flux Err} 
\\
\colhead{}  & \colhead{($\mu$m)} & \colhead{($\mu$m)} & \colhead{($\mu$m)} & \colhead{($\mu$m)} & \colhead{($\mathrm{erg\;s^{-1}\;cm^{-2}}$)} & \colhead{($\mathrm{erg\;s^{-1}\;cm^{-2}}$)}
}
\decimals
\startdata
H2 0-0 S(7) & 5.514 & 5.514 & 0.00162 & 0.000609 & 4.63E-17 & 6.63E-18 \\
H2 0-0 S(6) & 6.112 & 6.112 & 0.00162 & 0.000607 & 2.47E-17 & 6.07E-18 \\
PAH & 6.2 & 6.223 & 0.00195 & 0.097054 & 1.52E-15 & 9.73E-17 \\
H2 4-4 S(8) & 6.352 & 6.355 & 0.00145 & 0.000097 & 2.09E-17 & 6.70E-18 \\
H  I 24- 8 & 6.568 & 6.567 & 0.00122 & 0.000522 & 1.32E-17 & 6.15E-18 \\
H2 0-0 S(5) & 6.913 & 6.914 & 0.00211 & 0.000513 & 1.45E-16 & 6.97E-18 \\
{[ArII] 2P1/2-2P3/2} & 6.989 & 6.989 & 0.00200 & 0.001585 & 2.48E-17 & 5.67E-18 \\
H  I  6- 5 & 7.464 & 7.464 & 0.00191 & 0.000490 & 7.96E-17 & 6.14E-18 \\
H  I  8- 6 & 7.507 & 7.506 & 0.00172 & 0.000350 & 2.12E-17 & 5.80E-18 \\
PAH & 7.7 & 7.587 & 0.00905 & 0.179019 & 4.49E-15 & 4.90E-16 \\
PAH & 7.7 & 7.863 & 0.01005 & 0.327222 & 1.18E-14 & 8.93E-16 \\
H2 0-0 S(4) & 8.029 & 8.029 & 0.00247 & 0.000894 & 9.38E-17 & 1.22E-17 \\
PAH & 8.6 & 8.553 & 0.02586 & 0.375588 & 3.54E-15 & 7.75E-16 \\
H2 6-6 S(6) & 8.745 & 8.742 & 0.00175 & 0.002735 & 1.24E-17 & 7.38E-18 \\
{[ArIII] 3P1-3P2} & 8.996 & 8.996 & 0.00301 & 0.000416 & 2.69E-16 & 8.34E-18 \\
H2 0-0 S(3) & 9.670 & 9.670 & 0.00261 & 0.000018 & 4.81E-16 & 9.43E-18 \\
{[SIV] 2P3/2-2P1/2} & 10.516 & 10.516 & 0.00306 & 0.000310 & 1.74E-15 & 2.25E-17 \\
PAH & 11.2 & 11.276 & 0.00084 & 0.182943 & 1.16E-14 & 1.69E-16 \\
H2 0-0 S(2) & 12.285 & 12.286 & 0.00438 & 0.000627 & 5.71E-16 & 1.59E-17 \\
H  I  7- 6 & 12.379 & 12.379 & 0.00429 & 0.000231 & 1.13E-16 & 1.05E-17 \\
H  I 11- 8 & 12.394 & 12.394 & 0.00371 & 0.000382 & 2.85E-17 & 9.99E-18 \\
PAH & 12.7 & 12.754 & 0.00438 & 0.260863 & 2.60E-15 & 1.39E-16 \\
{[NeII] 2P1/2-2P3/2} & 12.821 & 12.821 & 0.00408 & 0.000398 & 6.42E-16 & 1.48E-17 \\
CO$_2$ + CO  & 15.1 & 15.117 & 0.00816 & 0.353013 & 1.79E-14 & 6.54E-16 \\
CO$_2$ + H$_2$O  & 15.25 & 15.271 & 0.00130 & 0.052995 & 9.59E-16 & 6.65E-17 \\
CO$_2$ + CH$_3$OH  & 15.4 & 15.428 & 0.02256 & 0.391601 & 8.14E-15 & 7.39E-16 \\
{[NeIII] 3P1-3P2} & 15.564 & 15.564 & 0.00626 & 0.000843 & 4.76E-15 & 3.75E-17 \\
H  I 10- 8 & 16.218 & 16.219 & 0.00620 & 0.001812 & 3.09E-17 & 9.20E-18 \\
PAH & 16.4 & 16.428 & 0.01292 & 0.168880 & 1.64E-15 & 4.00E-16 \\
H  I 12- 9 & 16.890 & 16.889 & 0.00666 & 0.000201 & 1.64E-17 & 9.44E-18 \\
H2 0-0 S(1) & 17.044 & 17.044 & 0.00564 & 0.000804 & 1.41E-15 & 1.72E-17 \\
{[PIII] 2P3/2-2P1/2} & 17.895 & 17.896 & 0.00635 & 0.000117 & 1.67E-17 & 9.88E-18 \\
{[SIII] 3P2-3P1} & 18.723 & 18.723 & 0.00827 & 0.000386 & 4.42E-15 & 1.17E-16 \\
H  I  8- 7 & 19.072 & 19.071 & 0.00963 & 0.001830 & 5.33E-17 & 5.54E-17 \\
{[ArIII] 3P0-3P1} & 21.842 & 21.845 & 0.00962 & 0.001211 & 1.00E-16 & 5.22E-17 \\
\enddata
\tablecomments{Column 1: Name of line. Column 2: Laboratory wavelength. Column 3: Measured wavelength. Column 4: Error in measured wavelength. Column 5: FWHM of line. Column 5: Measured flux. Column 6: Error in flux. All lines are in emission except CO$_2$ + CO, CO$_2$ + H$_2$O and CO$_2$ + CH$_3$OH.}
\label{tab:e_lines_Y533A}
\end{deluxetable*}

\begin{deluxetable*}{lcccccc}
\tablecaption{Spectral Features for Source Y533B}
\tablehead{
\colhead{Name} & \colhead{Lab Wave} & \colhead{Meas Wave}& \colhead{Meas Wave Err} & \colhead{FWHM} & \colhead{Flux} & \colhead{Flux Err} 
\\
\colhead{}  & \colhead{($\mu$m)} & \colhead{($\mu$m)} & \colhead{($\mu$m)} & \colhead{($\mu$m)} & \colhead{($\mathrm{erg\;s^{-1}\;cm^{-2}}$)} & \colhead{($\mathrm{erg\;s^{-1}\;cm^{-2}}$)}
}
\decimals
\startdata
H2 0-0 S(4) & 8.029 & 8.029 & 0.00227 & 0.000560 & 3.10E-16 & 1.81E-17 \\
{[ArIII] 3P1-3P2} & 8.996 & 8.996 & 0.00316 & 0.000102 & 2.53E-16 & 1.44E-17 \\
H2 0-0 S(3) & 9.670 & 9.670 & 0.00274 & 0.000210 & 1.30E-15 & 1.25E-17 \\
{[SIV] 2P3/2-2P1/2} & 10.516 & 10.516 & 0.00332 & 0.000316 & 1.60E-15 & 1.96E-17 \\
PAH & 11.2 & 11.277 & 0.00097 & 0.183048 & 1.46E-14 & 2.48E-16 \\
H2 0-0 S(2) & 12.285 & 12.286 & 0.00458 & 0.000723 & 9.01E-16 & 1.99E-17 \\
H  I  7- 6 & 12.379 & 12.379 & 0.00435 & 0.000288 & 1.41E-16 & 8.97E-18 \\
PAH & 12.7 & 12.762 & 0.03349 & 0.221322 & 2.65E-15 & 1.28E-15 \\
{[NeII] 2P1/2-2P3/2} & 12.821 & 12.821 & 0.00411 & 0.000378 & 8.31E-16 & 8.13E-18 \\
{[NeIII] 3P1-3P2} & 15.564 & 15.564 & 0.00557 & 0.000328 & 4.20E-15 & 6.64E-18 \\
H  I 10- 8 & 16.218 & 16.219 & 0.00587 & 0.001002 & 2.92E-17 & 6.76E-18 \\
PAH & 16.4 & 16.449 & 0.00415 & 0.153591 & 1.16E-15 & 9.97E-17 \\
H2 0-0 S(1) & 17.044 & 17.044 & 0.00525 & 0.000586 & 1.66E-15 & 6.08E-18 \\
{[FeII] a4F7/2-a4F9/2} & 17.946 & 17.946 & 0.00715 & 0.000111 & 4.28E-17 & 5.97E-18 \\
{[SIII] 3P2-3P1} & 18.723 & 18.723 & 0.00819 & 0.000187 & 4.19E-15 & 1.04E-16 \\
H  I  8- 7 & 19.072 & 19.071 & 0.00925 & 0.001736 & 4.50E-17 & 4.63E-17 \\
\enddata
\tablecomments{Column 1: Name of line. Column 2: Laboratory wavelength. Column 3: Measured wavelength. Column 4: Error in measured wavelength. Column 5: FWHM of line. Column 5: Measured flux. Column 6: Error in flux. All features are in emission.}
\label{tab:e_lines_Y533B}
\end{deluxetable*}

\begin{deluxetable*}{lcccccc}
\vspace{-.5in}
\tablecaption{Spectral Features for Source Y533C}
\tablehead{
\colhead{Name} & \colhead{Lab Wave} & \colhead{Meas Wave}& \colhead{Meas Wave Err} & \colhead{FWHM} & \colhead{Flux} & \colhead{Flux Err} 
\\
\colhead{}  & \colhead{($\mu$m)} & \colhead{($\mu$m)} & \colhead{($\mu$m)} & \colhead{($\mu$m)} & \colhead{($\mathrm{erg\;s^{-1}\;cm^{-2}}$)} & \colhead{($\mathrm{erg\;s^{-1}\;cm^{-2}}$)}
}
\decimals
\startdata
PAH & 11.2 & 11.285 & 0.00120 & 0.189657 & 6.97E-15 & 1.41E-16 \\
H2 0-0 S(2) & 12.285 & 12.286 & 0.00450 & 0.000698 & 3.70E-16 & 8.60E-18 \\
H  I  7- 6 & 12.379 & 12.379 & 0.00387 & 0.000224 & 9.09E-17 & 4.60E-18 \\
PAH & 12.7 &  12.756 & 0.02127 &  0.230388 & 2.00E-15 & 5.89E-16 \\
{[NeII] 2P1/2-2P3/2} & 12.821 & 12.821 & 0.00400 & 0.000509 & 6.63E-16 & 1.09E-17 \\
{[NeIII] 3P1-3P2} & 15.564 & 15.566 & 0.00621 & 0.000821 & 4.82E-15 & 3.45E-17 \\
H  I 10- 8 & 16.218 & 16.219 & 0.00553 & 0.000035 & 1.80E-17 & 3.60E-18 \\
PAH & 16.4 & 16.422 & 0.00369 & 0.156374 & 7.24E-16 & 5.43E-17 \\
H  I 12- 9 & 16.890 & 16.889 & 0.00619 & 0.000715 & 1.18E-17 & 3.87E-18 \\
H2 0-0 S(1) & 17.044 & 17.044 & 0.00574 & 0.000828 & 8.89E-16 & 8.87E-18 \\
{[PIII] 2P3/2-2P1/2} & 17.895 & 17.896 & 0.00560 & 0.000045 & 1.39E-17 & 4.22E-18 \\
{[FeII] a4F7/2-a4F9/2} & 17.946 & 17.946 & 0.00490 & 0.000891 & 6.94E-18 & 4.13E-18 \\
{[SIII] 3P2-3P1} & 18.723 & 18.723 & 0.00877 & 0.000543 & 4.02E-15 & 9.37E-17 \\
H  I  8- 7 & 19.072 & 19.071 & 0.00848 & 0.001738 & 8.43E-17 & 2.67E-17 \\
{[ArIII] 3P0-3P1} & 21.842 & 21.845 & 0.00835 & 0.003803 & 7.42E-17 & 3.25E-17 \\
\enddata
\tablecomments{Column 1: Name of line. Column 2: Laboratory wavelength. Column 3: Measured wavelength. Column 4: Error in measured wavelength. Column 5: FWHM of line. Column 5: Measured flux. Column 6: Error in flux. All features are in emission.}
\label{tab:e_lines_Y533C}
\end{deluxetable*}

\begin{deluxetable*}{lcccccc}
\tablecaption{Spectral Features for Source Y533D}
\tablehead{
\colhead{Name} & \colhead{Lab Wave} & \colhead{Meas Wave}& \colhead{Meas Wave Err} & \colhead{FWHM} & \colhead{Flux} & \colhead{Flux Err} 
\\
\colhead{}  & \colhead{($\mu$m)} & \colhead{($\mu$m)} & \colhead{($\mu$m)} & \colhead{($\mu$m)} & \colhead{($\mathrm{erg\;s^{-1}\;cm^{-2}}$)} & \colhead{($\mathrm{erg\;s^{-1}\;cm^{-2}}$)}
}
\decimals
\startdata
H2 0-0 S(2) & 12.285 & 12.286 & 0.00435 & 0.000645 & 4.71E-16 & 1.00E-17 \\
H  I  7- 6 & 12.379 & 12.379 & 0.00454 & 0.000084 & 1.09E-16 & 4.84E-18 \\
PAH & 12.7 & 12.748 & 0.02487 & 0.250114 & 2.34E-15 & 7.46E-16 \\
{[NeII] 2P1/2-2P3/2} & 12.821 & 12.821 & 0.00446 & 0.000640 & 6.37E-16 & 1.02E-17 \\
{[NeIII] 3P1-3P2} & 15.564 & 15.566 & 0.00594 & 0.000645 & 4.44E-15 & 3.37E-17 \\
H  I 10- 8 & 16.218 & 16.219 & 0.00396 & 0.000377 & 1.47E-17 & 3.41E-18 \\
PAH & 16.4 & 16.396 & 0.00774 & 0.202260 & 7.45E-16 & 9.05E-17 \\
H2 0-0 S(1) & 17.044 & 17.044 & 0.00594 & 0.000883 & 9.92E-16 & 9.21E-18 \\
{[PIII] 2P3/2-2P1/2} & 17.895 & 17.896 & 0.00551 & 0.000070 & 1.09E-17 & 4.14E-18 \\
{[FeII] a4F7/2-a4F9/2} & 17.946 & 17.946 & 0.00830 & 0.000009 & 1.73E-17 & 4.12E-18 \\
\enddata
\tablecomments{Column 1: Name of line. Column 2: Laboratory wavelength. Column 3: Measured wavelength. Column 4: Error in measured wavelength. Column 5: FWHM of line. Column 5: Measured flux. Column 6: Error in flux. All features are in emission.}
\label{tab:e_lines_Y533D}
\end{deluxetable*}

\vspace{-2.65in}

\subsubsection{Iron, Nickel and Sulfur}
\label{sec:fe_ni_s}

Of the five sources with complete MRS spectra, we observe fine-structure emission from iron in Y535, Y544A, Y544B, and Y532 at 5.34, 17.94, or 26.0~$\mu$m. We do not detect iron emission in source Y533A, suggesting that this young source is not as strong in UV radiation or shocks as its higher-mass counterparts. Iron lines can often arise alongside nickel lines in the presence of dissociative shock fronts from protostellar outflows \citep{bib:Neufeld2009, bib:Gieser2023, bib:Anderson2013}. Our most massive source Y535 shows three strong nickel emission lines ([\ion{Ni}{2}] at 6.64~$\mu$m, [\ion{Ni}{1}] at 7.51~$\mu$m, [\ion{Ni}{1}] 11.31$\mu$m) suggesting the presence of an outflow. A single nickel emission line each is seen in Y544A and Y544B, though in the case of Y544A, it may be attributable to an \ion{H}{1} line at a similar wavelength. IFU slices of both observations show that the iron and nickel emission closely match the YSO locations. 

In YSOs, atomic sulfur, [\ion{S}{1}] at 25.2$\mu$m, is associated with non-dissociative C-type shocks \citep{bib:holl89, bib:neuf07, bib:Anderson2013}. While we do not observe this line, we find sulfur in the form of [\ion{S}{3}] (18.7~$\mu$m) or [\ion{S}{4}] (10.5$\mu$m) in our five main targets. IFU slices of the [\ion{S}{3}] line show a morphology generally following the H$_2$ structure in Y535. In Y544A, Y532, and Y533A, the emission closely follows the filamentary structure observed in NIRCam and MIRI imaging.

\subsubsection{Neon, Argon and Chlorine}
\label{sec:ne_ar_cl}

With ionization energies above 41 eV, neon emission in the form of [\ion{Ne}{2}] and [\ion{Ne}{3}] indicates the presence of high-energy UV photons and temperatures above 1000K, either from the central star or from high-velocity shocks. Similarly, argon [\ion{Ar}{2}] emission at 6.98$\mu$m requires temperatures above 2000 K \citep{bib:vanHoof2018,bib:Gieser2023}. We find emission from both neon and argon in all five main targets in the form of [\ion{Ne}{2}] at 12.8 $\mu$m, [\ion{Ne}{3}] at 15.5$\mu$m, [\ion{Ar}{2}] at 6.98 $\mu$m and [\ion{Ar}{3}] at 8.9 $\mu$m and 21.8$\mu$m. Slices of the IFU cube containing Y533A show that argon and neon emission is overall weak, slightly tracing the filament, with a slight suggestion of a PDR front seen in [\ion{Ne}{2}] at 12.8~$\mu$m. Diffuse neon emission in Y544 and Y532 follows the [\ion{S}{3}] emission, again tracing out the filamentary structure seen in wide-field imaging. The most massive source in our sample, Y535, shows a bright structure to its southwest in the IFU slices targeting argon which we discuss further in \autoref{fig:outflows_O20}. 

Y535 is the only source to show emission from chlorine, bearing the [\ion{Cl}{2}] line at 14.37~$\mu$m. Fine-structure emission lines from chlorine ([\ion{Cl}{2}], [\ion{Cl}{3}] and [\ion{Cl}{4}]) have been seen in the collimated jets from Herbig-Haro (HH) object in the Orion Molecular Cloud \citep{bib:mend21.1, bib:mend21.2}. In the LMC, \cite{bib:nayak2024} observed [\ion{Cl}{2}] emission in five massive YSOs, and suggested low-velocity ($<$~30~km~s$^{-1}$) jets and bow shocks as a possible origin.

\section{Analysis}
\label{sec:analysis}

In this section, we further characterize the sizes and evolutionary states of our targets via their SEDs. We analyze the evidence for resolved protostellar outflows and for the detection of CO$_2$ ices in our sources. We examine the PAH content in our sources and the evidence of dust processing with evolution. Finally, we infer the accretion rates of our sources and discuss how these properties relate to their evolutionary states.

\subsection{Photometry and SED Fitting}
\label{sec:SED_fit}

We use a combination of MRS spectroscopy and archival photometry from \textit{Spitzer} and \textit{Herschel} and new photometry from JWST's NIRCam and MIRI to assess the physical properties of our target YSOs (\autoref{tab:photom_arhival} and \autoref{tab:Y533_photom}). 
Sources Y535, Y544A, Y544B, and Y532 were saturated in NIRCam and MIRI imaging \citep{bib:habel24}. For these sources, we consulted  near- to far-IR photometry taken with \textit{Spitzer} IRAC and MIPS and with \textit{Herschel} PACS and SPIRE photometry \citep{bib:sewilo2013, bib:Seale2014}. Adding to this, we convolved our MRS spectra with the throughput curves for nine MIRI imaging filters: F560W, F770W, F1000W, F1130W, F1280W, F1500W, F1800W, F2100W and F2550W to obtain mir-IR photometry (\autoref{tab:convol_photom}).

We fit this photometry to SED models in order to estimate the sources' radii, temperatures, luminosities and masses. We use the YSO model grid ``spubhmi'' from \cite{bib:robi17} which contains 10,000 model YSOs spanning a broad range of multiple protostellar parameters such as stellar radius (0.1-100 R$_{\odot}$), stellar temperature(2000 -30,000 K), disk mass(\num{e-8}-\num{e-1} M$_{\odot}$), envelope density (\num{10e-24}-\num{10e-16} g cm$^{-3}$),  cavity density (\num{10e-23}-\num{10e-20} g cm$^{-3}$), and cavity opening angle (0$^{\circ}$-60$^{\circ}$). We adopt aperture sizes for each photometric band based on the empirical FWHM of each instrument and filter (\autoref{tab:apertures}). For our convolved MIRI photometry, we derive a photometric error for each source based on the effective instrumental noise determined for each of the twelve MRS spectral bands by \cite{bib:law2025}. These errors range from $\sim$0.3-10\%; in practice, they represent the lower limit of the size of the uncertainty. We note that while this choice of error estimation causes the $\chi^2$ parameter (goodness of fit) to be relatively large, it does not strongly impact the parameters of the YSO model chosen by the fitter. We find that relaxing this error estimate to $\sim$20\% for all MRS photometric measurements does not significantly influence the choice of best-fit model.

\begin{figure*}[t]
    \centering
    \includegraphics[width=\linewidth]{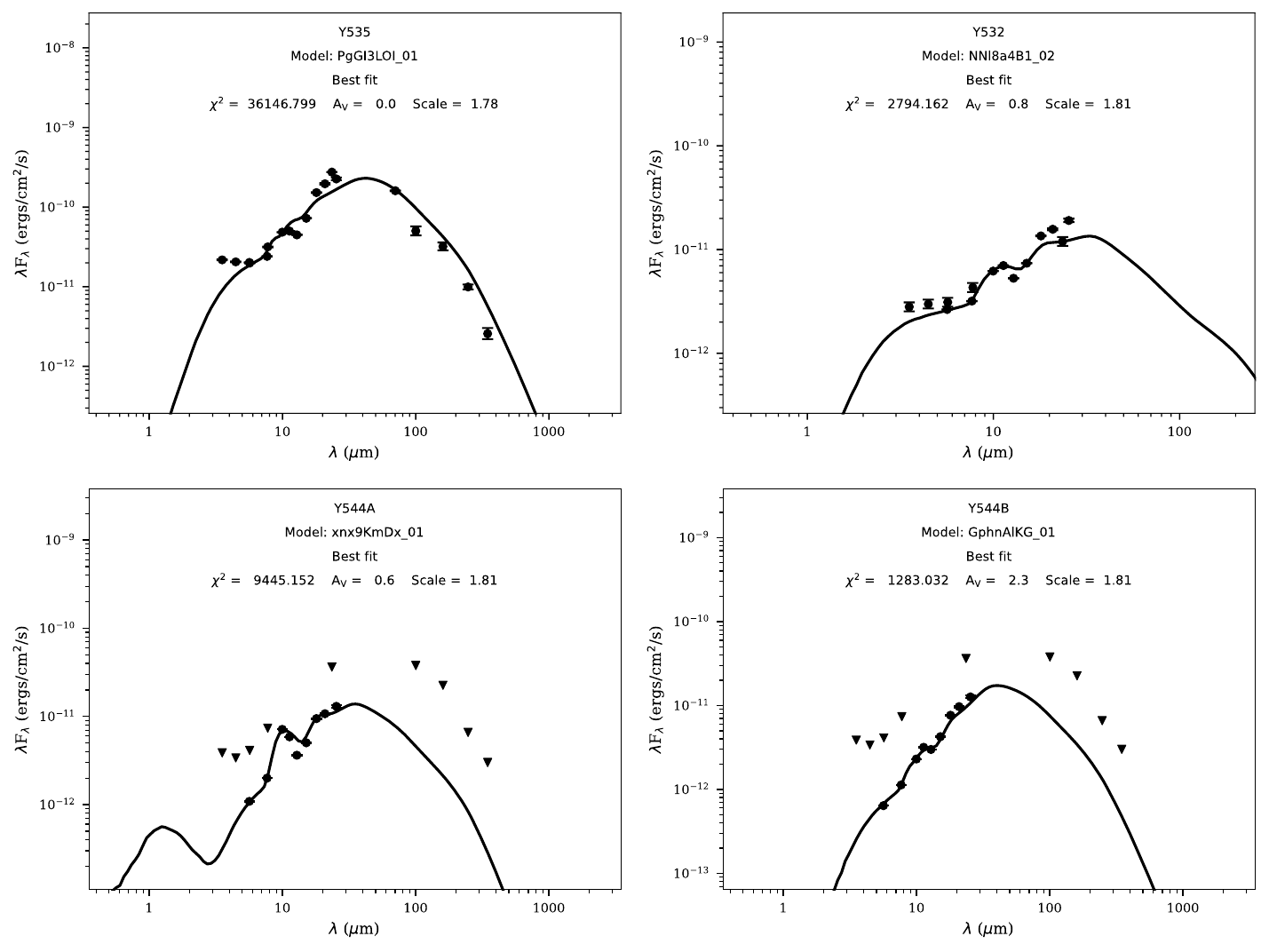}
    \caption{Best fit SEDs for sources Y535 (\textbf{top left}, Y532 (\textbf{top right}) and Y544A $\&$ B (\textbf{bottom left and right}). We combine \textit{Spitzer} IRAC and MIPS photometry, \textit{Herschel} PACS and SPIRE photometry with MRS spectroscopy convolved with nine MIRI imager filters (\autoref{tab:convol_photom}). Y532 does not have Herschel photometry. Archival photometry for Y535 and Y532 are fitted as data points (black dots) as both sources are isolated and dominate the mid-IR luminosity in their region. Because Y544A $\&$ B were not separately resolved by \textit{Spitzer} and \textit{Herschel}, we fit those measurements as upper limits (black triangles). The black line represents the YSO best-fit model. See \autoref{tab:fit_params} for the parameters of the best-fit YSO models.}
    \label{fig:sed_fits}
\end{figure*}

Because sources Y544A $\&$ B were not separately resolved by \textit{Spitzer} and \textit{Herschel}, we fit those photometric points as upper limits. The plotted SEDs (\autoref{fig:sed_fits}) show that MRS successfully disentangles these sources, with the convolved MIRI photometry sitting below the upper limits for each source. Y535 and Y532 are bright, isolated sources, thus, we assume they dominate the IR emission measured by \textit{Spitzer} and \textit{Herschel} and fit the archival photometry as data points. Y532 does not have \textit{Herschel} photometry.

\begin{figure*}[t]
    \centering
    \includegraphics[width=\linewidth]{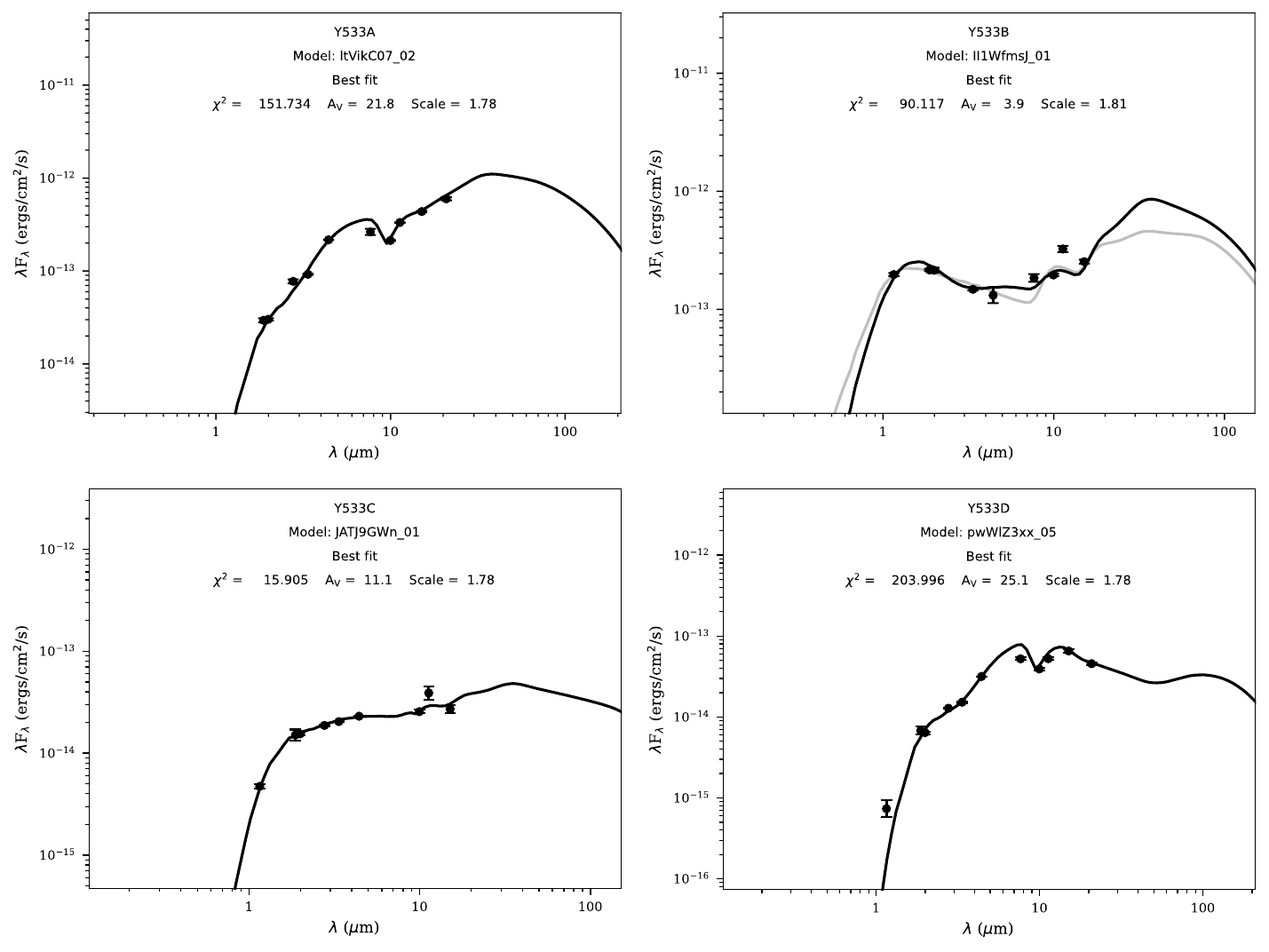}
    \caption{Best fit SEDS for sources Y533A, B, C an D. All photometry is from \citealt{bib:habel24} (see \autoref{tab:Y533_photom}) and spans eleven filters across NIRCam and MIRI from F115W to F2100W. The black line represents the best-fit YSO model. See \autoref{tab:fit_params} for the parameters of the best fit YSO models.}
    \label{fig:sed_fits_Y533}
\end{figure*}

For fitting the SEDs of YSOs captured in observation O23 (Y533A through E), we adopt the photometry from \cite{bib:habel24}, which spans eleven filters across NIRCam and MIRI (\autoref{tab:Y533_photom}). Source Y533E only has three photometric measurements and is excluded from this analysis. The remaining sources in observation O23 have nine or more measurements. The best fit model SEDs for these sources are shown in \autoref{fig:sed_fits_Y533}.

The best-fit models for sources Y535, Y544A, Y544B, Y532, Y533A, and Y533D are consistent with young, embedded YSOs, showing a characteristic rise in flux into the mid-IR. The fits from Y533B and Y533C indicate more evolved YSOs with a flatter continuum in near-IR and a smaller peak in the mid-IR. In all cases, the fitted SEDs closely trace the 10~$\mu$m silicate feature indicated by the photometry in either absorption or emission. We list the stellar radii, effective temperatures, luminosities, and stellar masses in \autoref{tab:fit_params}. At 18.0 M$_{\odot}$, we find Y535's mass to be consistent with previous estimates. Its temperature of 16190 K is, however, lower than previously inferred from the presence of H$_2$ and \ion{He}{1} emission lines in the near-IR. Such emission lines may be attributable to UV excitation or shocks from outflows impacting the surrounding interstellar material. Helium needs hard UV photons to ionize, with source temperatures in excess of 20,000 K required. However, helium may also be collisionally excited; \ion{He}{1} emission has been observed around low- and intermediate-mass T-Tauri and Herbig AeBe stars \citep{bib:Edwards2006, bib:Fischer2008, bib:Reiter2018}. Thus, such near-IR helium excitation is not a reliable indicator of the true temperature or evolutionary stage of the source itself.
Coupled with its overall continuum and the small amount of CO$_2$ ice absorption (see \autoref{sec:ice}, Y535's SED-derived parameters are consistent with it being a relatively young, massive YSO that is actively exciting its surroundings, but that has not fully cleared its envelope. 

Sources Y544A, Y544B, and Y532 have masses of roughly the same order (9.8, 9.7 and 9.6 M$_{\odot}$ respectively) Y544B and Y532 have the highest temperatures in our sample (both 20,000 K) consistent with their being more evolved than Y544A (10000 K). Y533A, B, C, and D are the least massive and coolest (7.1, 6.9, 2.9, and 4.0 M$_{\odot}$ and 8100, 7200, 6500, and 6700 K, respectively), consistent with their being the youngest, most embedded sources in our sample.

\begin{figure*}[h]
\centering
\vspace{-.25in}
\includegraphics[width=.9\linewidth]{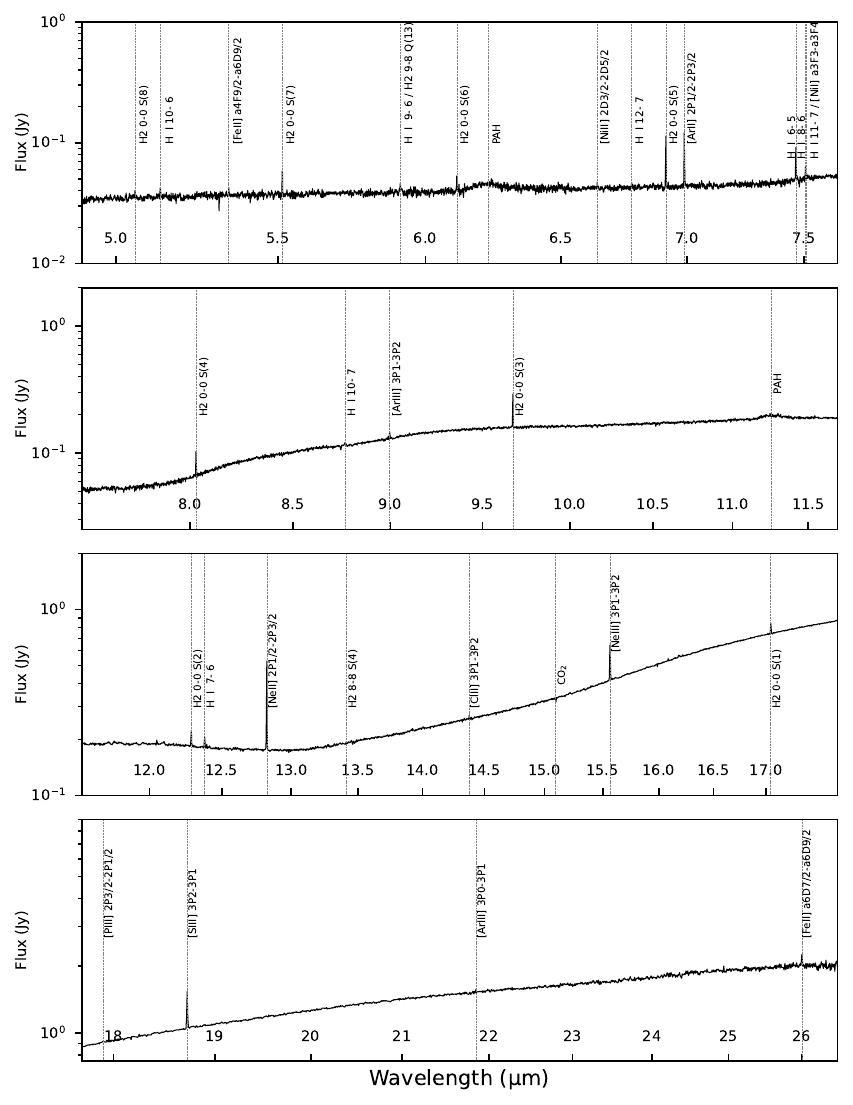}\hfill
\caption{Spectra for channels 1-4 of source \textbf{O20 Y535}. Full line list is given in \textbf{\autoref{tab:e_lines_Y535}}.}
\label{fig:spec_Y535}
\end{figure*}

\begin{figure*}
\centering
\vspace{-.25in}
\includegraphics[width=.9\linewidth]{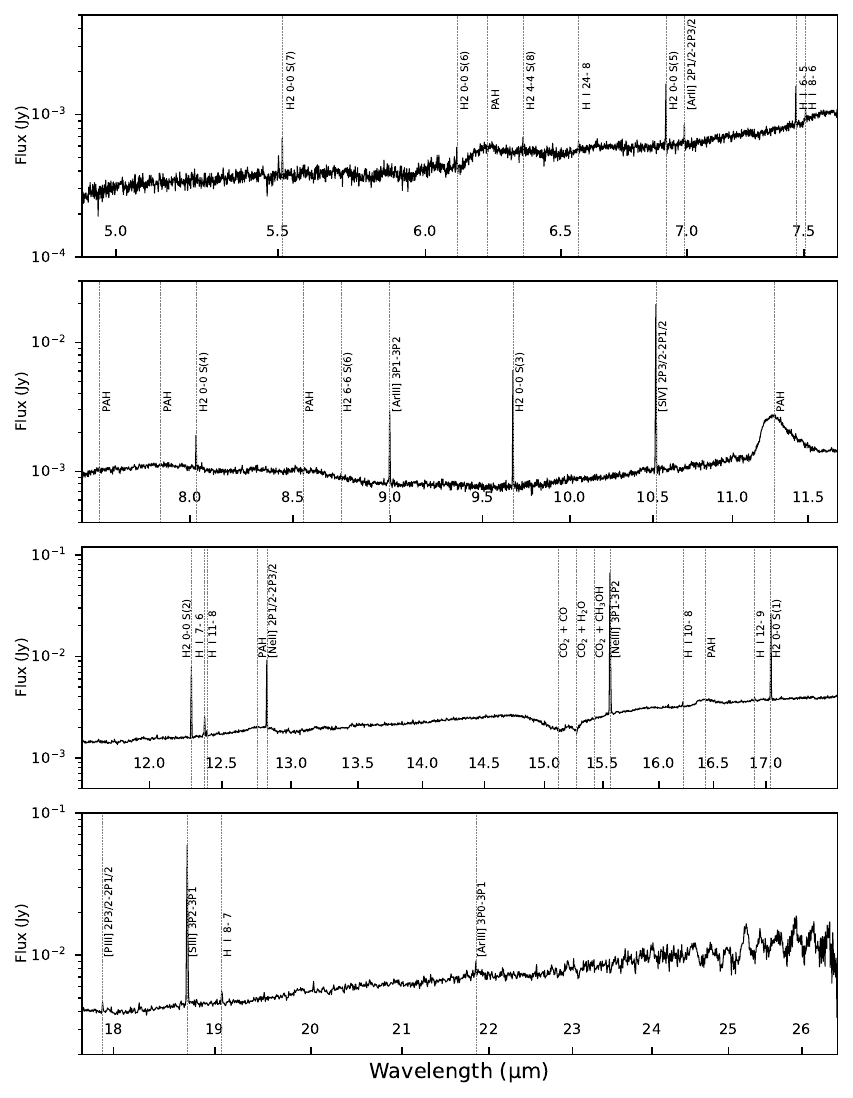}\hfill
\caption{Spectra for channels 1-4 of source \textbf{O23 Y533A}. The full line list is given in \textbf{\autoref{tab:e_lines_Y533A}}.}
\label{fig:spec_Y533A}
\end{figure*}

\subsection{Evidence for an Extended Jet Toward Y535}
\label{sec:jets}

\begin{figure*}
    \centering
    \includegraphics[width=\linewidth]{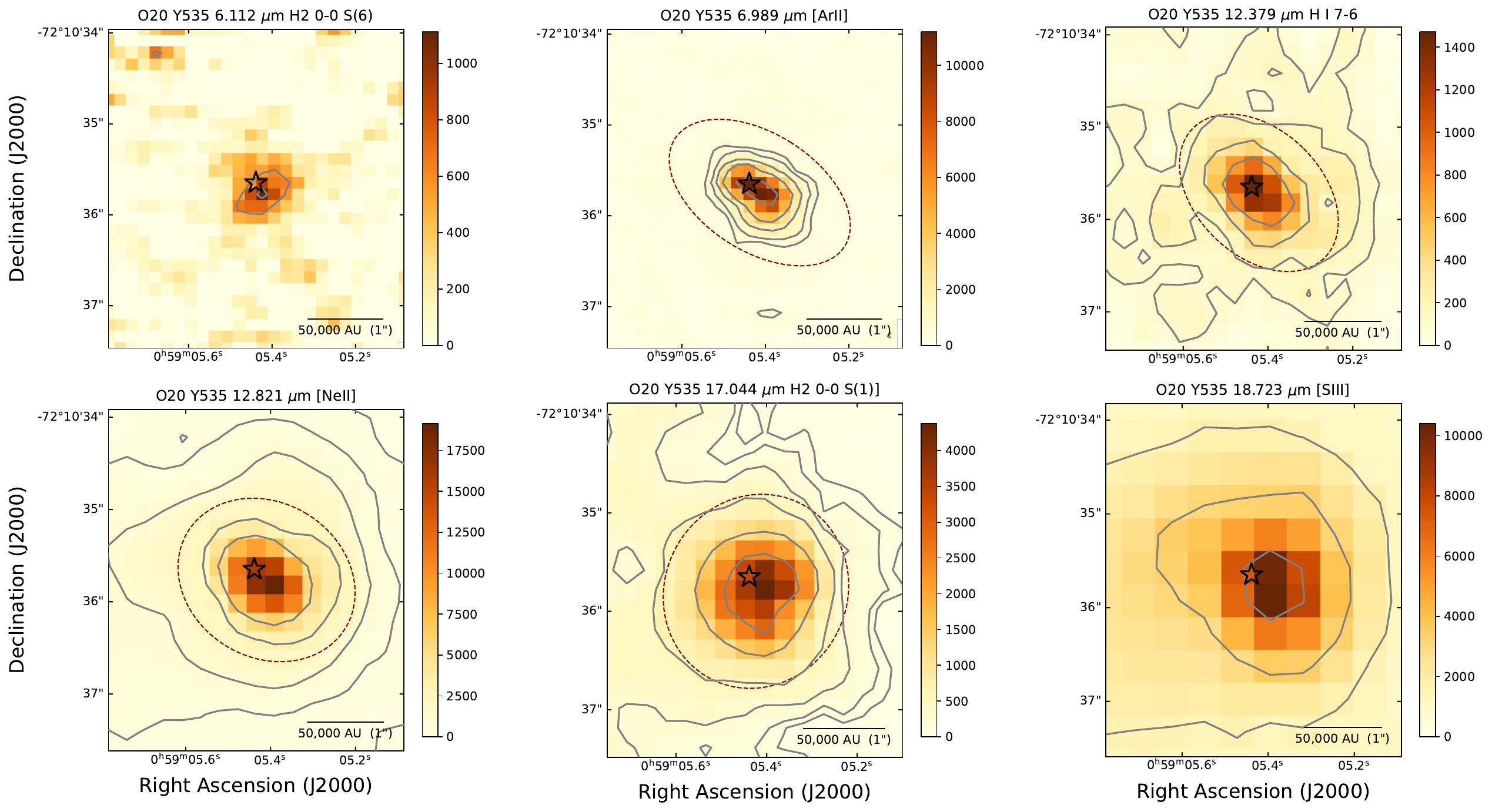}
    \caption{Extended emission tracing an outflow surrounding the source Y535 as seen in narrow-line emission. IFU cubes are continuum-subtracted to remove emission from the central source (marked with a star). H$_2$ 0-0 S(6) emission at 6.112~$\mu$m (\textbf{top left}), {[\ion{Ar}{2}]} emission at 6.989~$\mu$m (\textbf{top middle}), {\ion{H}{1}} emission at 12.379~$\mu$m (\textbf{top right}), {[\ion{Ne}{2}]} emission at 12.821~$\mu$m (\textbf{bottom left}), H$_2$ 0-0 S(1) emission at 17.044~$\mu$m (\textbf{bottom middle}) and {[\ion{S}{3}]} emission at 18.723~$\mu$m (\textbf{bottom right}). Cube slices are subtracted using the median continuum. Contours levels are drawn at 3, 5, 10, 20, 50, 100 and 500 times the standard deviation of the emission of a representative off-source patch. The continuum is the median of a sub-cube slice with a width of 0.05~$\mu$m centered at the given wavelength. Red ovals trace elliptical isophotes measuring the shape of the outflow (\autoref{tab:ellipticities}).}
    \label{fig:outflows_O20}
\end{figure*}

Protostellar jets and outflows have been commonly observed in galactic YSOs. These feedback mechanisms are thought to play an important roll in removing excess angular momentum from the protoplanetary disk during the accretion phase, allowing star formation to proceed \citep{bib:Ray_2021_jets}. Y535, the most luminous and massive YSO in our sample, shows evidence in its IFU cube of extended structure characteristic of a protostellar outflow or jet. We examine the morphology of this emission in the IFU cube slices, in particular slices centered on detected emission lines that are associated with YSO outflows. Slices at continuum wavelength show emission centered on the Y535. In general, extended structure is difficult to see against the strong continuum and line emission of the central source. Thus we construct and examine continuum-subtracted slices at several spectral features to isolate line emission. (\autoref{fig:outflows_O20}). To do this, first we find the median emission value of each spaxel in an IFU sub-cube centered on various emission features and spanning a width of 0.2$\mu$m, effectively creating a continuum map. Next we subtract this continuum map from the IFU slice centered on the emission line, leaving only line emission.

Molecular hydrogen is a common tracer of the local environment surrounding YSOs, potentially revealing shock-heating in outflows or UV-heating in photo-dissociation regions (PDRs),  \citep{bib:peet17, bib:knig21}.  In Y535, the majority of H$_2$  emission is tightly constrained about source Y535 (H$_2$ at 5.5~$\mu$m and 17.0~$\mu$m in \autoref{fig:slices_O20}).
When inspecting the continuum-subtracted H$_2$ emission, we see that its peak is slightly offset to the southwest (H$_2$ at 6.112 and 17.0~$\mu$m \autoref{fig:outflows_O20}), consistent with a protostellar outflow.

We find similar morphology in fine-structure line emission, which can stem from shock heating, making these lines strong tracers of outflows \citep{bib:drai93, bib:holl97}.
The continuum-subtracted 6.98~$\mu$m [\ion{Ar}{2}] IFU slice reveals a clear, narrow structure extending from Y535 to the southwest with a length of $\sim$0.5'' (\autoref{fig:outflows_O20}). We observe a similar morphology in 12.82~$\mu$m [\ion{Ne}{2}] and 18.72~$\mu$m [\ion{S}{3}] emission. Continuum-subtracted slices at other fine-structure emission lines detected in Y535 ([\ion{Fe}{2}], [\ion{Ni}{2}], [\ion{Ni}{1}], [\ion{Ar}{3}], [\ion{Cl}{3}], [\ion{Ne}{3}], [\ion{P}{3}]) contain overall less emission, and do not clearly show an organized structure.

Emission from hydrogen recombination lines also trace out a similar morphology, suggesting an ionized gas component in the outflow. \ion{H}{1} at 12.37~$\mu$m shows emission roughly centered on Y535, but elongated in the northeast and southwest directions.

At the distance of the SMC, the angular size of these structures suggests an outflow extending at least 0.15 pc or 30,000 AU. Such a size is approximately three times the canonical size of a protostellar envelope, and is consistent in extent with jets observed in galactic YSOs \citep{bib:Bally_2012_jets, bib:Qiu_2019_jets, bib:Ray_2021_jets, bib:habel21}. That this feature is traced by both H$_2$ and fine-structure emission suggests both low ($<$30~km~s$^{-1}$) and high-velocity ($>$70~km~s$^{-1}$) shocks are present. Comparing the morphology of the outflow feature between tracers, we see the suggestion of a wide, low-velocity wind (traced by 17.04~$\mu$m H$_2$ in \autoref{fig:outflows_O20}) and a narrower, high-velocity jet (traced by 6.98~$\mu$m [\ion{Ar}{2}] in \autoref{fig:outflows_O20}). To assess this, we fit elliptical isophotes about the emission structure in the continuum subtracted slices where possible for the detected emission lines in Y535. For each slice, we average the ellipticity of the fitted isotopes with semi-major axes extending from $\sim$0.5" to 1.0". We list the emission lines and ellipticities in \autoref{tab:ellipticities}. 
The generally larger (and thus flatter) ellipticity values for the \ion{H}{1} and atomic lines indicate a narrower structure in their emitting region than for the comparatively rounder H$_2$ structure. We note that the increasing PSF size toward the reddest MRS channels makes confidently assessing the shape of the outflow feature difficult at the longest wavelengths.

\begin{table}
    \centering
    \begin{tabular}{||lcccccr||}
    \hline
        Emmission Line  & & &  Wavelength  & &  & Ellipticity \\
    \hline
    \hline
    \multicolumn{7}{|c|}{Molecular Hydrogen Lines}\\
    \hline
        H$_2$~0-0 S(5)  & & &    6.914~$\mu$m  & & & 0.201 \\
        H$_2$~0-0 S(4)  & & &    8.029~$\mu$m  & & & 0.152 \\
        H$_2$~0-0 S(3)  & & &    9.670~$\mu$m  & & & 0.155 \\
        H$_2$~0-0 S(2)  & & &    12.286~$\mu$m & & & 0.116 \\
        H$_2$~0-0 S(1)  & & &    17.044~$\mu$m & & & 0.074  \\
    \hline
    \hline
    \multicolumn{7}{|c|}{Atomic Fine Structure \& Hydrogen Recombination Lines}\\
    \hline
        {[\ion{Ar}{2}] 2P1/2-2P3/2}    & & &  6.914~$\mu$m    & & & 0.413 \\
        \ion{H}{1}~6-5                 & & &  7.464~$\mu$m    & & & 0.431 \\
        \ion{H}{1}~7-6                 & & &  12.379~$\mu$m   & & & 0.318 \\
        {[NeII] 2P1/2-2P3/2}           & & &  12.821~$\mu$m   & & & 0.161 \\
        {[SIII] 3P2-3P1}               & & &  18.723~$\mu$m   & & & 0.114 \\
    \hline
    \end{tabular}
    \caption{Averaged ellipticity measurements of the outflow structure observed about Y535. Elliptical isophotes (\autoref{fig:outflows_O20}) fitted between 0.5'' and 1.0'' about extended emission show a narrower morphology for fine structure and hydrogen recombination lines than H$_2$ lines, suggesting both a narrow jet and a wide molecular wind component to the outflow.}
    \label{tab:ellipticities}
\end{table}

Inspecting the IFU cube about the wavelengths of the emission line tracers does not reveal distinct blue- or redshifted structures. Additionally, we note the same analysis does not reveal similar extended features about our other YSO sources; the continuum-subtracted slices do not reveal morphologies distinct from the filaments described in \autoref{sec:cubes}. JWST imaging reveals that Y535 is surrounded by several, nearly-as-bright sources in the near-IR (1.15~$\mu$m) within the immediate $\sim$1" \citep{bib:jones2023, bib:habel24}. These sources all fade rapidly toward the mid-IR. 
The lack of any clear mid-IR continuum emission at the location of the extended structure about Y535 makes a contaminating source unlikely. JWST's NIRSpec IFU, with a higher resolution than MIRI MRS, and a similar ability to detect near-IR emission lines tracing jets and outflows, could decisively rule out such a companion and further characterize the nature of the outflow.

\begin{table}
\setlength{\tabcolsep}{8pt}
    \centering
    \begin{tabular}{||l|cccc||}
    \hline
     Source ID    & \textbf{Y535} & \textbf{Y544A} & \textbf{Y544B} & \textbf{Y532}  \\
     \hline
     \hline
     \multicolumn{5}{|c|}{Convolved MIRI Photometry (AB Magnitudes)}\\
     \hline
     F560W   &  12.45 & 15.62 & 16.19 & 14.65   \\
     F770W   &  11.92 & 14.62 & 15.24 & 14.12    \\
     F1000W  &  10.88 & 12.96 & 14.18 & 13.11    \\
     F1130W  &  10.71 & 13.03 & 13.70 & 12.84     \\
     F1280W  &  10.69 & 13.42 & 13.63 & 13.01     \\
     F1500W  &  9.99  & 12.89 & 13.07 & 12.47    \\
     F1800W  &  8.99  & 12.01 & 12.24 & 11.62    \\
     F2100W  &  8.56  & 11.71 & 11.82 & 11.30   \\
     F2550W  &  8.19  & 11.29 & 11.32 & 10.87   \\

    \hline
    \end{tabular}
    \caption{MIRI photometry for four of our five primary targets produced by convolving MRS spectra with JWST MIRI imaging filter throughputs. Photometry data is presented in AB magnitudes.}
    \label{tab:convol_photom}
\end{table}

To date, there are few examples of resolved extragalactic protostellar outflows. With \textit{Spitzer} and \textit{HST}, \cite{bib:Chu_2005} made the first extragalactic identification of an HH object toward a YSO in the superbubble N51D within the LMC. They identified structures with a high [\ion{S}{2}]/H$\alpha$ ratio and a morphology suggestive of outflow knots. Also in the LMC, \cite{bib:ward2016} observed compact H$_2$ emission suggestive of a molecular outflow about a YSO in near-IR IFU spectroscopy using the VLT's Spectrograph for INtegral Field Observations in
the Near Infrared (SINFONI). Most recently, \cite{bib:McLeod_2024_kep_disk_MC}, discovered an optical, parsec-scale jet from a massive YSO in the LMC with the VLT's Multi Unit Spectroscopic Explorer (MUSE) IFU spectrograph. ALMA has also seen evidence for outflows around massive LMC YSOs in CO mapping \citep{bib:Fukui_2015, bib:Shimonishi_2016}. Until now, evidence of protostellar outflows in the Magellanic Clouds has been limited to the LMC. In this work, we suggest that the morphology detected about YSO Y535 is \textit{the first detection of a resolved protostellar outflow within the SMC, the first detected in the mid-IR, and is the most distant extragalactic protostellar outflow currently known.}

\begin{table}
    \centering
    \begin{tabular}{||l|rrrr||}
    \hline
        Source ID & Radius & Temperature  & Luminosity & Mass \\
         &  ($\mathrm{R_{\odot}}$) & (K)~~ & ($\mathrm{L_{\odot}}$)~~ & ($\mathrm{M_{\odot}}$) \\
         \hline
         \hline
       Y535  & 20 & 16000 & 25000 & 18.0\\
       Y544A & 16 & 10000 & 2900 & 9.8\\
       Y544B & 4.6 & 20000 & 2900 & 9.7\\
       Y532  & 4.5 & 20000 & 2800 & 9.6\\
       Y533A & 16 & 8100 & 950 & 7.1\\
       Y533B & 19 & 7200 & 850 & 6.9\\
       Y533C & 5.2 & 6500 & 44 & 2.9\\
       Y533D & 8.7 & 6700 & 130 & 4.0\\

         \hline
    \end{tabular}
    \caption{Table of YSO parameters as determined from SED fitting. Luminosity is calculated using the best-fit radius and best-fit temperature with the following formula: $\mathrm{L\;=\;4 \pi r^{2} \sigma T^{4}}$. Mass is calculated using $\mathrm{L \propto M^{3.5}}$.}
    \label{tab:fit_params}
\end{table}

\subsection{CO$_2$ Ice Detections}
\label{sec:ice}

Four of our YSOs display absorption in their spectra which we identify as the CO$_2$ 15.2~$\mu$m ice feature. The presence of solid-phase CO$_2$ ice absorption is an indicator that a YSO is still very young and still embedded within its envelope. In Y535, our most massive source, we find a slight, narrow absorption line centered on 15.09~$\mu$m (\autoref{fig:targ_spec_compare}, \autoref{fig:spec_Y535}). In three of the least massive sources in our sample, Y533A, Y533C and Y533D, this feature is seen as a broad doublet (\autoref{fig:targ_spec_compare_Y533}, \autoref{fig:spec_Y533A}). This feature is clearest in Y533A, where it extends from $\sim$14.8 to 15.8~$\mu$m with troughs centered at 15.15 and 15.27~$\mu$m.

The 15.2~$\mu$m absorption feature of CO$_2$ stems from its bending mode. The complex doublet profile of this feature results when CO$_2$ is present in mixtures with other molecules. \cite{bib:Pontoppidan_2008_co2decomp} showed that this absorption can be decomposed into multiple components attributed to CO$_2$ embedded in ice matrices with CO, H$_2$O and CH$_3$OH, or pure CO$_2$ ice alone \citep{bib:Brunken_2024_co2}. 
In an environment rich in H$_2$O, CO and CO$_2$ and heated by a central YSO, molecules of CH$_3$OH will form, in turn manifesting in the spectra as broadened absorption on the redder, 15.4~$\mu$m side \citep{bib:terw21}. CO$_2$ mixed with CO contributes to the blue side of the absorption, and pure CO$_2$ contributes a double trough in the center \citep{bib:Brunken_2024_co2, bib:Brunken2025_co2}.

\begin{figure}
    \centering
    \includegraphics[width=1\linewidth]{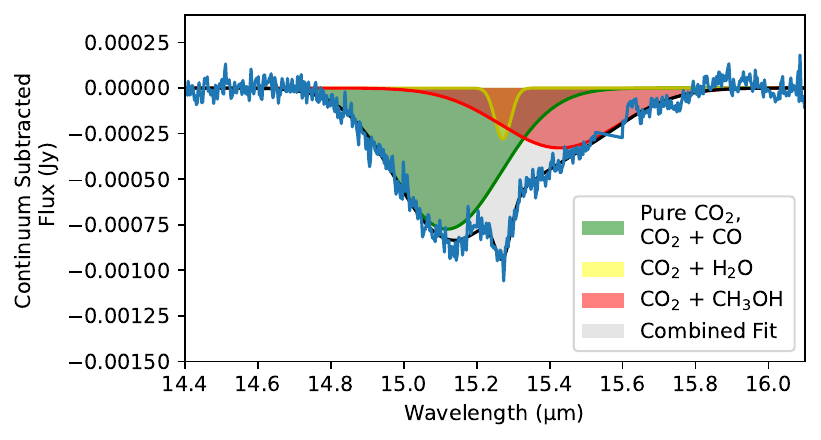}
    \caption{The 15.2~$\mu$m absorption feature found in the spectrum of Y533A. We find a best fit using a combination of three Gaussians. The spectral decomposition indicates the presence of CO$_2$ ice mixed with H$_2$O (\textbf{yellow}), CO$_2$ ice mixed with CH$_3$OH (\textbf{red}) and pure CO$_2$ along with CO$_2$ ice mixed with CO (\textbf{green}).}
    \label{fig:gaussains}
\end{figure}

The CO$_2$ absorption line for Y535 is weak, appearing as a single narrow feature. Thus we treat it similarly to the emission lines, fitting a single Gaussian profile. We report its emission parameters in \autoref{tab:e_lines_Y535}. In Y533A, which shows a doublet CO$_2$ feature, we measure the absorption profile by using a combination of three Gaussians fitted to the continuum-subtracted spectra, shown in \autoref{fig:gaussains}, and the parameters listed in \autoref{tab:e_lines_Y533A}. The first component (in green in \autoref{fig:gaussains}) is responsible for the majority of absorption in the bluer trough, and is consistent with pure CO$_2$ and CO$_2$ in a mixture with CO. The second (in yellow) fits the redder trough, tracing contributions from CO$_2$ and H$_2$O. The final component (in red) is responsible for tracing the red shoulder of the absorption profile created by the CO$_2$, CH$_3$OH ice mixture. Y533C and Y533D also show CO$_2$ doublet absorption, however we are unable to reliably fit these due to the significant noise in their spectra. The spectra of Y533E also shows a 15.2~$\mu$m absorption feature, but is likely contaminated by Y533A, thus a positive detection of CO$_2$ cannot be made.


The presence of the CO$_2$ ice feature in Y535 is slightly at odds with the other characteristics of this source which indicate that it is a Stage I YSO in a more evolved state than Y533A, Y533C and Y533D. However, observations of similarly evolved, massive YSOs in the LMC show ice absorption (e.g. HCN ice in the YSO ``Y6" in \citealt{bib:nayak2024}). Thus this absorption in Y535 may point to enhanced ice content relative to our other targets, or alternatively may be explained by an edge-on orientation of its disk causing additional absorption. 

The prominent absorption doublets seen in Y533A, Y533C and Y533D indicate that these sources are still surrounded in a dense, cool envelope and are at an earlier stage of evolution than our other ice-poor targets. These sources also show strong PAH emission relative to other sources (see \autoref{sec:pahs}), again suggesting denser, younger envelopes. Finally, Y533A, the only ice-bearing source with spectral coverage at 10~$\mu$m, is also the only source with a spectra showing tentative silicate absorption in that wavelength region, further indicating that it is the least evolved of the five main targets.


We detect no other spectral features linked to solid- or gas-phase ice absorption in any of our spectra. In contrast to our results, similar MRS observations of massive protostars in the LMC displayed numerous species of ice, and gas phase absorption such as CH$_4$, NH$_3$, CH$_3$OH, CH$_3$OCHO, and CO$_2$ \citep{bib:nayak2024}. Though the small sample sizes in both studies and little overlap in the mass range make direct comparison difficult, this may potentially point to a lower ice content in the SMC relative to YSOs in the LMC. As with the relative scarcity and weakness of PAH features (see \autoref{sec:pahs}) in our sample as compared with those found by \citeauthor{bib:nayak2024} in the LMC, this lower ice content may be linked to the lower metallicity of the SMC. We note that the profile we observe in the 15.2~$\mu$m absorption matches closely to those found by \citealt{bib:nayak2024}, which were also best fit by a combination of three Gaussian, suggesting CO$_2$ ice mixtures in similar ratios between the regions. This aligns with recent JWST NIRSpec observations of 1.8-4.1 M$_{\odot}$ YSOs in NGC 346, which found overall lower column densities of CO, CO$_2$, and H$_2$O than in the LMC, but at similar ratios (de Marchi et al., submitted).


\subsection{PAH Emission}
\label{sec:pahs}


Playing an important role in balancing photoionization with recombination, PAHs are commonly observed in dusty sources such as YSOs, evolved asymptotic giant branch stars, \ion{H}{2} regions, dusty filaments in star-forming regions, and planetary and reflection nebulae. In YSOs, PAH features are excited alongside fine-structure emission lines as the central ionizing source evolves and emits UV photons. In the mid-IR, PAHs are often seen in emission at 6.2, 7.7, 8.6, 11.2, 12.7, and 16.4~$\mu$m \citep{bib:hony01, bib:peet02, bib:shan16}. Mid-IR PAH features have various origins. Between 5 and 10~$\mu$m, PAH features originate from the pure CC stretching mode and the CH in-plane bending mode \citep{bib:jobl96,bib:hony01}; between 10 and 15~$\mu$m, they originate from the out-of-plane bending vibrations of aromatic hydrogen atoms \citep{bib:hony01}; between 15 and 20~$\mu$m they originate from CCC modes \citep{bib:smit07}. 

\subsubsection{Detected and Measured PAH Features}

We find PAH emission in each of our YSO spectra (\autoref{fig:targ_spec_compare} \& \autoref{fig:targ_spec_compare_Y533}). 
To measure their emission strengths, we first perform a local continuum subtraction via spline fitting after selecting reference points on the continuum away from the PAH features and after eliminating strong narrow emission lines. We next fit Gaussian profiles to the PAH emission features and extract their emission parameters (listed in Figures \ref{tab:e_lines_Y535}, \ref{tab:e_lines_Y544A}, \ref{tab:e_lines_Y544B}, \ref{tab:e_lines_Y532}, \ref{tab:e_lines_Y533A}, \ref{tab:e_lines_Y533B}, \ref{tab:e_lines_Y533C}, \ref{tab:e_lines_Y533D} and \ref{tab:e_lines_Y533E}). For the 7.7~$\mu$m PAH feature, we observe a slight double emission line profile with peaks at $\sim$7.6 and $\sim$7.8~$\mu$m and accordingly fit the emission using a combination of two Gaussians.


We observe emission from the 6.2~$\mu$m PAH feature in all five of our main targets (Y535, Y544A, Y544B, Y532, and Y533A). This asymmetric red rail has also been observed in 57 dusty sources (YSOs, planetary nebulae, and galaxies) by \cite{bib:peet02} and may be attributable to a combination of stretching modes 6.2~$\mu$m and bending modes at 6.3~$\mu$m. The detected peaks of the emission range from 6.223~$\mu$m to the slightly red-shifted 6.233~$\mu$m. \cite{bib:nayak2024} observed a similar spread in the peak of this PAH feature in massive YSOs in the N79 region of the LMC. 

\begin{figure}
  \centering
  {\includegraphics[width=0.45\textwidth]{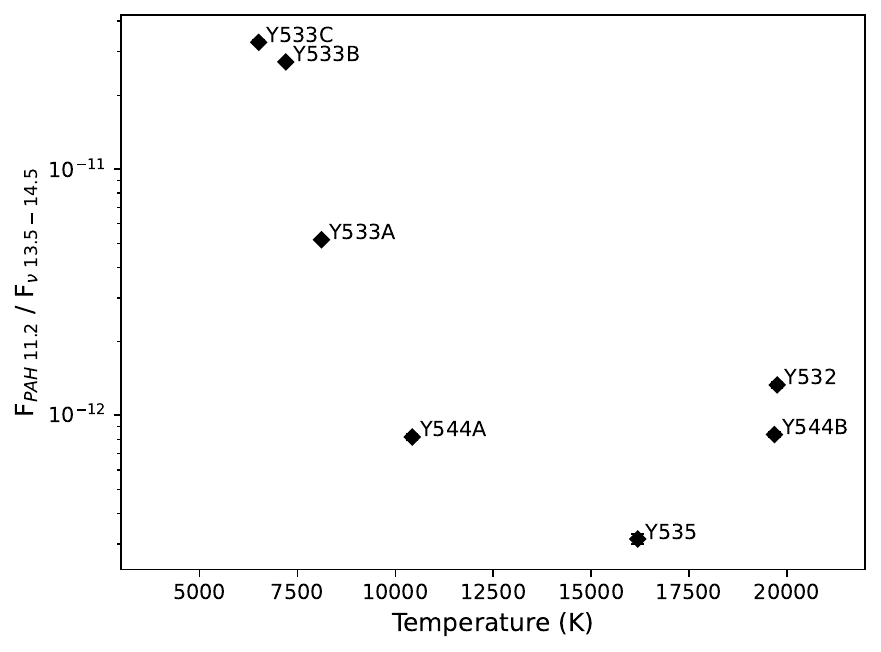}}
 {\includegraphics[width=0.45\textwidth]{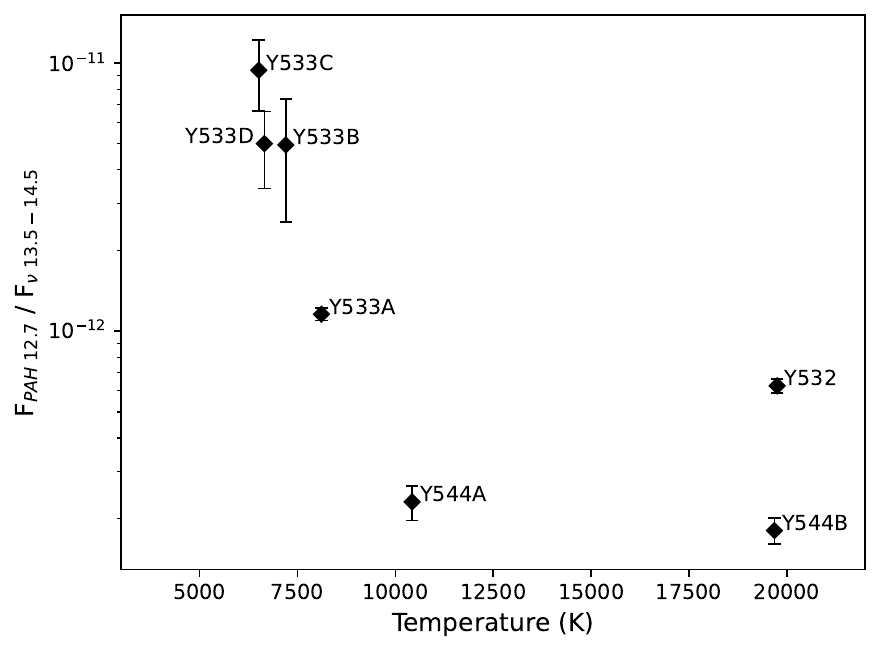}}
  \caption{YSO effective temperature versus 11.2 (\textbf{top}) and 12.7$\mu$m (\textbf{bottom}) PAH feature strength. Effective temperatures are determined through SED-fitting to model YSOs (see \autoref{sec:SED_fit}). PAH strength for each source is normalized by dividing the total emission by the median continuum emission between 13.5 and 14.5$\mu$m.}
  \label{fig:pah_strengths}
\end{figure}

Emission from the 7.7~$\mu$m PAH feature is seen in sources Y532 and Y533A, though more weakly than other PAH features. In both sources, this emission feature has a double-peaked profile. This profile is slightly most pronounced in source Y532. In Y533A, the two components are more difficult to separate visually. These two are the only sources out of the five with spectra covering 7.7~$\mu$m for which this feature is detected. 
Similarly, these two sources are the only ones in our sample with a discernible 8.6~$\mu$m emission. Additionally, they show the strongest overall PAH emission relative to their continua of these five sources (see \autoref{fig:pah_strengths}). Moreover, they show the greatest number of PAH emission lines in our sample. This points to a higher overall dust content in their envelopes.

We detect no emission owing to the 11.0~$\mu$m PAH feature in any source in our sample. In theory, the ionized PAH state that emits in the 5-10~$\mu$m should also give rise to emission at 11.0~$\mu$m \citep{bib:hudg14}. However, comparable MRS observations of massive LMC protostars showed that this emission feature is typically weaker than 8.6~$\mu$m \citep{bib:nayak2024}, and in our case may be below detectable levels. Additionally, \cite{bib:peet17} found 11.0 and 8.6~$\mu$m emission to be correlated, with the CH in-plane bending mode of the H atom generating the 8.6~$\mu$m line and the out-of-plane bending mode generating the 11.0~$\mu$m line. They also find a close correlation between the 7.6~$\mu$m and 11.0~$\mu$m lines. Given that we find only two weak detections of 7.6 and 8.6~$\mu$m in our sample, our results are consistent with the emission from these three lines being linked in origin, albeit with any 11.0~$\mu$m emission being below detectable levels. 


Each source in our sample displays a prominent 11.2~$\mu$m PAH feature in its spectra (except Y533D, which has an incomplete spectrum in this wavelength range as shown in \autoref{fig:targ_spec_compare_Y533}). For each source, this feature is the strongest of any present PAH emission. Similarly, all but one source (Y535) display the 12.7~$\mu$m feature, though fainter than 11.2~$\mu$m. Finally, the 16.4~$\mu$m feature is faintly present in five sources (Y544B and Y533A through E). Sources Y532 and Y544A also appear to have a hint of emission at this wavelength, but it cannot be confidently identified against the continuum noise.

\subsubsection{PAHs and YSO Evolution}

\vspace{-.1in}

PAH features across the mid-IR are linked to grains of different sizes, which in turn, can indicate the strength of the UV radiation emitted source, as well as the proximity of the dust grains in the local environment \citep{bib:baus08}. The largest grains emit at 7.8~$\mu$m, which are broken down into smaller grains emitting at 11.2~$\mu$m. Nearer to the central ionizing source, these are again reduced further, emitting at 12.7 and 16.4~$\mu$m. Finally, closest to the central YSO, the 6.2~$\mu$m emission is generated, and small grains emit at 7.6, 8.6, and 11.0~$\mu$m. 

The three most massive and luminous sources, Y535, Y544A and Y544B, show no evidence of the largest grains at 7.7~$\mu$m, and are consistent with bright YSOs with intense UV radiation that has processed the dust grains in their local environment. PAH occurrences in the remaining YSOs indicate they possess a mixture of grain sizes.

Comparing the relative strength of PAH features also reveals differences in dust content. To determine relative PAH strengths, we normalize the PAH emission by dividing by the median continuum level between 13.5 and 14.5~$\mu$m, which is relatively flat and free of strong emission lines in each spectra. In \autoref{fig:pah_strengths}, we plot the SED-derived YSO effective temperature vs the strengths for the 11.2 and 12.7~$\mu$m PAH features. We observe the strongest PAH emission in the coolest, and likely least evolved sources with some scatter for the hottest and most luminous. 

In general, we find PAH features are more numerous and stronger in our cooler, least evolved sources. This aligns with the expectation that such YSOs are still embedded in denser, and hence dustier protostellar envelopes, and are earlier in the process of breaking down the PAH grains than their hotter, more evolved and more luminous counterparts.


We note that the SMC sources in our sample are relatively PAH-poor when compared to the massive, LMC YSOs studied with JWST's MRS by \cite{bib:nayak2024}. Though the differing mass range and evolutionary states between the two samples make for an indirect comparison, we find no detectable 11.0~$\mu$m emission as was seen in the LMC, and overall weaker PAH features relative to the continua. This may indicate that our SMC YSOs are at a later evolutionary state, having reprocessed and destroyed the PAH molecules in their envelopes. Alternatively, the stronger presence of PAHs in the LMC's more massive, likely-UV-radiating YSOs may be attributable to the greater overall metallicity, and therefor dust content, in the LMC's N79 region (0.5~Z$_{\odot}$, \citealt{bib:Westerlund1997}) than in the SMC's NGC 346 (0.2~Z$_{\odot}$, \citealt{bib:Peimbert2000}).

\begin{figure}
  \centering
  {\includegraphics[width=0.45\textwidth]{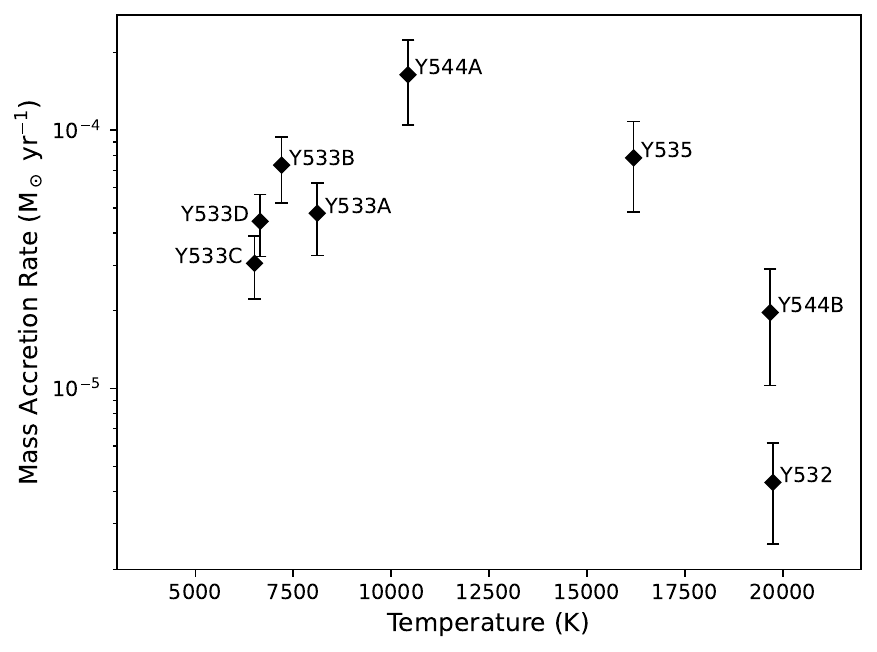}}
 {\includegraphics[width=0.45\textwidth]{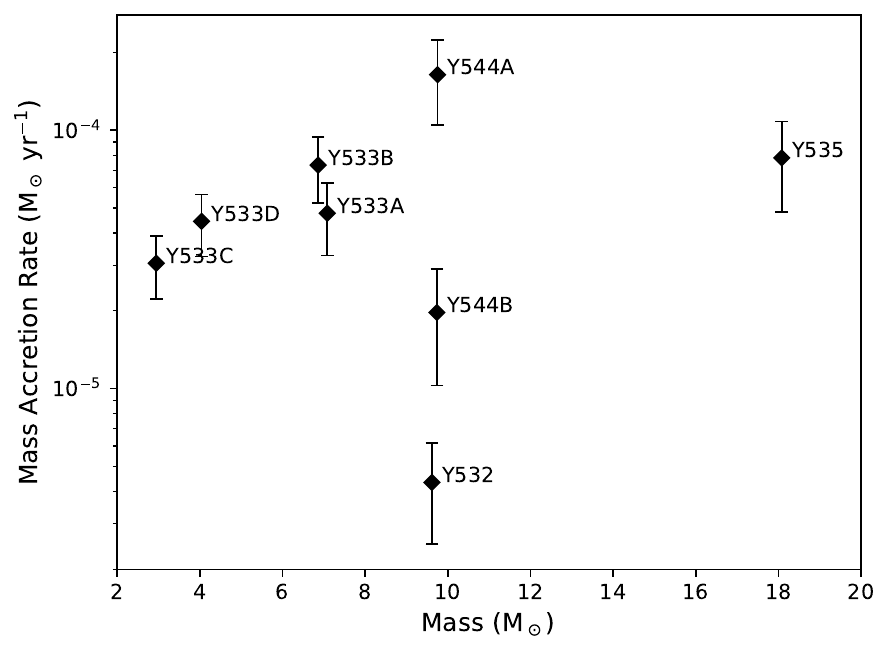}}
  \caption{YSO effective temperature (\textbf{top}) and mass (\textbf{bottom}) versus mass accretion rate. The error bars represent the upper and lower limits estimated using the relations in \autoref{eq:luminosity_alc} and \autoref{eq:mass_rate}, with points showing their averages.}
  \label{fig:acc_rates}
\end{figure}

\subsection{Accretion Rate Estimates from \ion{H}{1} Emission}
\label{sec:acc_rates}

Our MRS observations show hydrogen recombination lines are present in each object's spectra. Recombination lines are often indicators of active accretion, and as such may be used to infer the rate of mass accretion. The \ion{H}{1} (7-6) recombination line (Humphreys $\alpha$) at 12.37~$\mu$m is detected in all of our sources.

From this recombination line, we infer the accretion luminosity of our sources. Using \textit{Spitzer} IRS observations of T-tauri stars, \cite{bib:rigl15} found a close-to-quadratic relation between accretion luminosity and \ion{H}{1} (7-6) line luminosity. This relation is described by the equation:

\begin{equation}
    \mathrm{log} \frac{L_{HI(7-6)}}{L_{\odot}} = (0.48 \pm 0.09) \times \mathrm{log} \frac{ L_{acc}}{L_{\odot}} - (4.68 \pm 0.10).~
    \label{eq:luminosity_ric}
\end{equation}

However, this relation was calibrated for low-mass Milky Way T-Tauri stars and may differ for more-massive, embedded YSOs in low-metallicity regions like the LMC and the SMC. Additionally, this nearly-quadratic dependence is at odds with other studies which find a nearly linear relationship for multiple recombination lines. Using the VLT's X-shooter instrument, \cite{bib:alca14} observed the Brackett, Balmer and Paschen lines in 36 stellar and substellar YSOs and found a close-to-linear dependence (see Table 4 of their work). We find that adopting the 0.48 factor from \cite{bib:rigl15} leads to unphysically fast rates of mass accretion in our sources: as great as \num{7.16E-02}–\num{3.12E+01} M$_{\odot}~$yr$^{-1}$. At their upper limits, these accretion rates are significantly higher than the upper limit of mass accretion observed in Milky Way YSOs: $\sim$\num{E-04} M$_{\odot}~$yr$^{-1}$ \cite{bib:rigl15}. Thus in this work we report more conservative estimates of accretion luminosity and mass accretion rates by setting this factor to 1 (\autoref{eq:luminosity_alc}) in line with \cite{bib:alca14}. This same estimation method was used by \cite{bib:nayak2024} for several massive ($<$13~M$_{\odot}$) embedded YSOs in the LMC.

\begin{equation}
    \mathrm{log} \frac{L_{HI(7-6)}}{L_{\odot}} = \mathrm{log} \frac{ L_{acc}}{L_{\odot}} - (4.68 \pm 0.10).~
    \label{eq:luminosity_alc}
\end{equation}

We find \ion{H}{1} (7-6) emission line luminosities for our sources ranging from \num{4.71e-17}-\num{5e-16} $\mathrm{erg\;s^{-1}\;cm^{-2}}$. Accounting for the error in both our measured flux and in \autoref{eq:luminosity_alc}, we calculate the accretion luminosity range for each source (\autoref{tab:acc_rates}).
For Y544A, with the greatest accretion luminosity, we find \num{1.95E+03}-\num{4.15E+03} L$_{\odot}$; for Y532, our least luminous, we find a luminosity of \num{1.68E+02}-\num{4.16E+02} L$_{\odot}$.

\begin{table*}
\setlength{\tabcolsep}{15pt}
    \centering
    \begin{tabular}{||l|c|c||}
    \hline
        Source ID  & Accretion Luminosity & Accretion Rate \\
        & (using \autoref{eq:luminosity_alc}) & (using \autoref{eq:luminosity_alc}) \\
        & (L$_{\odot}$) & (M$_{\odot}~$yr$^{-1}$) \\

    \hline
    \hline

Y535		& 1.36E+03 -- 3.04E+03		& 4.83E-05 -- 1.08E-04   \\			
Y544A		& 1.95E+03 -- 4.15E+03		& 1.05E-04 -- 2.23E-04   \\
Y544B		& 6.82E+02 -- 1.93E+03		& 1.03E-05 -- 2.91E-05   \\
Y532		& 1.68E+02 -- 4.16E+02		& 2.50E-06 -- 6.17E-06   \\
Y533A		& 4.68E+02 -- 8.94E+02		& 3.28E-05 -- 6.26E-05   \\
Y533B		& 6.03E+02 -- 1.09E+03		& 5.23E-05 -- 9.42E-05   \\
Y533C		& 3.94E+02 -- 6.91E+02		& 2.22E-05 -- 3.89E-05   \\
Y533D		& 4.76E+02 -- 8.24E+02		& 3.25E-05 -- 5.63E-05   \\
Y533E		& 4.16E+02 -- 7.22E+02		& 8.21E-05 -- 1.43E-04   \\
\hline
    \end{tabular}
    \caption{The accretion luminosities and mass accretion rates for our sources calculated from the 12.37~$\mu$m \ion{H}{1} (7-6) emission line.}
    \label{tab:acc_rates}
\end{table*}

From the accretion luminosity, we can employ \autoref{eq:mass_rate}, following \cite{bib:gull98}, to obtain an estimate of the mass accretion rate $\dot{M}_{acc}$. 

\begin{equation}
    \mathrm{\dot{M}_{acc} \backsimeq  \frac{L_{acc}R_{star}}{G M_{star}} \left( 1-\frac{R_{star}}{R_{m}} \right)^{-1}}.
    \label{eq:mass_rate}
\end{equation}

This equation balances the accretion luminosity generated by shocks at the stellar surface with the gravitational energy lost by the material as it infalls from the inner magnetospheric disk radius $R_m$ to the stellar radius $R_{star}$. We further simplify this equation by setting the term in parentheses, which is on the order of unity, to 1. We adopt the stellar radii and masses derived by our SED fitting (\autoref{sec:SED_fit}, \autoref{tab:fit_params}) and the estimated accretion luminosities to calculate a range of accretion rates. 
Accordingly, we find that Y544A has the fastest accretion, ranging from 
\num{1.05E-04}–\num{2.23E-04} M$_{\odot}~$yr$^{-1}$,
and Y532 has the slowest at \num{2.50E-06}–\num{6.17E-06} M$_{\odot}~$yr$^{-1}$.
In \autoref{tab:acc_rates}, we list the estimated accretion luminosities and mass accretion rates for each source.

We compare these accretion rates to the effective temperatures and masses of our sources in \autoref{fig:acc_rates}.
Sources Y544A, Y544B and Y532 are nearly identical in mass at $\sim$10~M$_{\odot}$, but vary in temperature from $\sim$10,000K (Y544A) to $\sim$20,000K (Y544B and Y532). Among these three, we find slower accretion rates for the hotter sources, in line with the expectation that accretion lessens as YSOs evolve and heat up. This is further supported by comparing the hotter and more-massive Y535, with the cooler and lower-mass Y533A, Y533B, Y533C and Y533D, all five of which share similar accretion rates ($\sim$\num{2E-04}–\num{8E-04} M$_{\odot}~$yr$^{-1}$).

Relating the disk mass parameter output by our fitted SED models \citep{bib:robi17} to the mass accretion rate (and assuming a typical formation timescale of \num{E05}~yr) gives significantly lower estimates for accretion rates. Dividing the disk mass by the timescale yields rates of 
\num{1.16e-11}, \num{2.85e-11}, \num{4.305e-14}, \num{7.269e-12}, \num{9.66e-08 }, \num{2.00e-08}, \num{2.648e-08}, \num{5.36e-09} and \num{6.42e-11} M$_{\odot}~$yr$^{-1}$ for sources Y535, Y544A, Y544B, Y532, Y533A, Y533B, Y533C, Y533D and Y533E, 
respectively. The comparatively higher rates we find from Hydrogen recombination line emission suggest that our sources are at an early stage in their evolution when accretion proceeds more rapidly. Protostellar accretion is also observed to be variable, with short periods of increased activity leading to elevated luminosity. Alternatively, this may explain the high rates we observe, though with a single observation epoch, we cannot assess if this is occurring in our sources.

In their MRS observations of the massive YSO clusters in the LMC region N79, \cite{bib:nayak2024} noted that \ion{H}{1} (7-6) line luminosity may be inflated by environmental effects like strong winds and ionizing UV radiation for the most luminous sources. Such an effect may also be present for our targets especially in our brightest and most massive source, Y535. Comparing Y535 (18.0 $M_{\odot}$) to the massive YSO of \citeauthor{bib:nayak2024} (``Y1", $\sim13~M_{\odot}$), we observe \ion{H}{1} (7-6) line luminosity approximately 10$\%$ that of Y1, and accretion luminosity and accretion rates roughly an order of magnitude lower. Stellar winds can be variable and lessen over time as accretion diminishes, potentially owing to this difference, though both sources are still young and likely in the strongest stages of mass accretion and thus stellar wind activity. JWST's NIRSpec could further characterize additional hydrogen recombination lines in the near-IR to better constrain these accretion rates as well as help to calibrate the relation between mid-IR recombination line emission and accretion luminosity. Recent work has successfully used NIRSpec's multi-object spectrometer, (MOS) to characterize the the accretion rates of several $\sim$1 M$_{\odot}$ disk-bearing pre-main-sequence stars in NGC 346 by their H$\alpha$ emission, finding accretion rates of \num{1.7E-9}-\num{6.0E-6}~M$_{\odot}$yr$^{-1}$ \citep{bib:demarchi24_pms}.

\section{Summary and Conclusions}  
\label{sec:conclusion}

We conducted four JWST MIRI MRS observations in the low-metallicity SMC star forming region NGC 346, observing five intermediate- to high-mass YSOs over all four mid-IR wavelength channels from 4.90–27.90~$\mu$m, and four other likely-YSO sources over partial spans of this wavelength range. We summarize our findings:

\begin{itemize}

\item{ We extract full MRS spectra (4.9 to 27.9~$\mu$m) for five primary YSOs: Y535, Y544A, Y544B, Y532 and Y533A. For the three additional likely-YSOs (Y533B, Y533C and Y533D) partially captured by the FOV of one MRS pointing, we extract partial spectroscopy. We combine our MRS spectra with existing photometry from JWST, \textit{Spitzer} and \textit{Herschel} to construct SEDs extending across 1.1 to 350~$\mu$m. Via SED model fitting, we infer the masses of our five primary sources to be 18.0, 9.8, 9.7, 9.6 and 7.1~M$_{\odot}$, respectively, and the three partial sources to be 6.9, 2.9 and 4.0}~M$_{\odot}$, respectively.

\item{Each YSO spectrum shows a mix of line emission from atomic hydrogen recombination, molecular hydrogen and atomic fine-structure. \ion{H}{1} lines in each source indicate the presence of ongoing accretion. H$_2$ lines are likewise seen in all sources, suggesting either low-velocity shocks by outflows or UV heating in the local environment. We find H$_2$  17.03~$\mu$m/5.51~$\mu$m line ratios for sources Y535, Y544A, Y544B, Y532 and Y533A greater than unity, indicating these sources are still young and embedded in cool envelopes. Fine-structure emission (e.g. [\ion{Ne}{2}] 12.8~$\mu$m, [\ion{Ne}{3}] 15.5~$\mu$m, [\ion{Ar}{2}] 6.9~$\mu$m, [\ion{Ar}{3}] 8.9~$\mu$m \& 21.8~$\mu$m, [\ion{Fe}{2}] 17.9~$\mu$m \& 25.9~$\mu$m) indicate the presence of high-velocity ($>$~70~km~s$^{-1}$) shocks.}

\item{We detect extended fine-structure (e.g. 6.9~$\mu$m [\ion{Ar}{2}], 18.7~$\mu$m [\ion{S}{3}]), \ion{H}{1} and H$_2$ emission tracing a potential $\sim$~30,000 AU protostellar outflow about the source Y535. Line emission from Y535 also indicates the presence of both high- and low-velocity shocks from a jet or wind. We suggest that this is the first detection of a resolved protostellar outflow in the SMC, one of only a handful detected outside the Milky Way and the furthest yet detected.}

\item{We detect various PAH features in each YSO, detecting emission at 6.2, 7.7, 8.6, 11.2, 12.7 and 16.4~$\mu$m across our sample. Our more massive and more evolved sources show weaker and fewer PAH features as well as a preference for those arising from smaller grain sizes, suggesting that these sources are destroying the PAHs with strong radiation and stellar feedback. The young, less-evolved source Y533A, shows the strongest relative PAH emission, including that from larger grains, and potential 10~$\mu$m silicate in absorption, indicative of a young, dust-enriched envelope.}

\item{We detect the 15.2~$\mu$m CO$_2$ ice feature in absorption in four of our YSOs, Y535, Y533A, Y533C and Y533D. In Y533A, we find a doublet profile in this absorption. By fitting this feature with multiple components, we infer contributions from pure CO$_2$ ice, CO$_2$ mixed with CO, CO$_2$ mixed with H$_2$O and CO$_2$ mixed with CH$_3$OH. The relatively few molecular ice absorption features we detect in comparison to the higher-metallicity Milky Way and LMC may be a consequence of the SMC's lower metallicity.}

\item{We find \ion{H}{1} emission lines in the spectra of each YSO we observe. From the 12.37~$\mu$m \ion{H}{1} (7-6) line, we estimate accretion luminosities and mass accretion rates for each source. These rates span from \num{1.05E-04}–\num{2.23E-04} M$_{\odot}~$yr$^{-1}$ at the fastest in Y544A to \num{2.50E-06}–\num{6.17E-06} M$_{\odot}~$yr$^{-1}$ at the slowest in Y532. We find a trend toward faster rates of accretion in our coolest, least evolved sources.}

\end {itemize}

\clearpage
\section{Acknowledgements}
This work is based on observations made with the NASA/ESA/CSA James Webb Space Telescope. The data were obtained from the Mikulski Archive for Space Telescopes at the Space Telescope Science Institute, which is operated by the Association of Universities for Research in Astronomy, Inc., under NASA contract NAS 5-03127 for {\em JWST}. These observations are associated with program \#1227.
The data for the observations discussed in this work can be accessed via~\dataset[10.17909/qena-8a58]{http://dx.doi.org/10.17909/qena-8a58}.
NH and MM acknowledge that a portion of their research was carried out at the Jet Propulsion Laboratory, California Institute of Technology, under a contract with the National Aeronautics and Space Administration (80NM0018D0004).  NH and MM acknowledge support through NASA/{\em JWST} JPL Task Plan No. 71-209636.
OCJ acknowledges support from an STFC Webb fellowship. 
CN acknowledges support from an STFC studentship.
KF acknowledges support through the ESA Research Fellowship.
LL acknowledges support from the NSF through grant 2054178.
ASH is supported in part by an STScI Postdoctoral Fellowship.
ON acknowledges the NASA Postdoctoral Program at NASA Goddard Space Flight Center, administered by Oak Ridge Associated Universities under contract with NASA. Additionally, ON was supported by the director's discretionary fund as a postdoctoral fellow at STScI.  
© 2024 Jet Propulsion Laboratory. All rights reserved.

\vspace{5mm}
\facilities{JWST: NIRCAM, MIRI Imager, MIRI MRS; \textit{HST}: ACS; \textit{Spitzer}: IRAC, MIPS; \textit{Hershel}: PACS, SPIRE}

\software{
        astropy \citep{bib:Astropy2013,bib:astropy2018,bib:astropy2022},  
        }



\appendix


\section{Additional Tables}
\label{sec:appendix_tables}
In this section, we present additional tables. We list the aperture values adopted for various instruments and filters  used for YSO SED fitting. We include a table of the detected spectral features for source Y533E which is likely contaminated by emission from Y533A.

\begin{table*}[h]
\centering
\begin{tabular}{|cc||cc||cc||cc|}
\hline
NIRCam & Aperture & MIRI & Aperture & \textit{Spitzer}  & Aperture & \textit{Herschel} & Aperture\\
Filter & (Arcseconds) & Filter & (Arcseconds) & Filter & (Arcseconds) & Filter & (Arcseconds) \\
\hline
F115W & 0.040 & F560W  & 0.207 & IRAC I1  & 1.7  & PACS green & 8.8  \\
F187N & 0.064 & F770W  & 0.269 & IRAC I2  & 1.7  & PACS red   & 12.6 \\
F200W & 0.066 & F1000W & 0.328 & IRAC I3  & 1.9  & SPIRE PSW  & 18.3 \\
F277W & 0.092 & F1130W & 0.375 & IRAC I4  & 2.0  & SPIRE PMW  & 26.7 \\
F335M & 0.111 & F1280W & 0.420 & MIPS 24  & 6.0  &   -        &  -   \\
F444W & 0.145 & F1500W & 0.488 & MIPS 70  & 18.0 &   -        &  -   \\
   -  &  -    & F1800W & 0.591 &    -     &  -   &   -        &  -   \\
   -  &  -    & F2100W & 0.674 &    -     &  -   &   -        &  -   \\
   -  &  -    & F2550W & 0.803 &    -     &  -   &   -        &  -   \\
\hline
\end{tabular}
\caption{Table of the aperture values adopted for the YSO SED fitting in this work.}
\label{tab:apertures}
\end{table*}

\begin{deluxetable*}{lcccccc}
\tablecaption{Spectral Features for Source Y533E}
\tablehead{
\colhead{Name} & \colhead{Lab Wave} & \colhead{Meas Wave}& \colhead{Meas Wave Err} & \colhead{FWHM} & \colhead{Flux} & \colhead{Flux Err} 
\\
\colhead{}  & \colhead{($\mu$m)} & \colhead{($\mu$m)} & \colhead{($\mu$m)} & \colhead{($\mu$m)} & \colhead{($\mathrm{erg\;s^{-1}\;cm^{-2}}$)} & \colhead{($\mathrm{erg\;s^{-1}\;cm^{-2}}$)}
}
\decimals
\startdata
H2 0-0 S(7) & 5.514 & 5.514 & 0.00171 & 0.016073 & 2.23E-17 & 1.48E-18 \\
H2 7-6 O(8) & 6.602 & 6.602 & 0.00207 & 0.002356 & 1.60E-17 & 1.28E-18 \\
H2 0-0 S(5) & 6.913 & 6.914 & 0.00198 & 0.000740 & 7.07E-17 & 2.04E-18 \\
{[ArII] 2P1/2-2P3/2} & 6.989 & 6.989 & 0.00206 & 0.000103 & 2.60E-17 & 1.49E-18 \\
H  I  6- 5 & 7.464 & 7.464 & 0.00201 & 0.000165 & 8.57E-17 & 2.30E-18 \\
H  I  8- 6 & 7.507 & 7.506 & 0.00220 & 0.000664 & 2.68E-17 & 1.62E-18 \\
H2 0-0 S(4) & 8.029 & 8.029 & 0.00260 & 0.000625 & 5.90E-17 & 2.97E-18 \\
H  I 10- 7 & 8.765 & 8.765 & 0.00359 & 0.002202 & 2.07E-18 & 1.90E-18 \\
{[ArIII] 3P1-3P2} & 8.996 & 8.996 & 0.00315 & 0.000522 & 2.95E-16 & 5.18E-18 \\
H2 0-0 S(3) & 9.670 & 9.670 & 0.00267 & 0.000079 & 3.10E-16 & 4.84E-18 \\
{[SIV] 2P3/2-2P1/2} & 10.516 & 10.516 & 0.00309 & 0.000330 & 1.81E-15 & 1.99E-17 \\
PAH & 11.2 & 11.281 & 0.00084 & 0.19173 & 6.70E-15 & 9.42E-17 \\
{[NiI] a3F2-a3F3} & 11.314 & 11.315 & 0.00304 & 0.002455 & 1.20E-17 & 5.38E-18 \\
H2 0-0 S(2) & 12.285 & 12.286 & 0.00433 & 0.000633 & 3.92E-16 & 8.50E-18 \\
H  I  7- 6 & 12.379 & 12.379 & 0.00386 & 0.000279 & 9.54E-17 & 4.39E-18 \\
H  I 11- 8 & 12.394 & 12.394 & 0.00426 & 0.000267 & 1.68E-17 & 3.76E-18 \\
H  I 14- 9 & 12.594 & 12.594 & 0.00318 & 0.000027 & 2.12E-18 & 3.62E-18 \\
PAH & 12.7 & 12.758 & 0.02718 & 0.220077 & 1.48E-15 & 5.83E-16 \\
{[NeII] 2P1/2-2P3/2} & 12.821 & 12.821 & 0.00414 & 0.000411 & 5.87E-16 & 8.97E-18 \\
HeII 11-10 & 13.136 & 13.134 & 0.00429 & 0.000853 & 7.29E-18 & 3.48E-18 \\
{[NeIII] 3P1-3P2} & 15.564 & 15.564 & 0.00649 & 0.001046 & 4.90E-15 & 3.65E-17 \\
H  I 10- 8 & 16.218 & 16.219 & 0.00541 & 0.000603 & 2.31E-17 & 4.56E-18 \\
PAH & 16.4 & 16.428 & 0.00350 & 0.146197 & 7.63E-16 & 5.81E-17 \\
H2 0-0 S(1) & 17.044 & 17.044 & 0.00564 & 0.000823 & 9.42E-16 & 1.07E-17 \\
{[PIII] 2P3/2-2P1/2} & 17.895 & 17.896 & 0.00602 & 0.000061 & 2.19E-17 & 6.03E-18 \\
{[SIII] 3P2-3P1} & 18.723 & 18.723 & 0.00845 & 0.000356 & 4.34E-15 & 9.48E-17 \\
H  I  8- 7 & 19.072 & 19.071 & 0.00955 & 0.001859 & 8.02E-17 & 3.30E-17 \\
{[ArIII] 3P0-3P1} & 21.842 & 21.845 & 0.00952 & 0.001747 & 1.34E-16 & 3.83E-17 \\
\enddata
\tablecomments{Column 1: Name of line. Column 2: Laboratory wavelength. Column 3: Measured wavelength. Column 4: Error in measured wavelength. Column 5: FWHM of line. Column 5: Measured flux. Column 6: Error in flux. All features are in emission.}
\label{tab:e_lines_Y533E}
\end{deluxetable*}


\FloatBarrier
\bibliography{ngc346_mrs.bib}{}
\bibliographystyle{aasjournal}



\end{document}